\documentclass[a4paper,twocolumn]{aastex63}
\usepackage{graphicx}
\pdfoutput=1

\def\kms{\hbox{km s$^{-1}$}}
\def\VLSR{\hbox{$V_{\rm LSR}$}}
\def\R32_10{\hbox{$R_{3\mbox{--}2/1\mbox{--}0}$}}

\received{}
\revised{}
\accepted{}

\submitjournal{ApJS}

\shorttitle{High Velocity Dispersion Compact Clouds}
\shortauthors{Oka et al.}
\usepackage{amsmath}
\usepackage[dvipdfmx]{}
\usepackage{graphicx}
\usepackage{here}
\usepackage{longtable}
\usepackage[online]{threeparttablex}
\usepackage{color}

\begin{document}
\title{Catalog of High Velocity Dispersion Compact Clouds in the Central Molecular Zone of Our Galaxy 
}

\author[0000-0002-5566-0634]{Tomoharu Oka}
\affiliation{Department of Physics, Faculty of Science and Technology, Keio University, 3-14-1 Hiyoshi, Kohoku-ku, Yokohama, Kanagawa 223-8522, Japan}
\affiliation{School of Fundamental Science and Technology, Graduate School of Science and Technology, Keio University, 3-14-1 Hiyoshi, Kohoku-ku, Yokohama, Kanagawa 223-8522, Japan}

\author{Asaka Uruno}
\affiliation{School of Fundamental Science and Technology, Graduate School of Science and Technology, Keio University, 3-14-1 Hiyoshi, Kohoku-ku, Yokohama, Kanagawa 223-8522, Japan}

\author[0000-0003-2735-3239]{Rei Enokiya}
\affiliation{Department of Physics, Faculty of Science and Technology, Keio University, 3-14-1 Hiyoshi, Kohoku-ku, Yokohama, Kanagawa 223-8522, Japan}

\author{Taichi Nakamura}
\affiliation{School of Fundamental Science and Technology, Graduate School of Science and Technology, Keio University, 3-14-1 Hiyoshi, Kohoku-ku, Yokohama, Kanagawa 223-8522, Japan}

\author{Yuto Yamasaki}
\affiliation{School of Fundamental Science and Technology, Graduate School of Science and Technology, Keio University, 3-14-1 Hiyoshi, Kohoku-ku, Yokohama, Kanagawa 223-8522, Japan}

\author{Yuto Watanabe}
\affiliation{School of Fundamental Science and Technology, Graduate School of Science and Technology, Keio University, 3-14-1 Hiyoshi, Kohoku-ku, Yokohama, Kanagawa 223-8522, Japan}

\author[0000-0003-2636-0365]{Sekito Tokuyama}
\affiliation{School of Fundamental Science and Technology, Graduate School of Science and Technology, Keio University, 3-14-1 Hiyoshi, Kohoku-ku, Yokohama, Kanagawa 223-8522, Japan}

\author[0000-0002-9255-4742]{Yuhei Iwata}
\affiliation{School of Fundamental Science and Technology, Graduate School of Science and Technology, Keio University, 3-14-1 Hiyoshi, Kohoku-ku, Yokohama, Kanagawa 223-8522, Japan}
\affiliation{Center for Astronomy, Ibaraki University, 2-1-1 Bunkyo, Mito, Ibaraki 310-8512, Japan}

\begin{abstract}
This study developed an automated identification procedure for compact clouds with broad velocity widths in the spectral line data cubes of highly crowded regions. The procedure was applied to the CO {\it J}=3--2 line data, obtained using the James Clerk Maxwell Telescope, to identify 184 high velocity dispersion compact clouds (HVCCs), which is a category of peculiar molecular clouds found in the central molecular zone of our galaxy. A list of HVCCs in the area $-1\fdg 4\! \leq\! l \! \leq\! +2\fdg 0$, $-0\fdg 25\! \le\! b\! \le\! +0\fdg 25$ was presented with their physical parameters, CO {\it J}=3--2/{\it J}=1--0 intensity ratios, and morphological classifications. Consequently, the list provides several intriguing sources that may have been driven by encounters with point-like massive objects, local energetic events, or cloud-to-cloud collisions.
\end{abstract}

\section{Introduction} 
High velocity dispersion compact clouds (HVCCs)\footnote{These were referred to as `high-velocity compact clouds' in our previous papers.  The term originated from its first example, CO 0.02--0.02, which has a large systemic velocity as well as a large velocity dispersion \citep{Oka99}.  Here we rename it learning that many of them do not have large systemic velocities.}
are a category of peculiar molecular clouds found in the central molecular zone (CMZ) of our galaxy \citep[e.g.,][]{Oka98b}. Approximately 120 HVCCs have been identified in the CO {\it J}=1--0 data obtained using the 45 m telescope at Nobeyama Radio Observatory (NRO) as well as in the CO {\it J}=3--2 data obtained using the Atacama Submillimeter Telescope Experiment (ASTE) \citep[e.g.,][]{Nagai08}. They are characterized by their compact appearance ($d\!<\! 5$ pc) and extraordinary broad velocity widths ($\Delta V\!\ge\! 50$ \kms ). 
Their compact sizes differ sharply from those of the seven wide line clouds \citep[WLCs; e.g.,][]{Kumar97, Liszt06}, which were previously identified at Galactic longitudes between $-6\arcdeg$ and $+6\arcdeg$, suggesting different formation mechanisms.
Most of HVCCs have no apparent associated energy sources. Some of them exhibit very high CO {\it J}=3--2/{\it J}=1--0 intensity ratios \citep[$\ge\! 1.5$;][]{Oka07, Oka12}, suggesting that they have been influenced by certain type of heating and/or compression processes.

Few HVCCs are associated with clearly expanding shell structures. CO 1.27+0.01 \citep{Oka01a} and CO--1.21--0.12 \citep{Oka12, Tsujimoto18} are representatives of such ``shell-type" HVCCs. Such HVCCs possess enormous kinetic energy ($\gtrsim\! 10^{50}$ erg) and multiple expanding shells, indicating that it was formed by a series of supernova explosions in the embedded massive stellar cluster. HVCC CO 0.02--0.02 is associated with a clear emission cavity in its southwest, wherein a group of point-like infrared sources is included \citep{Oka08}. This could be another shell-type HVCC accelerated by supernovae, although the cavity does not exhibit a clear expanding motion \citep{Oka99}.

\begin{figure*}[htbp]
\begin{center}
\includegraphics[scale=0.8, width=15cm]{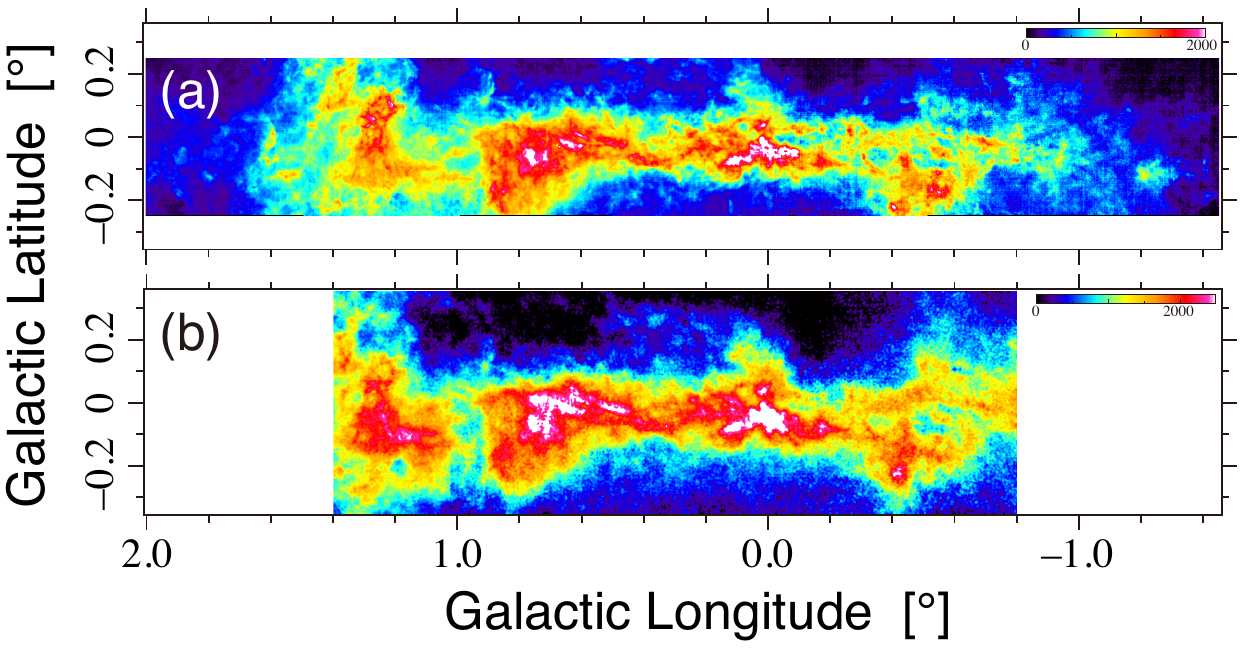}
\end{center}
\label{fig:data}
\caption{The CO data sets used in this study. Integrated intensity ($\int T_{\mathrm{R}}^* dV$) maps of (a) CO {\it J}=3--2, and (b) CO {\it J}=1--0 line emission. The emission was integrated over velocities between $\VLSR\! =\! -220$ and $+220\,\kms $.}
\end{figure*} 

Further, HVCCs with featureless spatial and position-velocity distributions were observed. Certain of such ``simple-type" HVCCs could be explained by gravitational interaction with an invisible, massive compact object. CO--0.40--0.22 \citep[originally CO--0.41--0.23 in][]{Oka07, Oka12} is the first example of such a gravitationally kicked HVCC, which may harbor an intermediate-mass black hole (IMBH) with a mass of $10^5\,M_{\sun}$ \citep{Oka16, Oka17}. Subsequently, HCN--0.009--0.044 \citep{Takekawa17, Takekawa19a}, CO--0.31+0.11 \citep{Takekawa19b}, and HCN--0.085--0.094 \citep{Takekawa20} were considered as candidates for HVCCs that include IMBHs. Compact broad-velocity-width features were also detected in the galactic disk. The ``Bullet" in the W44 molecular cloud \citep{Sashida13} exhibits a clear `Y'-shape in position-velocity maps, indicating that it may have been accelerated by a high-velocity plunge of a compact massive object \citep{Yamada17, Nomura18}. Further, \citet{Yokozuka21} reported the discovery of a compact ($d\!\simeq\! 3$ pc) broad-velocity width ($\Delta V\!\simeq\! 25$ \kms ) molecular feature without an apparent driving source at $(l, b)\!\simeq\!(+16\fdg 134, -0\fdg 553)$, which may be an analog of HVCCs. 

Many studies on HVCCs and related matters have been published to date. In particular, recent important results regarding the origin of HVCCs \citep[for example,][]{Oka16, Yamada17, Takekawa17, Takekawa19a} have increased attention towards HVCCs. However, no statistical study based on a systematic search for HVCCs has yet been performed, with the only study by \citet{Nagai08} being based on spatially undersampled CO {\it J}=1--0 data \citep{Oka98b} and by eye identification of HVCCs. To provide a complete list of HVCCs in the CMZ, an automated identification procedure was developed. This paper presents the details of the HVCC identification procedure and subsequent results obtained after applying this procedure to the CO {\it J}=3--2 line data cube, wherein warm/dense molecular gas was highlighted more than in the CO {\it J}=1--0 line. In this study, $D_{\rm GC}\! =\! 8.3$ kpc was employed as the distance to the Galactic center \citep{Gillessen09, Gravity22}. 

\begin{figure*}[htbp]
\begin{center}
\includegraphics[scale=0.8, width=15cm]{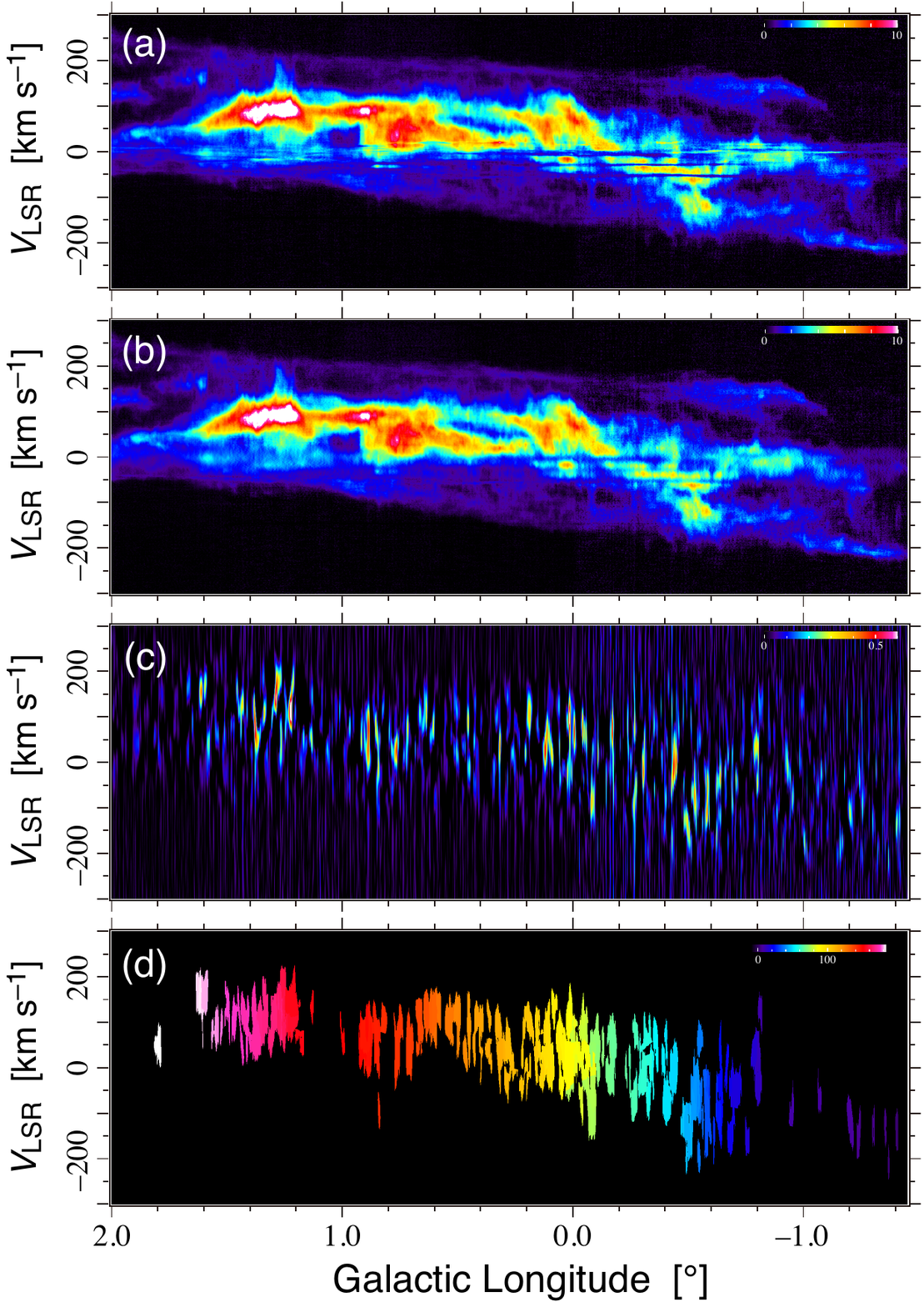}
\caption{Longitude-velocity maps of CO {\it J}=3--2 emission averaged over latitudes.  (a) Map of the original CO {\it J}=3--2 data.  (b) Map of the ``pressed" data.  (c) Map of the ``pressed" and ``unsharp masked" data.  (d) HVCC candidates identified in the ``pressed" and ``unsharp masked" data with the modified CLUMPFIND algorithm.  Color represents the identification number in Table \ref{tb:catalog}.}
\label{fig:lvd}
\end{center}
\end{figure*}

\section{Data}
\subsection{James Clerk Maxwell Telescope}
To identify HVCCs in the CMZ, the CO {\it J}=3--2 (345.796 GHz) line data obtained using the James Clerk Maxwell Telescope (JCMT) by the JCMT galactic plane survey (JPS) team \citep{Parsons18} was utilized. The observations were conducted in the periods ranging from July to September 2013, the month of July, 2014, and from March and June 2015. The data utilized covered the area $-1\fdg 4\! \leq\! l\! \leq\! +2\fdg 0$ and $-0\fdg 25\! \leq\! b\! \leq\! +0\fdg 25$ (Figure \ref{fig:data}a). 

The Hetrodyne Array Receiver Program \citep[HARP][]{Buckle09}, which is a single sideband array receiver with a $4\!\times\! 4$ SIS receptors spaced every $30\arcsec$, was employed to obtain the CO {\it J}=3--2 image of the CMZ. In addition, the auto-correlation spectral imaging system (ACSIS) was also employed as a receiver backend with the 1800 MHz bandwidth and 0.97 MHz spectral resolution mode.

The data were reduced using Starlink software package and the resulting data cube had an angular resolution of $16\farcs 6$ and a velocity resolution of 0.897 \kms.  
The resultant data grid is $6\arcsec\! \times\! 6\arcsec\! \times\! 0.897$ \kms .
The RMS noise was approximately $\Delta T^{*}_{\rm A} \! = \! 0.4$ K. The data, originally obtained in the $T^{*}_{\rm A}$ scale, were converted to the $T^{*}_{\rm R}$ scale using the forward spillover and scattering efficiency of $\eta _{\rm fss}\! =\! 0.71$. Subsequently, $T^{*}_{\rm R}$ was multiplied by 1.107 to ensure consistency with the intensity scale of CO {\it J}=3--2 data obtained with ASTE \citep{Oka12, Tokuyama19}. 

\subsection{Nobeyama Radio Observatory 45 m Telescope}
The CO {\it J}=1--0 (115.271 GHz) line data, used to calculate the CO {\it J}=1--0 luminosity and CO {\it J}=3--2/{\it J}=1--0 intensity ratio (\R32_10), was obtained from observations using the 45 m telescope at NRO in January, 2011 \citep{Tokuyama19}. The mapped area is $-0\fdg 8\! \leq\! l\! \leq\! +1\fdg 4$ and $-0\fdg 35\! \leq\! b\! \leq\! +0\fdg 35$ (Figure \ref{fig:data}b).

The 25-BEam array receiver system \citep[BEARS;][]{Sunada00, Yamaguchi00}, a $5\!\times\!5$ focal-plane array SIS receiver was used for the NRO 45 m observations. The data obtained were reduced using the NOSTAR reduction package, resulting in a $7\farcs 5\! \times\! 7\farcs 5\! \times\! 2$ \kms\ grid data cube with an RMS noise of $\Delta T^{*}_{\rm A}\!=\! 1$ K (1$\sigma$), respectively. Further, the antenna temperatures were converted to $T^{*}_{\rm R}$ divided by $\eta_{\rm fss}\!=\! 0.71$ \citep[see][]{Tokuyama19}.

\section{Identification Procedure}\label{sec: id}
An automated identification procedure was developed to extract HVCCs from a spectral line data cube. The procedure comprised three major steps: (1) reducing the disk gas emission/absorption, (2) highlighting spatially compact, broad-velocity-width features, and (3) automated identification of emission clumps. In this section, each step of the procedure has been described.

\subsection{Pressing Method}
Because we are in the middle of the Galactic disk, the spectral line data toward the CMZ are severely contaminated by emission and absorption from low-density disk gas. Such undesirable contaminations appear in the longitude-velocity ({\it l--V}) maps as horizontal, narrow-velocity-width emission/absorption stripes at velocities ranging from $\VLSR\!=\!-60$ \kms\ to $+20$ \kms (Figure \ref{fig:lvd}a). These stripes result in the broad-velocity-width emission features from the CMZ being dismembered in the corresponding velocity range. Thus, to facilitate thorough identification of HVCCs, an appropriate treatment for reducing the effect of disk gas contamination is required.

The ``pressing method" was developed by \citet{Sofue95} to reduce the disk gas contamination from a CO data cube. The method is outlined as follows: 

\begin{enumerate}
\item Smoothen the original data cube $F_0$ in the \VLSR\ direction using a Gaussian function with ${\rm FWHM}\! =\! x_1$, and write the result in $F_1$. 
\item Subtract $F_1$ from $F_0$, $\rightarrow\, F_2$. 
\item Smoothen $F_2$ in the $l, b$ direction using a Gaussian function with ${\rm FWHM}\! =\! x_2$, $\rightarrow\, F_3$.
\item Subtract $F_3$ from $F_0$, $\rightarrow\, F_4$.
\item Replace $F_0$ by $F_4$, thereafter repeat steps 1--4 till an acceptable $F_4$ is achieved.
\end{enumerate}

\noindent
Only when the disk gas contaminations were sufficiently reduced, it is considered acceptable.

This method was applied to data in the velocity range of $-55+10 \left(l/\arcdeg\right)\!\leq\!\left(\VLSR/\kms \right)\! \leq\! 25$, where the disk gas contamination is significant \citep{Oka98b, Oka07, Oka12}. Steps 1--4 were repeated 50 times with $x_1\! =\! 20\,\kms$ and $x_2\! =\! 0\fdg 025$ for the JCMT CO {\it J}=3--2 data cube. Consequently, the ``pressed" data replaced the original data in the corresponding velocity range. The data obtained clearly exhibit a remarkable reduction in the horizontal stripes in the {\it l--V} maps (Figure \ref{fig:lvd}b).

\begin{figure}
\begin{center}
\includegraphics[scale=0.8, width=8cm]{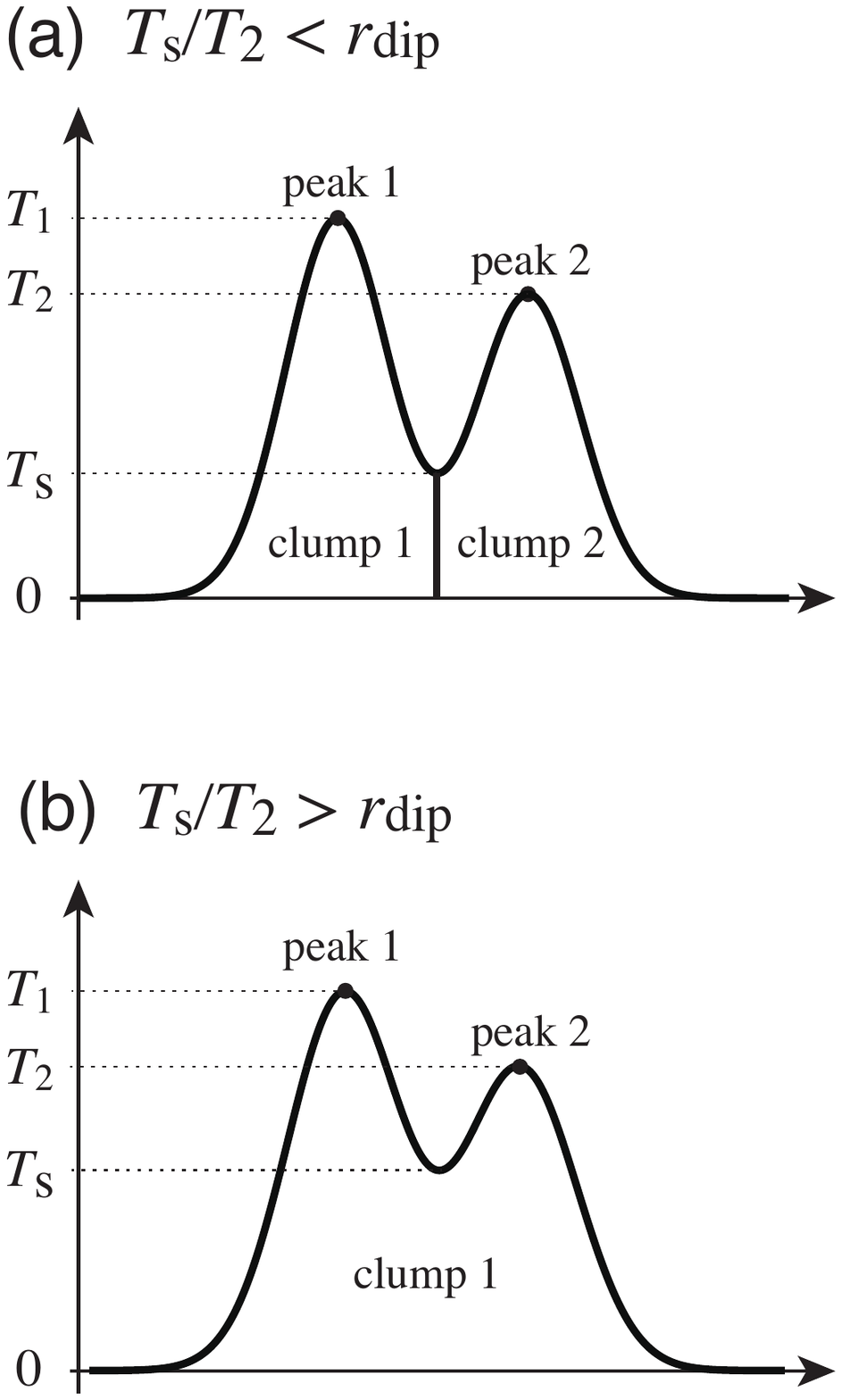}
\caption{One-dimensional, schematic drawings explaining the meaning of $r_{\rm dip}$ introduced in the modified CLUMPFIND algorithm:  (a) two clumps case,  (b) one clump case. }
\label{fig:modcfind}
\end{center}
\end{figure}

\subsection{Unsharp Masking}
Unsharp masking is an image-sharpening method that highlights compact structures. This method was employed to highlight spatially compact and broad-velocity-width features in a data cube. The method was applied to the pressed data ($F_4$) as follows: 

\begin{enumerate}
\item Smoothen $F_4$ in the \VLSR\ direction using a Gaussian function with ${\rm FWHM}\! =\! y_1$ yielding $F_5$.
\item Smoothen $F_5$ in the $l, b$ direction using a Gaussian function with ${\rm FWHM}\! =\! y_2$, $\rightarrow\, F_6$.
\item Subtract $F_6$ from $F_5$, $\rightarrow\, F_7$.
\end{enumerate}

The data cube obtained from the unsharp masking is $F_7$. Subsequently, $y_1\! =\! 75\,\kms$ and $y_2\! =\! 0\fdg 04$ were employed for the JCMT CO {\it J}=3--2 data cube. The latitude-integrated {\it l--V} map obtained from the ``unsharp-masked" CO {\it J}=3--2 data is shown in Figure \ref{fig:lvd}c. Spatially compact and broad-velocity-width features that were found in the pressed data (Figure \ref{fig:lvd}b) are highlighted in the unsharp masked data (Figure \ref{fig:lvd}c).

\subsection{Modified CLUMPFIND}
There exist several algorithms that can identify clouds in spectral line datasets. CLUMPFIND is a cloud identification algorithm developed by \citet{Williams94}, that is extensively used because of its convenience and good performance. CLUMPFIND defines a clump as a lump of pixels, which has a single local peak of spectral line intensity, with intensities greater than a particular threshold ($T_{\rm min}$). If the lump has multiple peaks, it is divided into several clumps, each of which contains a single peak. Consequently, CLUMPFIND identifies as many clumps as the number of intensity peaks detected by it. However, this property results in excessive division of clouds in certain cases, particularly in crowded regions such as the CMZ.

The CLUMPFIND algorithm was modified to avoid the excessive-division problem (Figure \ref{fig:modcfind}). A new parameter, $r_{\rm dip}$, which serves as a threshold for determining whether to bifurcate the lump, was introduced. If a lump of pixels with intensities greater than $T_{\rm min}$ includes two local peaks, $T_1$ and $T_2$ ($T_1\! >\! T_2$), the modified CLUMPFIND refers to the saddle point intensity ($T_{\rm s}$). Subsequently, if the saddle point is deeper than the threshold, $T_{\rm s}/T_2\! <\! r_{\rm dip}$, the lump is divided into two separate clumps else, the lump is defined as a single clump.

This modified CLUMPFIND algorithm was applied to the pressed and unsharp-masked data cubes ($F_7$). For the JCMT CO {\it J}=3--2 data, the parameter set employed was as follows: $T_{\rm min}\!=\! 1.3$ K, $\Delta T\!=\! 0.5$ K (peak search interval), $r_{\rm dip}\!=\! 0.5$, and $N_{\rm min}\!= \! 200$ (the minimum number of pixels). Consequently, 352 HVCC candidates were identified. The {\it l--V} distribution of the identified candidates is shown in Figure \ref{fig:lvd}d. A total of 168 candidates were dropped by the HVCC criterion, $\sigma_{V}\!\ge\! 20\,\kms$, 
the resulting 184 candidates were accepted as HVCCs. 

\subsection{Performance Verification}
To check the performance of our HVCC identification method, we applied it to the data from the CO {\it J}=3--2 High-resolution Survey of the Galactic Plane \citep[COHRS;][]{Dempsey13}.  We utilized the data in the range of $33\fdg 253\!\leq\! l\!\leq\! 36\fdg 781$ and $-0\fdg 300\!\leq\! b\!\leq\! +0\fdg 300$, which has an approximately same angular extent as that of the CMZ data utilized in this study.  The angular resolution and velocity resolution are also the same as those of the CMZ data.  The pressing method was skipped since the target in this performance check is the emission from the disk part of our Galaxy. Parameters in the unsharp masking and modified CLUMPFIND are the same as those employed in the CMZ analyses except for $T_{\rm min}$ and $\Delta T$.  We employed $T_{\rm min}\!=\! 0.65$ K and $\Delta T\!=\! 0.25$ K here, since the RMS noise of the COHRS data was a factor of 2 lower than that of the CMZ data.

Consequently, seven compact broad velocity width features were identified (HVCC candidates; Table \ref{tab:GPtest}), all of which were dropped by the HVCC criterion, $\sigma_{V}\!\ge\! 20\,\kms$.  These seven candidates show double-sided wing features in the original COHRS data.  Inspecting the SIMBAD database, we confirmed that all of these candidates are associated with infrared bubbles, H{\sc ii} regions, or young stellar objects within a $36\arcsec$ radius.  Thus, it is most likely that the HVCC candidates identified in the Galactic plane with our method are bipolar molecular outflows driven by young stellar objects.  This result verifies our HVCC identification method, providing strong confirmation that the HVCCs found in the CMZ are real.

\begin{deluxetable*}{lcccccl}[hbtp]
\tablecaption{Compact broad velocity width features identified in the Galactic plane. \label{tab:GPtest}}
\tablewidth{0pt}
\tablehead{
\colhead{id} & \colhead{$l$} & \colhead{$b$} & \colhead{$\VLSR$} &
\colhead{$\sqrt{\sigma_{l} \sigma_{b}}$} & \colhead{$\sigma_{V}$} &\colhead{SIMBAD objects} \\
\colhead{} & \colhead{(\arcdeg)} & \colhead{(\arcdeg)} & \colhead{(\kms)} &
\colhead{(\arcsec)} & \colhead{(\kms)} & \colhead{} 
}
\startdata
1 & 35.593 & 	$-0.030$ &  51.1 &  	8.9 &  	17.9 &  	bubble, H{\sc ii} \\
2 & 33.919 & 	0.110 &  	107.1 &  	11.8 &  14.7 &  	H{\sc ii}, YSO, bubble \\
3 & 34.601 &  	0.245 &  	$-30.9$ &  	5.2 & 	7.2  &  	H{\sc ii} \\
4 & 34.263 &  	0.154 &  	59.1 &  	13.1 &  12.7 &  	H{\sc ii}, SFR \\
5 & 33.396 &  	0.012 &  	107.1 &  	5.8 &  	14.2 &  	H{\sc ii} \\
6 & 33.813 &  	$-0.186$ &  46.1 &  	5.3 &  	11.1 &  	H{\sc ii}, YSO, bubble  \\
7 & 34.471 &  	0.249 &  	64.1 &  	5.9 &  	11.2 &  	YSO \\
\enddata
\end{deluxetable*}

\begin{figure}
\begin{center}
\includegraphics[scale=0.6, width=8cm]{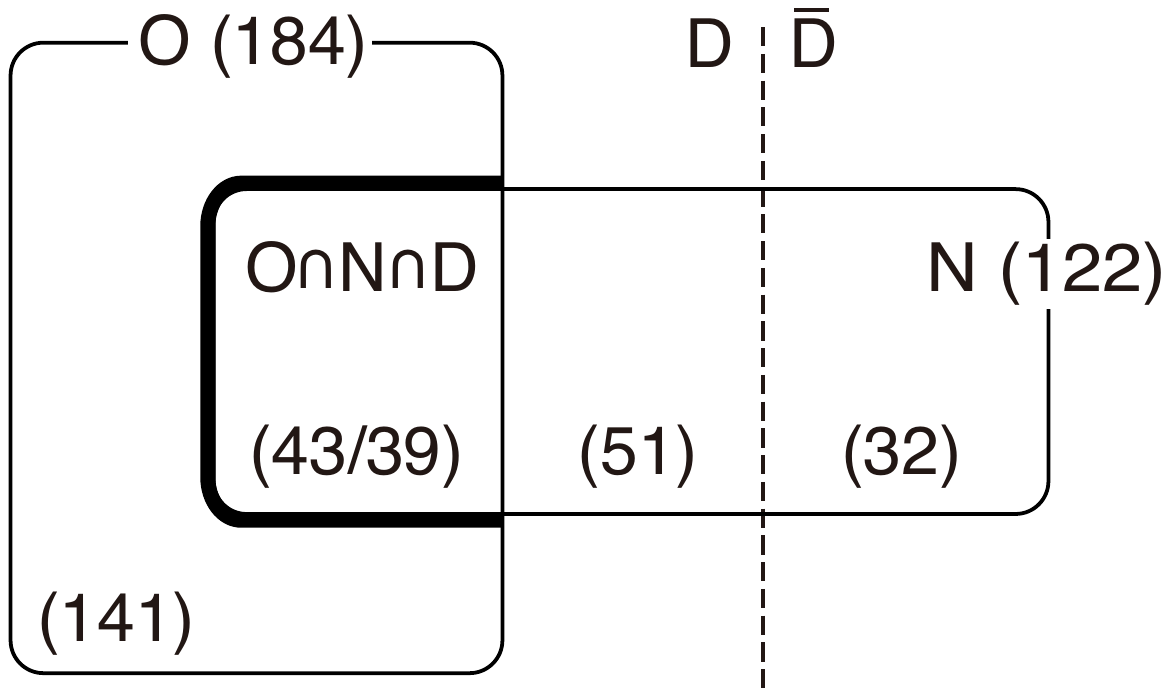}
\caption{Venn diagram showing the set relationship between the HVCCs identified in this study (symbol ``{\textsf O}") and those of \citet[][symbol ``{\textsf N}"]{Nagai08}.  The number in the parentheses denotes that of HVCCs which belong to each set.  The symbol ``{\textsf D}" denotes the data coverage of this study.  The number of HVCCs in ${\textsf O}\!\cap\!{\textsf N}\!\cap\!{\textsf D}$ is 39 in {\textsf N}, while it is 43 in {\textsf O} (see text).}
\label{fig:venn}
\end{center}
\end{figure}

\section{Results and Discussion}
\subsection{Comparison with Previous Studies}
The 184 HVCCs identified by the automated procedure in the JCMT CO {\it J}=3--2 data are listed in Table \ref{tb:catalog} (Appendix \ref{hvcclist}). The list contains prominent HVCCs that have already been investigated to a certain extent, for example, CO 0.02--0.02 \citep{Oka99, Oka08}, CO 1.27+0.01 \citep{Oka01a}, CO--0.40--0.22 \citep{Oka16, Oka17}, and CO--1.21--0.12 \citep{Oka12, Tsujimoto18}. However, the identification procedure failed to detect several previously identified HVCCs, CO--0.31+0.11 \citep{Takekawa19b}, HCN--0.009--0.044 \citep{Takekawa17, Takekawa19a}, and HCN--0.085--0.094 \citep{Takekawa17, Takekawa20}. Such identification failures occurred primarily due to their faintness in CO {\it J}=3--2 or overlapping with unrelated clouds. 

Consequently, the correspondence between HVCCs identified in this study and those in a previous study \citep{Nagai08} were examined. Figure \ref{fig:venn} shows the set relationship between the HVCCs identified in this study and \citet{Nagai08}. In the data coverage of this study (${\textsf D}$ in Figure \ref{fig:venn}), there are 90 Nagai-identified HVCCs (${\textsf N}\!\cap\!{\textsf D}$). Further our automated procedure successfully identified 39 Nagai-HVCCs as 43 HVCCs, because CO 0.01--0.03 was divided into four HVCCs while CO--0.30--0.07 was divided into two in the current list. Thus, the detection rate of Nagai-HVCCs (${\textsf O}\!\cap\!{\textsf N}\!\cap\!{\textsf D}/{\textsf N}\!\cap\!{\textsf D}$) is $43.3$ \%.
Remaining 141 HVCCs in the list are previously unknown HVCCs.

\subsection{Classification of HVCCs}
The identified HVCCs were categorized into five classes by inspecting three-dimensional morphology of each cloud: simple-type, shell-type, wing-type, bridge-type, and complex-type. These morphological classes are expected to reflect their nature and origin. The descriptions of the five morphological classes are as follows: \\

\noindent
{\bf Simple-type:} This type of HVCC is generally isolated in the {\it l-b-V} space, which has simple kinematics. A prototypical example of this is CO--0.40--0.22, which is considered to have originated from a gravitational kick of a molecular cloud due to a massive, invisible object \citep{Oka16}. Two other gravitationally kicked HVCCs that belong to this category are HCN--0.009--0.044 \citep{Takekawa17, Takekawa19a} and HCN--0.085--0.094 \citep{Takekawa20}.\\

\noindent
{\bf Shell-type:} This type of HVCC has a shell/arc-shaped morphology and expanding kinematics, suggesting an origin related to local explosive events. The most promising candidate for the driving source of this type of HVCC is a supernova explosion(s) \citep{Oka01a, Oka08}. Prototypical examples are CO 1.27+0.01 \citep{Oka01a} and CO--1.21--0.12 \citep{Oka12, Tsujimoto18}. HVCCs found in the $l\!=\! +1\fdg 3$ region primarily belong to this category.\\

\noindent
{\bf Wing-type:} Molecular outflows from young stellar objects are often observed as high-velocity ``wings" of line profiles. This type of HVCC appears as a compact broad-velocity-width emission that arises from a parent normal-velocity-width cloud in position-velocity maps. Supernova/molecular cloud interactions also cause wing emission \citep[e.g., ][]{Seta98, Seta04}. \\

\noindent
{\bf Bridge-type:} This type of HVCC connects two molecular clouds with different velocities in the position-velocity space. Such features can be observed in regions of cloud-to-cloud collisions. CO--0.71--0.02 located at the root of the Pigtail molecular cloud \citep{Matsumura12} may belong to this category. \\

\noindent
{\bf Complex-type:} This type of HVCC generally comprises a few components with complex morphologies and/or kinematics. These can belong to other types of HVCCs, severely contaminated by physically unrelated gases. A certain few could be the result of chance gathering of clouds in the {\it l-b-V} space. \\

Although this morphological classification is still based on a working hypothesis, it could serve as a helpful guideline for investigating the nature and origin of HVCCs. However, unique classification was difficult for approximately half of the 184 identified HVCCs. Thus, to describe each HVCC properly, a two-step classification method was introduced. If an HVCC possesses an attribute that does not belong to the primary class (type 1), it was described as the secondary class (type 2). The following discussion based on morphological classification uses only the primary class (type 1). As a result of by-eye inspection, the number of simple, shell, wing, bridge, and complex-type HVCCs were 35, 28, 57, 18, and 46, respectively. However, three HVCCs were part of the circumnuclear disk (CND) of the galaxy, and thus need to be treated separately. 


\subsection{Distribution of HVCCs}
\begin{figure*}
\begin{center}
\includegraphics[scale=0.8, width=15cm]{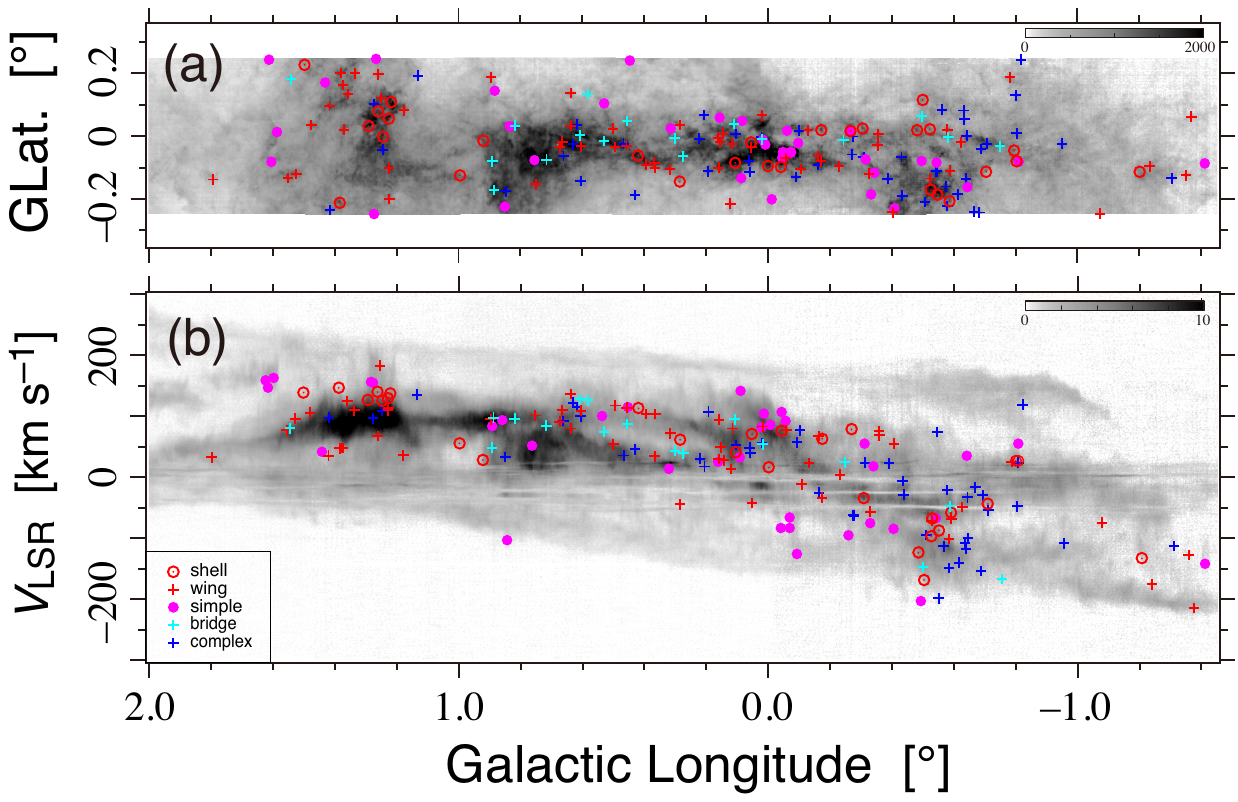}
\caption{(a) Spatial distribution of identified HVCCs superposed on the velocity-integrated map of CO {\it J}=3--2 emission.  (b) Longitude-velocity distribution of identified HVCCs superposed on the latitude-integrated map of CO {\it J}=3--2 emission.  
Red dot-circles denote shell-type, red crosses are wing-type, magenta filled-circles are simple-type, cyan crosses are bridge-type, and blue crosses are complex-type HVCCs.}
\label{fig:dist}
\end{center}
\end{figure*}

Figure \ref{fig:dist} shows the spatial and longitudinal velocity distributions of the identified HVCCs, which spread over the CMZ, not necessarily preferring the intense ridge of CO emission, which is the midplane of interstellar matter. Moreover, they are also not biased towards giant H{\sc ii} regions such as Sgr B2 suggesting that the dominant origin of HVCCs is irrelevant to the star formation activity. 

The longitude-velocity distribution of the identified HVCCs exhibited a gradual increase in \VLSR\ with increasing longitude. It roughly followed the {\it l--V} behavior of the 120-pc molecular ring \citep{Sofue95}, whereas the latitudinal distribution was wider. This behavior indicates that their parent gas is in the CMZ and predominantly rotates with the Galactic rotation. Further, very few HVCCs were identified in the 200-pc expanding molecular ring \citep[EMR;][]{Scoville72, Kaifu72}, which may be partly due to the employed $T_{\rm min}$, which is larger than the typical CO intensity in the EMR.

The distributions of the HVCCs exhibit a minor dependence on their morphological class. The shell- and wing-type HVCCs are associated with intense ridges in CO emissions. Concentrations of shell-type HVCCs at $l\! =\! +1\fdg3$ corresponds to a proto-superbubble candidate in the literature \citep{Oka01a, Tsujimoto21}. The bridge-type HVCCs favor velocities between the CO emission ridges in the {\it l-V} plane. However, the simple-type HVCCs are rather scattered in the {\it l-b-V} space compared to the other types. These trends are consistent with the hypothesis regarding the origin of the HVCCs described above.

\begin{figure}
\begin{center}
\includegraphics[scale=0.7, width=8cm]{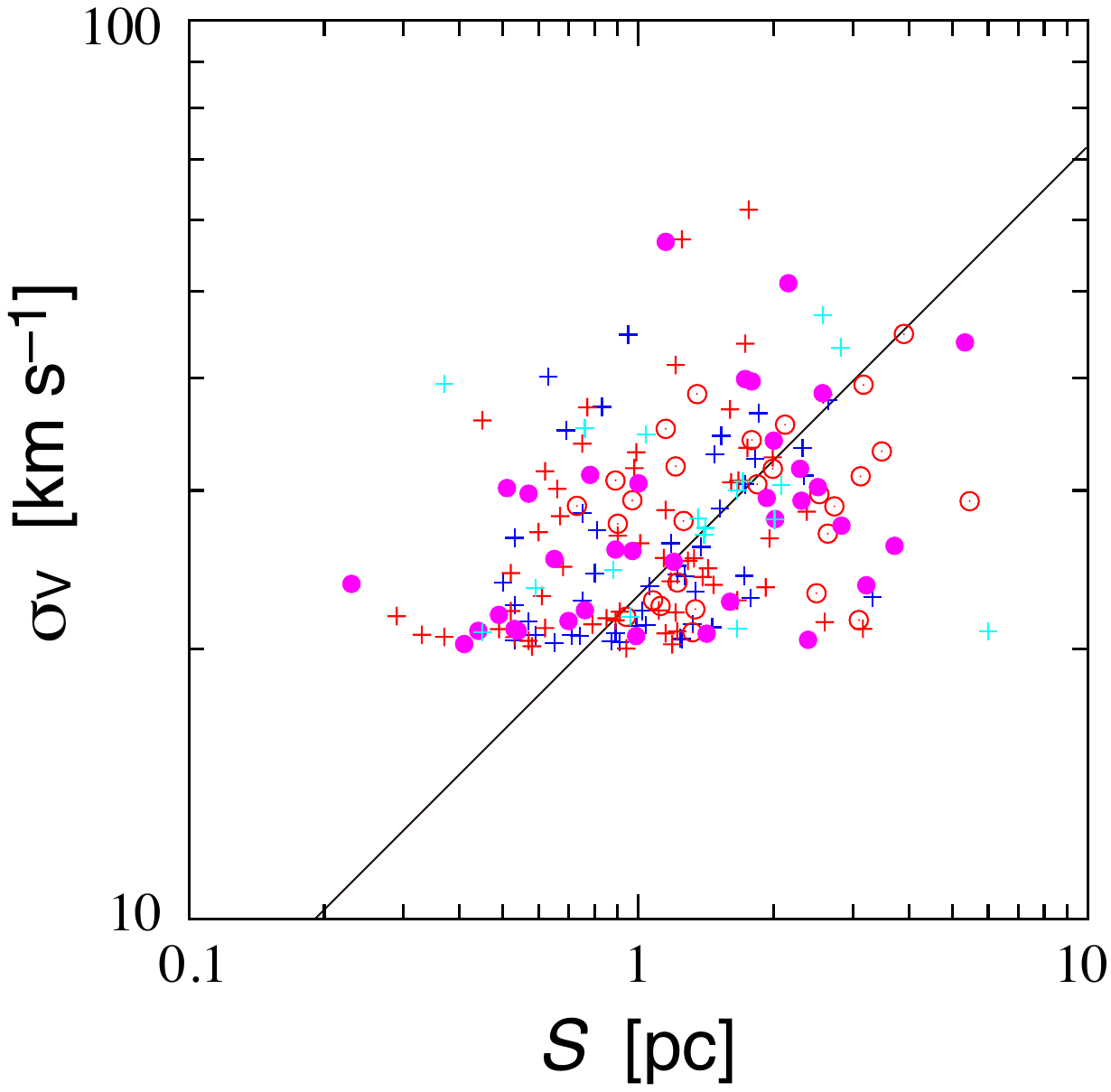}
\caption{A plot of size ($S$) versus velocity dispersion ($\sigma_{\rm V}$ for 184 identified HVCCs.  Symbols are the same as in Figure \ref{fig:dist}.  The solid line represents the best-fit line for all identified HVCCs, $(\sigma_{\rm V}/\kms)\!=\! 22.9\, (S/{\rm pc})^{0.5}$. }
\label{fig:S-sv}
\end{center}
\end{figure}

\begin{figure}
\begin{center}
\includegraphics[scale=0.7, width=8cm]{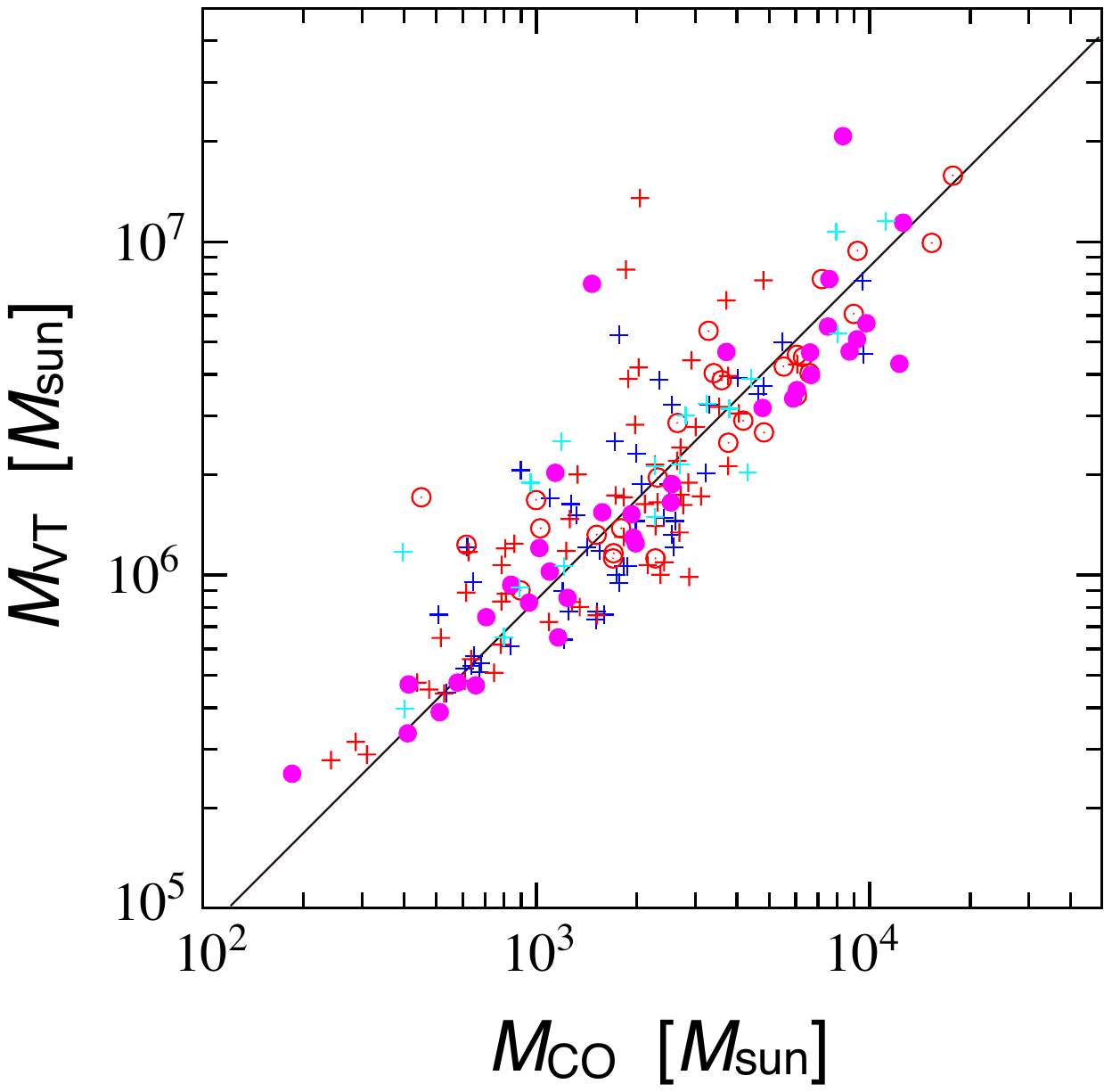}
\caption{A plot of molecular mass ($M_{\rm CO}$) versus virial mass ($M_{\rm VT}$) for 184 identified HVCCs.  Symbols are the same as in Figure \ref{fig:dist} and \ref{fig:S-sv}.  The solid line represents the best-fit line for all identified HVCCs, $(M_{\rm VT}/M_{\sun})\!=\! 834\, (M_{\rm CO}/M_{\sun})$. }
\label{fig:Mco-Mvt}
\end{center}
\end{figure}

\subsection{Physical Parameters}
We calculated the size parameter ($S$), velocity dispersion ($\sigma_V$), CO {\it J}=3--2 luminosity ($L_{\rm CO}$), molecular gas mass ($M_{\rm CO}$), virial theorem mass ($M_{\rm VT}$), kinetic energy ($E_{\rm kin}$), expansion time ($t_{\rm exp}$), kinetic power ($P$), and CO {\it J}=3--2/{\it J}=1--0 ratio ($R_{3\mbox{--}2/1\mbox{--}0}$) for each HVCC (see Appendix \ref{hvcclist}).  These are listed in Table \ref{tb:param}. 

The average size of the HVCCs obtained was 1.4 pc, and the average velocity dispersion was 28 \kms. Compared to the average velocity dispersion of a few \kms\ for molecular clouds in the CMZ with a size of 3 pc \citep{Shetty12}, HVCCs exhibit larger velocity dispersions than typical molecular clouds in the CMZ. The $S$-$\sigma_{\rm V}$ plot for the identified HVCCs is shown in Figure \ref{fig:S-sv}. 
Although almost no correlation between $S$ and $\sigma_{\rm V}$ is evident, it may still be useful to diagnose the internal kinetic state of clouds in comparison with those in other environments.  In order to make comparisons with previous studies \citep[e.g.,][]{Oka98a, Oka01b, Solomon87} easier, we employed the $S$-$\sigma_{\rm V}$ relation with the exponent fixed to $0.5$.  
Applying least-squares fitting to $\sigma_{\rm V}\!=\! A S^{0.5}$ yields a size line width coefficient of $A\!=\!(22.9\pm 0.6)$ km s$^{-1}$ pc$^{-0.5}$ for all identified HVCCs.

The averages of the molecular gas mass ($M_{\rm CO}$) and virial theorem mass ($M_{\rm VT}$) were obtained as $10^{3.5}$ $M_{\odot}$ and $10^{6.4}$ $M_{\odot}$, respectively. Consequently, this yielded an average virial parameter of $\alpha\!\sim\!1000$, indicating that the HVCCs are gravitationally unbound. Figure \ref{fig:Mco-Mvt} shows the $M_{\rm CO}$-$M_{\rm VT}$ plot for the identified HVCCs. Applying the least-squares fitting to $M_{\rm VT}\!=\!\alpha M_{\rm CO}$ yielded a virial parameter, $\alpha\!=\! 834\pm 34$ for all identified HVCCs.

The average kinetic energy ($E_{\rm kin}$) obtained was $10^{50.0}$ erg. Because a supernova explosion ejects a kinetic energy of (1--3)\!$\times\! 10^{50}$ erg into the interstellar space, the average kinetic energy of HVCCs can be said to roughly correspond to a single supernova explosion. The expansion time ($t_{\rm exp}$) average was $10^{4.7}$ years, while the average kinetic power ($P$) was $10^{37.8}$ erg s$^{-1}$. 

The average CO {\it J}=3--2/{\it J}=1--0 ratio was $0.99\pm 0.02$, which is substantially higher than the average CO {\it J}=3--2/{\it J}=1--0 ratio reported inside the CMZ (0.71; Oka et al. 2012). As the CO {\it J}=3--2/{\it J}=1--0 ratio is sensitive to density and temperature, the ratio is considered a good indicator of excitation inside molecular gas. The high average of CO {\it J}=3--2/{\it J}=1--0 ratio inside the HVCCs suggests that these clouds are in a highly excited state. 

The average values of the physical parameters for each morphological class (Table \ref{tab:average}) were compared. On average, shell-type HVCCs have a slightly larger size, molecular mass, virial mass, kinetic energy, and expansion time compared to the other types. Moreover, bridge, simple, and shell-type HVCCs have a larger velocity dispersion than wing and complex HVCCs. The simple-type HVCCs possess a larger kinetic power and higher CO {\it J}=3--2/{\it J}=1--0 ratio. Such tendencies in the average values may reflect their origins. The top seven highest-ratio HVCCs are of the simple-type, and four of them belong to the CND. Among them clump C1 \citep{Oka11} might also be related to the CND. The other two, id39 and id46 in the list (Tables \ref{tb:catalog} and \ref{tb:param}), are targets for future investigation. 

The size-line width coefficient ($A$) and virial parameter ($\alpha$) for each morphological class are listed in Table \ref{tab:coeff}. The size-line width coefficient ranges between 20.4 and 25.2, with the maximum at wing-type and minimum at shell-type. This tendency is also apparent in the average expansion time, which is the shortest for wing-type and longest for shell-type. The virial parameter ranges between 759 and 1040, with the maximum at wing-type and minimum at complex-type, with the largest virial parameter indicating that most wing-type HVCCs are gravitationally extreme-unbound.

\begin{deluxetable*}{l|cccccccc}[hbtp]
\tablecaption{Average value of parameters for each morphological class \label{tab:average}}
\tablewidth{0pt}
\tablehead{
\colhead{type} & \colhead{$S$} & \colhead{$\sigma_{\rm V}$} &
\colhead{$M_{\rm CO}$} & \colhead{$M_{\rm VT}$} & 
\colhead{$E_{\rm kin}$} & \colhead{$t_{\rm exp}$} &
\colhead{$P_{\rm kin}$} & \colhead{$R_{3\mbox{--}2/1\mbox{--}0}$} \\
\colhead{} & \colhead{(pc)} & \colhead{(\kms)} &
\colhead{($10^3$ $M_{\sun}$)} & \colhead{($10^6$ $M_{\sun}$)} &
\colhead{($10^{50}$ erg)} & \colhead{($10^4$ yr)} & 
\colhead{($10^{37}$ erg s$^{-1}$)} & \colhead{} 
}
\startdata
simple & $1.6\pm 0.2$ & $29\pm 1$ & $3.8\pm 0.6$ & $3.3\pm 0.7$ & $1.3\pm 0.3$ & $5.3\pm 0.6$ & $8.3\pm 2.7$ & $1.20\pm 0.08$ \\
shell & $2.0\pm 0.2$ & $29\pm 1$ & $4.7\pm 0.8$ & $3.9\pm 0.6$ & $1.6\pm 0.4$ & $6.7\pm 0.7$ & $7.2\pm 1.7$ & $0.98\pm 0.03$ \\
wing & $1.2\pm 0.2$ & $27\pm 1$ & $1.9\pm 0.2$ & $2.1\pm 0.3$ & $0.53\pm 0.08$ & $4.3\pm 0.3$ & $4.5\pm 0.7$ & $0.89\pm 0.02$ \\
bridge & $1.7\pm 0.3$ & $30\pm 2$ & $3.3\pm 0.7$ & $3.2\pm 0.7$ & $1.1\pm 0.4$ & $5.8\pm 1.3$ & $7.1\pm 2.5$ & $0.91\pm 0.04$ \\
complex & $1.2\pm 0.1$ & $26\pm 1$ & $2.2\pm 0.3$ & $1.9\pm 0.2$ & $0.58\pm 0.11$ & $4.5\pm 0.3$ & $4.2\pm 0.7$ & $0.98\pm 0.03$ \\
\hline
all & $1.4\pm 0.1$ & $28\pm 1$ & $2.9\pm 0.2$ & $2.6\pm 0.2$ & $0.91\pm 0.11$ & $5.0\pm 0.2$ & $5.8\pm 0.7$ & $0.99\pm 0.02$ \\
\enddata
\tablecomments{Errors are standard deviations for each type.}
\end{deluxetable*}

\begin{deluxetable}{l|cc}[hbtp]
\tablecaption{Size-line width coefficient and virial parameter for each morphological class\label{tab:coeff}}
\tablewidth{0pt}
\tablehead{
\colhead{type} & \colhead{$A$} & \colhead{$\alpha$} \\
\colhead{} & \colhead{(km s$^{-1}$ pc$^{-0.5}$)} & \colhead{}
}
\startdata
simple & $22.7\pm 1.4$ & $795\pm 94$ \\
shell & $20.4\pm 1.1$ & $800\pm 36$ \\
wing & $25.2\pm 1.2$ & $1040\pm 120$ \\
bridge & $21.9\pm 2.2$ & $965\pm 65$ \\
complex & $23.6\pm 1.0$ & $759\pm 47$ \\
\hline
all & $22.9\pm 0.6$ & $834\pm 34$ \\
\enddata
\tablecomments{Errors are uncertainties of best-fit parameters.}
\end{deluxetable}

\section{Summary}
A total of 184 HVCCs were identified in the CO {\it J}=3--2 line data of the central molecular zone of our galaxy. An automated identification procedure, comprising a pressing method, unsharp masking, and modified CLUMPFIND, was developed and subsequently utilized. Consequently, a list of HVCCs was presented with their physical parameters, CO {\it J}=3--2/{\it J}=1--0 intensity ratios, and morphological classifications.

The identified HVCCs have a size-line width coefficient of $22.9\pm 0.6$ and a virial parameter of $834\pm 34$. Further, the average CO {\it J}=3--2/{\it J}=1--0 intensity ratio for HVCCs was $0.99\pm 0.02$, which is significantly higher than that of the CMZ (0.71). The physical parameters of the identified HVCCs were found to be weakly dependent on their classifications, while the CO {\it J}=3--2/{\it J}=1--0 intensity ratio exhibit a sharp difference between HVCCs in the nuclear environment and the others.

The presented HVCC catalog provides a number of intriguing sources that deserve detailed study in the future. Thus, it may be useful when searching for invisible massive objects, deeply embedded protostars and supernova remnants, and sites of cloud-to-cloud collisions. \\
\\

The results in this paper are based on observations at the Nobeyama Radio Observatory (NRO) and James Clerk Maxwell Telescope (JCMT). The Nobeyama 45-m radio telescope is operated by the Nobeyama Radio Observatory, a division of the National Astronomical Observatory of Japan. 

The James Clerk Maxwell Telescope is operated by the East Asian Observatory on behalf of The National Astronomical Observatory of Japan, the Academia Sinica Institute of Astronomy and Astrophysics, the Korea Astronomy and Space Science Institute, the National Astronomical Observatories of China, and the Chinese Academy of Sciences (Grant No. XDB09000000), with additional funding support from the Science and Technology Facilities Council of the United Kingdom and participating universities in the United Kingdom and Canada.

We are grateful to the NRO staff and all the members of the JCMT team for operation of the telescope. T.O. acknowledges support from JSPS Grant-in-Aid for Scientific Research (A) No. 20H00178.


\appendix

\section{List of Identified HVCCs}
\label{hvcclist}
This section presents the list of identified HVCCs with their $(l, b, \VLSR)$ locations and various physical parameters. The physical parameters were calculated using standard definitions \citep[e.g.,][]{Solomon87, Oka01b} and have also been included in the catalog (Table \ref{tb:catalog}). 

The size of the cloud $S$ can be calculated using the following formula:
\begin{equation}
S = D\: {\rm tan} \sqrt{\sigma_{l} \sigma_{b}} 
\end{equation} 
where $\sigma_{\rm x}$ is the dispersion in the x-direction, and $D$ represents the distance to the cloud. In addition, $D\!=\!D_{\rm GC}$ was employed here. 

The total CO luminosity $L_{\rm CO}$ was calculated with
\begin{equation}
L_{\rm CO} \equiv D^2 \int I_{\rm CO} dl db
\end{equation}
where
\begin{equation}
I_{\rm CO} = \int T_R^{*}(\mbox{CO})\: dV. 
\end{equation}

The molecular gas mass $M_{\rm CO}$ was calculated with 
\begin{equation}
M_{\rm CO} = \mu m_{\rm H_2} X_{\rm CO} L_{\rm CO}
\end{equation}
where $\mu$ is the mean molecular weight ($=\! 1.36$), $m_{\rm H_2}$ is the mass of molecular hydrogen, and $X_{\rm CO}$ represents the CO-to-$\rm H_{2}$ conversion factor. The total luminosity of CO {\it J}=3--2 emissions was used to perform the molecular mass calculation.

The conversion factor is typically defined as $X_{\rm CO}\!\equiv\! N({\rm H_{2}})/I_{\rm CO 1\mbox{--}0}$, where $N(\rm H_{2})$ is the column density of molecular hydrogen. However, the value of $X_{\rm CO}$ at the Galactic Center is somewhat controversial. The standard value is $3.0\!\times\! 10^{20}$ [cm$^{-2}$ (K \kms )$^{-1}$] \citep{Young91}. However, in the CMZ, many authors have calculated the value of $X_{\rm CO}$ to be significantly smaller than the standard value \citep[e.g.,][]{Arimoto96, Oka98a}. This study employed $X_{\rm CO 1\mbox{--}0}\!=\! 1.0\!\times\! 10^{20}$ [cm$^{-2}$ (K \kms )$^{-1}$] \citep{Arimoto96} divided by the total CO {\it J}=3--2/CO {\it J}=1--0 luminosity ratio of the CMZ \citep[0.71;][]{Oka12}, $X_{\rm CO 3\mbox{--}2} \!=\! 1.4\!\times \! 10^{20}$ [cm$^{-2}$ (K \kms )$^{-1}$]. 

The virial theorem mass $M_{\rm VT}$ was calculated as follows:
\begin{equation}
M_{\rm VT} = 3 f_p \frac {S \sigma_{V} ^2} {G}.
\end{equation}
where $G$ is the gravitational constant, and $f_p$ is the projection factor. To maintain consistency with previous studies \citep[e.g.,][]{Oka98a, Oka01b}, the $f_p\! =\! 2.9$ [$\rho (r)\!\propto\! r^{-1}$ case] was employed.

The kinetic energy $E_{\rm kin}$ and expansion time $t_{\rm exp}$ were derived using the following formulae: 
\begin{equation}
E_{\rm kin} = \frac {3} {2} M_{\rm CO} \sigma_V^2
\end{equation}
and 
\begin{equation}
t_{\rm exp} = \frac{S} {\sigma_V}. 
\end{equation}
From $E_{\rm kin}$ and $t_{\rm exp}$, the kinetic power was calculated with 
\begin{equation}
P_{\rm kin} = \frac{E_{\rm kin}}{t_{\rm exp}}.
\end{equation}

The CO {\it J}=3--2/{\it J}=1--0 ratio of each HVCC is derived as 
\begin{equation}
\R32_10 = \frac {\sum I_{\rm CO 3\mbox{--}2}} {\sum I_{\rm CO 1\mbox{--}0}}
\end{equation}
$\sum I_{\rm CO}$ is the summation of the velocity-integrated CO emissions for pixels belonging to the HVCC. 

\clearpage

\begin{ThreePartTable}
\begin{TableNotes}
  \item[\textsc{Note}---] Reference: {\bf a:} \citet{Nagai08},  {\bf b:} \citet{Oka99}, {\bf c:} \citet{Oka01a}, {\bf d:} \citet{Oka08}, {\bf e:} \citet{Oka11}, {\bf f:} \citet{Oka12}, {\bf g:} \citet{Oka16}, {\bf h:} \citet{Matsumura12}, {\bf i:} \citet{Tsujimoto18}, {\bf j:} \citet{Tsujimoto21}.  
\end{TableNotes}

\begin{longtable}{lcccllll}
\label{tb:catalog} \\
\insertTableNotes
\endlastfoot
\caption{Catalog of High-velocity Compact Clouds}\\
\hline\hline
id & $l$ & $b$ & $\VLSR $ & type 1 & type 2 & previous ID & Reference \\
 & ($\arcdeg$) & ($\arcdeg$) & ($\kms$) & &  &  &  \\ \hline
\endfirsthead
\hline\hline
id & $l$ & $b$ & $\VLSR $ & type 1 & type 2 & previous ID & Reference \\
 & ($\arcdeg$) & ($\arcdeg$) & ($\kms$) & &  &  &  \\ \hline
\endhead
\hline
\endfoot
1	&	$-$1.401 	&	$-$0.082 	&	$-$140.5 	&	simple	&		&	CO--1.41--0.08	&	{\bf a}	\\	
2	&	$-$1.366 	&	0.064 	&	$-$207.4 	&	wing	&		&		&		\\	
3	&	$-$1.349 	&	$-$0.121 	&	$-$121.9 	&	wing	&	complex	&		&		\\	
4	&	$-$1.302 	&	$-$0.136 	&	$-$113.5 	&	complex	&		&		&		\\	
5	&	$-$1.231 	&	$-$0.092 	&	$-$169.3 	&	wing	&	complex	&		&		\\	
6	&	$-$1.199 	&	$-$0.109 	&	$-$130.4 	&	shell	&	complex	&	CO--1.21--0.12/S1/$l\!=\!-1\fdg2$	&	{\bf a}; {\bf i}	\\	
7	&	$-$1.071 	&	$-$0.242 	&	$-$69.4 	&	wing	&	shell	&		&		\\	
8	&	$-$0.949 	&	$-$0.027 	&	$-$109.2 	&	complex	&	shell	&		&		\\	
9	&	$-$0.816 	&	0.239 	&	115.1 	&	complex	&	wing	&	CO--0.81+0.21	&	{\bf a}	\\	
10	&	$-$0.802 	&	$-$0.076 	&	27.1 	&	shell	&		&		&		\\	
11	&	$-$0.802 	&	0.006 	&	21.2 	&	complex	&		&		&		\\	
12	&	$-$0.801 	&	$-$0.077 	&	54.2 	&	simple	&	bridge	&		&		\\	
13	&	$-$0.799 	&	0.126 	&	$-$49.1 	&	complex	&		&	CO--0.80+0.11	&	{\bf a}	\\	
14	&	$-$0.794 	&	$-$0.042 	&	27.9 	&	shell	&	complex	&	CO--0.80--0.04	&	{\bf a}	\\	
15	&	$-$0.781 	&	0.188 	&	28.8 	&	wing	&	complex	&	CO--0.79+0.18	&	{\bf a}	\\	
16	&	$-$0.749 	&	$-$0.029 	&	$-$163.4 	&	bridge	&		&		&		\\	
17	&	$-$0.706 	&	$-$0.027 	&	$-$55.9 	&	bridge	&	complex	&	CO--0.71--0.02/Pigtail	&	{\bf a}; {\bf h}	\\	
18	&	$-$0.704 	&	$-$0.109 	&	$-$42.3 	&	shell	&		&		&		\\	
19	&	$-$0.687 	&	$-$0.044 	&	$-$31.3 	&	complex	&		&		&		\\	
20	&	$-$0.682 	&	$-$0.244 	&	$-$154.1 	&	complex	&	wing	&	CO--0.67--0.22	&	{\bf a}	\\	
21	&	$-$0.664 	&	$-$0.242 	&	$-$18.6 	&	complex	&		&	CO--0.65--0.25	&	{\bf a}	\\	
22	&	$-$0.641 	&	$-$0.002 	&	$-$100.8 	&	complex	&	bridge	&		&		\\	
23	&	$-$0.639 	&	$-$0.137 	&	$-$34.7 	&	complex	&	bridge	&		&		\\	
24	&	$-$0.637 	&	$-$0.156 	&	33.9 	&	simple	&	complex	&		&		\\	
25	&	$-$0.634 	&	0.051 	&	$-$118.5 	&	complex	&	shell	&		&		\\	
26	&	$-$0.631 	&	0.078 	&	$-$109.2 	&	complex	&	shell	&	CO--0.63+0.07	&	{\bf a}	\\	
27	&	$-$0.622 	&	$-$0.017 	&	$-$44.0 	&	wing	&	complex	&		&		\\	
28	&	$-$0.612 	&	$-$0.186 	&	$-$140.5 	&	complex	&		&		&		\\	
29	&	$-$0.587 	&	$-$0.106 	&	$-$62.7 	&	wing	&		&		&		\\	
30	&	$-$0.586 	&	$-$0.202 	&	$-$55.9 	&	shell	&	complex	&		&		\\	
31	&	$-$0.582 	&	$-$0.001 	&	$-$45.7 	&	bridge	&		&		&		\\	
32	&	$-$0.579 	&	$-$0.224 	&	$-$149.0 	&	complex	&	wing	&		&		\\	
33	&	$-$0.579 	&	0.023 	&	$-$95.7 	&	wing	&	shell	&		&		\\	
34	&	$-$0.574 	&	$-$0.162 	&	$-$22.9 	&	complex	&		&		&		\\	
35	&	$-$0.562 	&	0.081 	&	$-$114.3 	&	complex	&	bridge	&		&		\\	
36	&	$-$0.547 	&	$-$0.181 	&	$-$84.7 	&	shell	&		&	CO--0.54$-$0.17	&	{\bf a}	\\	
37	&	$-$0.546 	&	$-$0.114 	&	$-$198.1 	&	complex	&	wing	&	CO--0.55--0.13	&	{\bf a}	\\	
38	&	$-$0.542 	&	$-$0.114 	&	71.1 	&	complex	&		&		&		\\	
39	&	$-$0.537 	&	$-$0.079 	&	$-$66.9 	&	simple	&		&		&		\\	
40	&	$-$0.526 	&	$-$0.167 	&	$-$66.0 	&	shell	&		&	CO--0.54--0.17	&	{\bf a}	\\	
41	&	$-$0.524 	&	$-$0.131 	&	$-$66.0 	&	wing	&	complex	&		&		\\	
42	&	$-$0.522 	&	0.024 	&	$-$94.0 	&	shell	&	complex	&		&		\\	
43	&	$-$0.506 	&	$-$0.211 	&	$-$95.7 	&	complex	&	bridge	&		&		\\	
44	&	$-$0.499 	&	0.118 	&	$-$165.1 	&	shell	&	wing	&	CO--0.51+0.12	&	{\bf a}	\\	
45	&	$-$0.496 	&	0.066 	&	$-$143.9 	&	bridge	&	complex	&		&		\\	
46	&	$-$0.489 	&	$-$0.074 	&	$-$200.7 	&	simple	&		&		&		\\	
47	&	$-$0.481 	&	0.021 	&	$-$121.1 	&	shell	&	complex	&		&		\\	
48	&	$-$0.434 	&	$-$0.191 	&	$-$30.5 	&	complex	&		&		&		\\	
49	&	$-$0.431 	&	$-$0.069 	&	$-$8.5 	&	complex	&	bridge	&		&		\\	
50	&	$-$0.402 	&	$-$0.237 	&	58.4 	&	wing	&		&		&		\\	
51	&	$-$0.402 	&	$-$0.222 	&	$-$83.8 	&	simple	&	shell	&	CO--0.40--0.22/CO--0.41--0.23	&	{\bf f}; {\bf g}	\\	
52	&	$-$0.386 	&	$-$0.137 	&	20.3 	&	complex	&		&		&		\\	
53	&	$-$0.356 	&	0.009 	&	72.0 	&	wing	&	bridge	&		&		\\	
54	&	$-$0.354 	&	$-$0.026 	&	78.7 	&	wing	&	complex	&	CO--0.36--0.02	&	{\bf a}	\\	
55	&	$-$0.337 	&	$-$0.111 	&	17.8 	&	simple	&	complex	&		&		\\	
56	&	$-$0.327 	&	$-$0.116 	&	$-$51.6 	&	wing	&		&		&		\\	
57	&	$-$0.326 	&	$-$0.179 	&	$-$74.5 	&	simple	&	bridge	&		&		\\	
58	&	$-$0.309 	&	$-$0.069 	&	54.2 	&	simple	&		&	CO--0.30--0.07	&	{\bf a}	\\	
59	&	$-$0.309 	&	$-$0.069 	&	21.2 	&	complex	&		&	CO--0.30--0.07	&	{\bf a}	\\	
60	&	$-$0.306 	&	0.028 	&	$-$32.2 	&	shell	&		&		&		\\	
61	&	$-$0.274 	&	$-$0.004 	&	$-$64.3 	&	complex	&	bridge	&		&		\\	
62	&	$-$0.272 	&	$-$0.061 	&	$-$63.5 	&	complex	&		&		&		\\	
63	&	$-$0.266 	&	0.019 	&	78.7 	&	shell	&	wing	&		&		\\	
64	&	$-$0.257 	&	0.019 	&	$-$94.8 	&	simple	&		&	CO--0.27+0.04	&	{\bf a}	\\	
65	&	$-$0.244 	&	$-$0.009 	&	25.4 	&	bridge	&		&		&		\\	
66	&	$-$0.229 	&	$-$0.092 	&	8.5 	&	wing	&	bridge	&		&		\\	
67	&	$-$0.172 	&	0.021 	&	63.5 	&	shell	&		&	CO--0.18+0.02	&	{\bf a}	\\	
68	&	$-$0.171 	&	$-$0.076 	&	$-$28.8 	&	wing	&	complex	&		&		\\	
69	&	$-$0.162 	&	$-$0.094 	&	$-$27.1 	&	complex	&	wing	&		&		\\	
70	&	$-$0.161 	&	$-$0.061 	&	71.1 	&	wing	&	bridge	&	CO--0.17--0.07	&	{\bf a}	\\	
71	&	$-$0.129 	&	0.021 	&	27.1 	&	wing	&	bridge	&	CO--0.13+0.03	&	{\bf a}	\\	
72	&	$-$0.106 	&	$-$0.101 	&	$-$7.6 	&	wing	&		&		&		\\	
73	&	$-$0.101 	&	0.014 	&	73.7 	&	complex	&	wing	&		&		\\	
74	&	$-$0.091 	&	$-$0.132 	&	54.2 	&	complex	&	wing	&		&		\\	
75	&	$-$0.091 	&	$-$0.016 	&	$-$124.5 	&	simple	&		&		&		\\	
76	&	$-$0.067 	&	$-$0.047 	&	$-$82.1 	&	simple	&		&	CND	&		\\	
77	&	$-$0.067 	&	$-$0.046 	&	$-$66.0 	&	simple	&		&	CND	&		\\	
78	&	$-$0.057 	&	$-$0.067 	&	80.4 	&	wing	&		&		&		\\	
79	&	$-$0.054 	&	0.021 	&	90.6 	&	simple	&	wing	&		&		\\	
80	&	$-$0.042 	&	$-$0.094 	&	76.2 	&	shell	&	wing	&	CO--0.04--0.08	&	{\bf a}	\\	
81	&	$-$0.042 	&	$-$0.046 	&	105.8 	&	simple	&		&	CO--0.03--0.05/CND	&	{\bf a}	\\	
82	&	$-$0.039 	&	$-$0.064 	&	$-$82.1 	&	simple	&		&	C1/CND	&	{\bf a}; {\bf e}	\\	
83	&	$-$0.006 	&	$-$0.194 	&	84.7 	&	simple	&		&		&		\\	
84	&	$-$0.001 	&	$-$0.092 	&	16.9 	&	shell	&	wing	&	CO 0.00--0.11	&	{\bf a}	\\	
85	&	0.016 	&	$-$0.021 	&	102.4 	&	simple	&	shell	&	CO 0.01--0.03/CO 0.02--0.02	&	{\bf a}; {\bf b}; {\bf d}	\\	
86	&	0.019 	&	$-$0.007 	&	54.2 	&	bridge	&		&	CO 0.01$-$0.03/CO 0.02--0.02	&	{\bf a}; {\bf b}; {\bf d}	\\	
87	&	0.019 	&	0.069 	&	86.4 	&	wing	&		&		&		\\	
88	&	0.021 	&	$-$0.006 	&	53.3 	&	complex	&	bridge	&	CO 0.01--0.03/CO 0.02--0.02	&	{\bf a}; {\bf b}; {\bf d}	\\	
89	&	0.023 	&	$-$0.006 	&	80.4 	&	simple	&	shell	&	CO 0.01--0.03/CO 0.02--0.02	&	{\bf a}; {\bf b}; {\bf d}	\\	
90	&	0.053 	&	$-$0.032 	&	$-$37.3 	&	wing	&	bridge	&		&		\\	
91	&	0.053 	&	$-$0.017 	&	71.1 	&	shell	&	simple	&		&		\\	
92	&	0.059 	&	$-$0.116 	&	37.3 	&	complex	&		&		&		\\	
93	&	0.061 	&	$-$0.081 	&	45.7 	&	complex	&	bridge	&		&		\\	
94	&	0.089 	&	0.053 	&	139.7 	&	simple	&		&		&		\\	
95	&	0.093 	&	$-$0.127 	&	31.3 	&	simple	&		&	CO 0.10--0.11	&	{\bf a}	\\	
96	&	0.104 	&	0.004 	&	50.0 	&	complex	&	wing	&		&		\\	
97	&	0.106 	&	$-$0.081 	&	41.5 	&	shell	&	bridge	&		&		\\	
98	&	0.109 	&	0.041 	&	94.8 	&	bridge	&		&		&		\\	
99	&	0.116 	&	$-$0.021 	&	83.0 	&	wing	&		&		&		\\	
100	&	0.121 	&	$-$0.211 	&	17.8 	&	wing	&		&	CO 0.10--0.24	&	{\bf a}	\\	
101	&	0.144 	&	0.011 	&	31.3 	&	wing	&		&		&		\\	
102	&	0.154 	&	$-$0.017 	&	52.5 	&	wing	&	complex	&		&		\\	
103	&	0.159 	&	$-$0.097 	&	97.4 	&	wing	&		&	CO 0.18--0.10	&	{\bf a}	\\	
104	&	0.161 	&	$-$0.002 	&	33.9 	&	wing	&		&		&		\\	
105	&	0.163 	&	0.063 	&	24.6 	&	simple	&		&		&		\\	
106	&	0.193 	&	$-$0.114 	&	104.1 	&	complex	&	wing	&		&		\\	
107	&	0.206 	&	0.064 	&	15.2 	&	complex	&		&		&		\\	
108	&	0.221 	&	$-$0.011 	&	27.9 	&	complex	&	bridge	&		&		\\	
109	&	0.274 	&	$-$0.061 	&	39.8 	&	bridge	&	complex	&		&		\\	
110	&	0.284 	&	$-$0.141 	&	61.8 	&	shell	&		&		&		\\	
111	&	0.284 	&	0.038 	&	$-$39.8 	&	wing	&		&		&		\\	
112	&	0.301 	&	$-$0.004 	&	44.0 	&	bridge	&		&		&		\\	
113	&	0.316 	&	$-$0.102 	&	75.4 	&	wing	&	complex	&		&		\\	
114	&	0.321 	&	0.029 	&	13.5 	&	simple	&	complex	&		&		\\	
115	&	0.364 	&	$-$0.097 	&	106.7 	&	wing	&	shell	&		&		\\	
116	&	0.366 	&	$-$0.081 	&	38.1 	&	wing	&	shell	&		&		\\	
117	&	0.393 	&	$-$0.087 	&	106.7 	&	wing	&	bridge	&		&		\\	
118	&	0.418 	&	$-$0.057 	&	112.6 	&	shell	&		&		&		\\	
119	&	0.429 	&	$-$0.189 	&	43.2 	&	complex	&	shell	&	CO 0.44--0.19	&	{\bf a}	\\	
120	&	0.453 	&	0.244 	&	113.5 	&	simple	&		&		&		\\	
121	&	0.454 	&	0.051 	&	87.2 	&	bridge	&	wing	&		&		\\	
122	&	0.456 	&	$-$0.027 	&	116.8 	&	wing	&		&		&		\\	
123	&	0.466 	&	$-$0.027 	&	33.0 	&	complex	&		&		&		\\	
124	&	0.493 	&	$-$0.031 	&	120.2 	&	wing	&		&		&		\\	
125	&	0.499 	&	0.026 	&	57.6 	&	wing	&	simple	&		&		\\	
126	&	0.529 	&	$-$0.012 	&	74.5 	&	bridge	&		&		&		\\	
127	&	0.534 	&	0.108 	&	98.2 	&	simple	&	complex	&		&		\\	
128	&	0.581 	&	0.136 	&	125.3 	&	bridge	&		&		&		\\	
129	&	0.603 	&	$-$0.029 	&	111.8 	&	wing	&		&		&		\\	
130	&	0.604 	&	$-$0.144 	&	96.5 	&	complex	&	bridge	&		&		\\	
131	&	0.606 	&	0.006 	&	127.0 	&	bridge	&		&		&		\\	
132	&	0.616 	&	0.033 	&	111.8 	&	complex	&	bridge	&		&		\\	
133	&	0.629 	&	$-$0.029 	&	118.5 	&	complex	&	wing	&		&		\\	
134	&	0.636 	&	0.036 	&	83.0 	&	wing	&		&		&		\\	
135	&	0.636 	&	0.139 	&	138.9 	&	wing	&	complex	&		&		\\	
136	&	0.659 	&	$-$0.066 	&	88.9 	&	complex	&	shell	&		&		\\	
137	&	0.663 	&	$-$0.014 	&	112.6 	&	wing	&		&		&		\\	
138	&	0.671 	&	$-$0.032 	&	94.0 	&	wing	&		&		&		\\	
139	&	0.714 	&	$-$0.071 	&	84.7 	&	bridge	&		&		&		\\	
140	&	0.749 	&	$-$0.149 	&	104.1 	&	wing	&	complex	&		&		\\	
141	&	0.761 	&	$-$0.071 	&	50.8 	&	simple	&	shell	&		&		\\	
142	&	0.816 	&	0.036 	&	95.7 	&	bridge	&		&		&		\\	
143	&	0.841 	&	0.036 	&	$-$102.4 	&	simple	&		&	CO 0.83+0.03	&	{\bf a}	\\	
144	&	0.846 	&	$-$0.177 	&	30.5 	&	complex	&		&	CO 0.84--0.18	&	{\bf a}	\\	
145	&	0.856 	&	$-$0.219 	&	91.4 	&	simple	&		&		&		\\	
146	&	0.884 	&	$-$0.167 	&	97.4 	&	bridge	&	complex	&		&		\\	
147	&	0.888 	&	0.148 	&	82.1 	&	simple	&	complex	&		&		\\	
148	&	0.889 	&	$-$0.076 	&	49.1 	&	bridge	&		&	CO 0.88--0.07	&	{\bf a}	\\	
149	&	0.893 	&	0.189 	&	98.2 	&	wing	&		&		&		\\	
150	&	0.918 	&	$-$0.012 	&	28.8 	&	shell	&		&		&		\\	
151	&	0.994 	&	$-$0.121 	&	55.9 	&	shell	&		&		&		\\	
152	&	1.129 	&	0.188 	&	131.2 	&	complex	&	shell	&	CO 1.16+0.21	&	{\bf a}	\\	
153	&	1.173 	&	0.084 	&	39.8 	&	wing	&	complex	&	CO 1.17+0.08	&	{\bf a}	\\	
154	&	1.216 	&	0.111 	&	136.3 	&	shell	&		&	CO 1.23+0.13/$l\!=\!+1\fdg3$	&	{\bf a}; {\bf c}; {\bf j}	\\	
155	&	1.221 	&	$-$0.196 	&	116.8 	&	wing	&	bridge	&		&		\\	
156	&	1.223 	&	0.059 	&	128.7 	&	shell	&		&	$l\!=\!+1\fdg3$	&	{\bf c}; {\bf j}	\\	
157	&	1.223 	&	$-$0.099 	&	112.6 	&	wing	&		&		&		\\	
158	&	1.241 	&	0.001 	&	124.5 	&	shell	&		&	$l\!=\!+1\fdg3$	&	{\bf c}; {\bf j}	\\	
159	&	1.243 	&	$-$0.046 	&	105.0 	&	complex	&		&	CO 1.25$-$0.03	&	{\bf a}	\\	
160	&	1.248 	&	0.124 	&	184.6 	&	wing	&		&	$l\!=\!+1\fdg3$ 	&	{\bf c}; {\bf j}	\\	
161	&	1.256 	&	0.081 	&	138.9 	&	shell	&		&	$l\!=\!+1\fdg3$	&	{\bf c}; {\bf j}	\\	
162	&	1.256 	&	0.198 	&	71.1 	&	wing	&	shell	&		&		\\	
163	&	1.271 	&	0.098 	&	93.1 	&	complex	&	shell	&	N3/$l\!=\!+1\fdg3$ 	&	{\bf j}	\\	
164	&	1.271 	&	0.249 	&	152.4 	&	simple	&	wing	&		&		\\	
165	&	1.278 	&	$-$0.241 	&	154.1 	&	simple	&	bridge	&	CO 1.28--0.24	&	{\bf a}	\\	
166	&	1.288 	&	0.034 	&	126.2 	&	shell	&		&	$l\!=\!+1\fdg3$	&	{\bf c}; {\bf j}	\\	
167	&	1.331 	&	0.201 	&	112.6 	&	wing	&	bridge	&		&		\\	
168	&	1.354 	&	0.136 	&	127.0 	&	wing	&	shell	&		&		\\	
169	&	1.368 	&	0.023 	&	51.6 	&	wing	&		&		&		\\	
170	&	1.371 	&	0.164 	&	51.6 	&	wing	&		&		&		\\	
171	&	1.378 	&	0.203 	&	50.8 	&	wing	&		&	CO 1.39+0.20	&	{\bf a}	\\	
172	&	1.381 	&	$-$0.207 	&	145.6 	&	shell	&		&	CO 1.38--0.21	&	{\bf a}	\\	
173	&	1.413 	&	$-$0.236 	&	94.0 	&	complex	&		&		&		\\	
174	&	1.414 	&	0.098 	&	38.9 	&	wing	&	simple	&		&		\\	
175	&	1.434 	&	0.173 	&	40.6 	&	simple	&	shell	&		&		\\	
176	&	1.473 	&	0.038 	&	108.4 	&	wing	&		&		&		\\	
177	&	1.494 	&	0.228 	&	138.0 	&	shell	&		&		&		\\	
178	&	1.521 	&	$-$0.117 	&	99.1 	&	wing	&		&		&		\\	
179	&	1.538 	&	0.183 	&	80.4 	&	bridge	&		&		&		\\	
180	&	1.548 	&	$-$0.129 	&	80.4 	&	wing	&	bridge	&		&		\\	
181	&	1.591 	&	0.018 	&	160.0 	&	simple	&		&	CO 1.60--0.01/G1.6--0.025	&	{\bf a}	\\	
182	&	1.608 	&	$-$0.077 	&	143.9 	&	simple	&	bridge	&	G1.6--0.025	&		\\	
183	&	1.616 	&	0.246 	&	156.6 	&	simple	&		&		&		\\	
184	&	1.789 	&	$-$0.134 	&	36.4 	&	wing	&	complex	&		&		\\	
\hline
\end{longtable}

\clearpage

\begin{TableNotes}
  \item[\textsc{Note}---] $R_{3\mbox{--}2/1\mbox{--}0}$ of HVCCs located outside of the CO {\it J}=1--0 survey coverage \citep{Tokuyama19} left blank. 
\end{TableNotes}

\begin{longtable}{lccccccccc}
\label{tb:param} \\
\insertTableNotes
\endlastfoot
\caption{Physical Parameters of High-velocity Compact Clouds}\\
\hline\hline
id & $S$ & ${\sigma}_{V}$ & $L_{\rm CO}$ & $M_{\rm CO}$ & $M_{\rm VT}$ & $E_{\rm kin}$ & $t_{\rm exp}$ & $P_{\rm kin}$ & $R_{3\mbox{--}2/1\mbox{--}0}$   \\
 & (pc) & (\kms) & {\footnotesize ($10^2$ K \kms pc$^2$)} & ($10^2$ $M_{\sun}$) & ($10^5$ $M_{\sun}$) & ($10^{48}$ erg) & ($10^4$ yr) & ($10^{36}$ erg s$^{-1}$) &  \\ \hline
\endfirsthead
\hline\hline
id & $S$ & ${\sigma}_{V}$ & $L_{\rm CO}$ & $M_{\rm CO}$ & $M_{\rm VT}$ & $E_{\rm kin}$ & $t_{\rm exp}$ & $P_{\rm kin}$ & $R_{3\mbox{--}2/1\mbox{--}0}$   \\
 & (pc) & (\kms) & {\footnotesize ($10^2$ K \kms pc$^2$)} & ($10^2$ $M_{\sun}$) & ($10^5$ $M_{\sun}$) & ($10^{48}$ erg) & ($10^4$ yr) & ($10^{36}$ erg s$^{-1}$) &  \\ \hline
\endhead
\hline
\endfoot
1	&	0.97 	&	25.7 	&	6.4 	&	19.6 	&	12.9 	&	38.5 	&	3.7 	&	33.2 	&		\\
2	&	0.33 	&	20.7 	&	1.0 	&	3.1 	&	2.9 	&	4.0 	&	1.6 	&	8.0 	&		\\
3	&	0.61 	&	22.9 	&	1.7 	&	5.2 	&	6.5 	&	8.1 	&	2.6 	&	9.8 	&		\\
4	&	0.80 	&	24.3 	&	2.1 	&	6.5 	&	9.5 	&	11.4 	&	3.2 	&	11.2 	&		\\
5	&	0.62 	&	31.5 	&	2.8 	&	8.6 	&	12.4 	&	25.4 	&	1.9 	&	41.9 	&		\\
6	&	1.84 	&	30.5 	&	19.8 	&	60.5 	&	34.5 	&	167.8 	&	5.9 	&	90.4 	&		\\
7	&	0.62 	&	21.1 	&	2.1 	&	6.4 	&	5.6 	&	8.4 	&	2.9 	&	9.3 	&		\\
8	&	1.06 	&	23.5 	&	5.1 	&	15.5 	&	11.8 	&	25.4 	&	4.4 	&	18.3 	&		\\
9	&	0.99 	&	21.2 	&	3.9 	&	12.0 	&	8.9 	&	16.1 	&	4.5 	&	11.2 	&	0.85 	\\
10	&	0.90 	&	27.5 	&	5.9 	&	18.0 	&	13.8 	&	40.7 	&	3.2 	&	40.3 	&	0.83 	\\
11	&	1.48 	&	32.9 	&	8.3 	&	25.5 	&	32.4 	&	82.3 	&	4.4 	&	59.4 	&	0.81 	\\
12	&	1.42 	&	20.8 	&	6.5 	&	19.9 	&	12.4 	&	25.6 	&	6.7 	&	12.2 	&	0.94 	\\
13	&	1.32 	&	21.3 	&	8.5 	&	25.9 	&	12.1 	&	35.1 	&	6.1 	&	18.4 	&	1.03 	\\
14	&	1.08 	&	22.6 	&	5.5 	&	17.0 	&	11.2 	&	25.9 	&	4.7 	&	17.6 	&	0.90 	\\
15	&	1.22 	&	20.9 	&	7.1 	&	21.6 	&	10.7 	&	28.2 	&	5.7 	&	15.7 	&	0.78 	\\
16	&	0.96 	&	21.7 	&	2.9 	&	8.9 	&	9.2 	&	12.6 	&	4.3 	&	9.2 	&	1.17 	\\
17	&	2.01 	&	27.8 	&	12.4 	&	38.0 	&	31.5 	&	87.9 	&	7.1 	&	39.4 	&	1.22 	\\
18	&	2.49 	&	23.1 	&	15.8 	&	48.3 	&	26.7 	&	76.6 	&	10.5 	&	23.0 	&	1.28 	\\
19	&	0.95 	&	44.7 	&	7.7 	&	23.4 	&	38.4 	&	139.7 	&	2.1 	&	213.3 	&	1.18 	\\
20	&	1.46 	&	21.1 	&	8.3 	&	25.5 	&	13.2 	&	33.9 	&	6.8 	&	15.9 	&	1.21 	\\
21	&	0.69 	&	35.0 	&	3.6 	&	11.0 	&	17.0 	&	40.0 	&	1.9 	&	66.1 	&	1.12 	\\
22	&	0.83 	&	37.2 	&	6.5 	&	20.0 	&	23.1 	&	82.2 	&	2.2 	&	119.5 	&	1.01 	\\
23	&	1.24 	&	20.5 	&	6.1 	&	18.8 	&	10.6 	&	23.7 	&	5.9 	&	12.7 	&	1.04 	\\
24	&	0.51 	&	30.2 	&	2.7 	&	8.4 	&	9.3 	&	22.8 	&	1.6 	&	44.0 	&	1.10 	\\
25	&	1.02 	&	22.0 	&	5.7 	&	17.4 	&	10.0 	&	25.2 	&	4.5 	&	17.6 	&	1.06 	\\
26	&	0.53 	&	20.4 	&	1.8 	&	5.4 	&	4.4 	&	6.7 	&	2.5 	&	8.5 	&	1.08 	\\
27	&	0.45 	&	35.9 	&	2.0 	&	6.3 	&	11.7 	&	24.1 	&	1.2 	&	62.2 	&	1.20 	\\
28	&	0.53 	&	26.6 	&	1.7 	&	5.1 	&	7.6 	&	10.7 	&	2.0 	&	17.3 	&	1.04 	\\
29	&	0.85 	&	21.6 	&	4.4 	&	13.5 	&	8.0 	&	18.9 	&	3.8 	&	15.6 	&	1.28 	\\
30	&	1.26 	&	27.7 	&	7.5 	&	23.1 	&	19.6 	&	52.9 	&	4.4 	&	37.7 	&	1.08 	\\
31	&	0.76 	&	35.2 	&	3.1 	&	9.6 	&	18.9 	&	35.5 	&	2.1 	&	53.4 	&	1.17 	\\
32	&	1.52 	&	28.6 	&	5.6 	&	17.2 	&	25.1 	&	42.0 	&	5.2 	&	25.8 	&	1.19 	\\
33	&	1.15 	&	28.5 	&	9.3 	&	28.6 	&	18.9 	&	69.3 	&	3.9 	&	55.9 	&	0.94 	\\
34	&	1.53 	&	34.5 	&	15.7 	&	48.1 	&	36.8 	&	170.7 	&	4.3 	&	124.7 	&	1.20 	\\
35	&	0.59 	&	20.7 	&	2.2 	&	6.7 	&	5.1 	&	8.6 	&	2.8 	&	9.8 	&	1.02 	\\
36	&	1.15 	&	35.1 	&	8.7 	&	26.5 	&	28.6 	&	97.6 	&	3.2 	&	97.1 	&	1.18 	\\
37	&	0.74 	&	20.7 	&	3.9 	&	12.1 	&	6.4 	&	15.4 	&	3.5 	&	13.9 	&	1.25 	\\
38	&	1.25 	&	20.5 	&	6.0 	&	18.3 	&	10.6 	&	22.9 	&	5.9 	&	12.2 	&	1.45 	\\
39	&	0.23 	&	23.6 	&	0.6 	&	1.8 	&	2.5 	&	3.1 	&	0.9 	&	10.5 	&	1.68 	\\
40	&	1.79 	&	34.1 	&	18.0 	&	55.2 	&	42.2 	&	191.8 	&	5.1 	&	118.6 	&	1.13 	\\
41	&	1.75 	&	33.4 	&	12.3 	&	37.7 	&	39.5 	&	125.9 	&	5.1 	&	78.2 	&	1.14 	\\
42	&	0.94 	&	21.7 	&	2.9 	&	9.0 	&	9.0 	&	12.6 	&	4.2 	&	9.4 	&	0.92 	\\
43	&	1.22 	&	24.7 	&	4.3 	&	13.2 	&	15.1 	&	24.1 	&	4.8 	&	15.9 	&	1.11 	\\
44	&	1.35 	&	38.4 	&	11.2 	&	34.1 	&	40.2 	&	150.1 	&	3.4 	&	138.6 	&	0.86 	\\
45	&	1.66 	&	30.0 	&	9.2 	&	28.1 	&	30.1 	&	75.7 	&	5.4 	&	44.6 	&	0.85 	\\
46	&	0.44 	&	20.9 	&	1.7 	&	5.1 	&	3.9 	&	6.7 	&	2.0 	&	10.4 	&	2.18 	\\
47	&	1.12 	&	22.3 	&	7.4 	&	22.8 	&	11.2 	&	33.8 	&	4.9 	&	21.9 	&	0.86 	\\
48	&	0.57 	&	21.4 	&	2.0 	&	6.1 	&	5.2 	&	8.4 	&	2.6 	&	10.3 	&	0.88 	\\
49	&	2.34 	&	31.2 	&	31.4 	&	96.1 	&	46.0 	&	278.4 	&	7.3 	&	120.1 	&	1.07 	\\
50	&	0.68 	&	24.7 	&	2.6 	&	7.9 	&	8.3 	&	14.3 	&	2.7 	&	16.9 	&	0.69 	\\
51	&	2.00 	&	34.1 	&	28.4 	&	87.0 	&	46.8 	&	301.4 	&	5.7 	&	166.9 	&	1.20 	\\
52	&	0.89 	&	20.8 	&	4.1 	&	12.5 	&	7.8 	&	16.1 	&	4.2 	&	12.2 	&	0.88 	\\
53	&	0.98 	&	31.7 	&	4.3 	&	13.3 	&	20.0 	&	40.0 	&	3.0 	&	41.9 	&	0.91 	\\
54	&	0.57 	&	20.4 	&	2.0 	&	6.1 	&	4.8 	&	7.6 	&	2.7 	&	8.8 	&	0.85 	\\
55	&	0.99 	&	20.7 	&	4.0 	&	12.4 	&	8.5 	&	15.8 	&	4.7 	&	10.7 	&	1.03 	\\
56	&	0.91 	&	22.0 	&	2.0 	&	6.2 	&	8.8 	&	8.9 	&	4.0 	&	7.0 	&	1.10 	\\
57	&	1.20 	&	25.0 	&	6.3 	&	19.3 	&	15.2 	&	35.9 	&	4.7 	&	24.2 	&	1.14 	\\
58	&	2.29 	&	31.7 	&	15.1 	&	37.2 	&	46.6 	&	57.8 	&	7.1 	&	12.9 	&	0.96 	\\
59	&	3.32 	&	22.8 	&	12.2 	&	46.3 	&	34.9 	&	138.8 	&	14.2 	&	62.3 	&	0.95 	\\
60	&	2.73 	&	28.8 	&	19.8 	&	60.5 	&	45.7 	&	149.7 	&	9.2 	&	51.3 	&	1.24 	\\
61	&	0.50 	&	23.7 	&	2.1 	&	6.5 	&	5.7 	&	10.9 	&	2.1 	&	16.6 	&	1.22 	\\
62	&	1.04 	&	21.2 	&	5.8 	&	17.7 	&	9.5 	&	23.9 	&	4.8 	&	15.9 	&	1.39 	\\
63	&	1.34 	&	22.1 	&	5.0 	&	15.2 	&	13.2 	&	22.2 	&	5.9 	&	11.9 	&	1.04 	\\
64	&	2.02 	&	27.9 	&	15.6 	&	47.8 	&	31.7 	&	110.8 	&	7.1 	&	49.7 	&	1.38 	\\
65	&	0.37 	&	39.4 	&	1.3 	&	4.0 	&	11.7 	&	18.4 	&	0.9 	&	63.3 	&	1.05 	\\
66	&	0.37 	&	20.6 	&	0.9 	&	2.9 	&	3.1 	&	3.6 	&	1.7 	&	6.6 	&	1.05 	\\
67	&	1.99 	&	31.7 	&	21.6 	&	66.2 	&	40.4 	&	198.6 	&	6.1 	&	102.7 	&	1.15 	\\
68	&	1.47 	&	23.5 	&	7.6 	&	23.1 	&	16.5 	&	38.2 	&	6.1 	&	19.9 	&	1.17 	\\
69	&	1.27 	&	24.1 	&	7.9 	&	24.2 	&	14.8 	&	41.8 	&	5.2 	&	25.7 	&	1.18 	\\
70	&	1.96 	&	26.5 	&	9.8 	&	30.1 	&	27.8 	&	63.2 	&	7.2 	&	27.8 	&	0.71 	\\
71	&	0.67 	&	28.1 	&	2.6 	&	7.9 	&	10.7 	&	18.5 	&	2.3 	&	25.3 	&	1.04 	\\
72	&	0.75 	&	33.8 	&	5.7 	&	17.3 	&	17.3 	&	59.0 	&	2.2 	&	86.4 	&	0.77 	\\
73	&	1.78 	&	22.8 	&	6.8 	&	20.7 	&	18.7 	&	31.9 	&	7.6 	&	13.2 	&	0.91 	\\
74	&	1.82 	&	32.5 	&	13.2 	&	40.3 	&	39.0 	&	127.0 	&	5.5 	&	73.6 	&	0.72 	\\
75	&	0.49 	&	21.8 	&	2.2 	&	6.6 	&	4.7 	&	9.4 	&	2.2 	&	13.7 	&	1.88 	\\
76	&	5.33 	&	43.8 	&	24.8 	&	83.2 	&	206.8 	&	368.0 	&	11.9 	&	178.7 	&	1.93 	\\
77	&	2.57 	&	38.5 	&	27.2 	&	75.9 	&	77.1 	&	434.9 	&	6.5 	&	116.0 	&	2.19 	\\
78	&	0.29 	&	21.7 	&	0.8 	&	2.4 	&	2.8 	&	3.4 	&	1.3 	&	8.3 	&	1.00 	\\
79	&	0.41 	&	20.2 	&	1.3 	&	4.1 	&	3.4 	&	5.0 	&	2.0 	&	8.2 	&	0.89 	\\
80	&	3.10 	&	21.5 	&	19.8 	&	41.8 	&	29.0 	&	69.0 	&	14.0 	&	16.4 	&	0.75 	\\
81	&	3.22 	&	23.5 	&	13.7 	&	60.7 	&	35.9 	&	83.9 	&	13.4 	&	18.9 	&	1.69 	\\
82	&	0.78 	&	31.2 	&	5.2 	&	15.8 	&	15.4 	&	45.8 	&	2.5 	&	59.2 	&	1.89 	\\
83	&	0.53 	&	21.0 	&	1.4 	&	4.1 	&	4.7 	&	5.5 	&	2.4 	&	7.1 	&	1.00 	\\
84	&	5.46 	&	29.2 	&	30.1 	&	92.1 	&	94.0 	&	234.1 	&	18.3 	&	40.6 	&	0.83 	\\
85	&	2.16 	&	51.0 	&	41.3 	&	126.5 	&	113.6 	&	981.4 	&	4.1 	&	751.1 	&	1.13 	\\
86	&	6.00 	&	20.9 	&	26.3 	&	80.4 	&	52.9 	&	104.8 	&	28.0 	&	11.9 	&	0.74 	\\
87	&	0.89 	&	21.5 	&	4.2 	&	12.8 	&	8.3 	&	17.7 	&	4.0 	&	13.9 	&	0.66 	\\
88	&	2.65 	&	37.8 	&	31.2 	&	95.4 	&	76.2 	&	406.1 	&	6.8 	&	188.3 	&	0.87 	\\
89	&	2.39 	&	20.5 	&	3.7 	&	11.4 	&	20.2 	&	14.3 	&	11.4 	&	4.0 	&	0.79 	\\
90	&	2.37 	&	28.4 	&	6.2 	&	18.9 	&	38.7 	&	45.6 	&	8.2 	&	17.7 	&	1.06 	\\
91	&	2.12 	&	35.5 	&	10.7 	&	32.9 	&	53.9 	&	123.5 	&	5.8 	&	67.2 	&	0.80 	\\
92	&	0.71 	&	20.7 	&	2.7 	&	8.4 	&	6.1 	&	10.7 	&	3.3 	&	10.2 	&	0.73 	\\
93	&	1.85 	&	36.6 	&	17.9 	&	54.8 	&	49.9 	&	218.8 	&	4.9 	&	140.6 	&	0.83 	\\
94	&	0.76 	&	22.0 	&	2.3 	&	7.1 	&	7.5 	&	10.3 	&	3.4 	&	9.7 	&	0.88 	\\
95	&	2.31 	&	29.2 	&	21.9 	&	66.9 	&	39.7 	&	170.1 	&	7.7 	&	69.9 	&	0.80 	\\
96	&	1.38 	&	26.0 	&	8.2 	&	25.0 	&	18.7 	&	50.3 	&	5.2 	&	30.8 	&	0.71 	\\
97	&	2.53 	&	29.7 	&	20.7 	&	63.2 	&	45.0 	&	166.3 	&	8.3 	&	63.5 	&	0.86 	\\
98	&	1.41 	&	27.3 	&	7.4 	&	22.7 	&	21.2 	&	50.3 	&	5.0 	&	31.6 	&	0.73 	\\
99	&	0.94 	&	20.0 	&	5.0 	&	15.2 	&	7.6 	&	18.1 	&	4.6 	&	12.6 	&	0.74 	\\
100	&	1.15 	&	20.8 	&	7.7 	&	23.6 	&	10.0 	&	30.4 	&	5.4 	&	17.9 	&	0.91 	\\
101	&	1.76 	&	61.6 	&	6.7 	&	20.5 	&	134.7 	&	231.7 	&	2.8 	&	263.3 	&	0.81 	\\
102	&	0.52 	&	24.3 	&	2.6 	&	7.8 	&	6.2 	&	13.8 	&	2.1 	&	20.9 	&	0.75 	\\
103	&	2.60 	&	21.4 	&	8.9 	&	27.1 	&	24.1 	&	37.0 	&	11.9 	&	9.9 	&	0.96 	\\
104	&	0.53 	&	20.6 	&	1.6 	&	4.8 	&	4.5 	&	6.0 	&	2.5 	&	7.6 	&	0.81 	\\
105	&	1.00 	&	30.5 	&	8.3 	&	25.5 	&	18.7 	&	71.0 	&	3.2 	&	70.6 	&	1.12 	\\
106	&	1.72 	&	24.1 	&	10.6 	&	32.3 	&	20.2 	&	56.0 	&	7.0 	&	25.5 	&	0.88 	\\
107	&	2.32 	&	33.4 	&	5.8 	&	17.7 	&	52.4 	&	58.9 	&	6.8 	&	27.5 	&	1.04 	\\
108	&	0.63 	&	40.2 	&	2.9 	&	9.0 	&	20.6 	&	43.2 	&	1.5 	&	89.0 	&	0.83 	\\
109	&	1.71 	&	30.7 	&	10.6 	&	32.5 	&	32.6 	&	91.5 	&	5.4 	&	53.3 	&	0.86 	\\
110	&	1.22 	&	23.7 	&	12.3 	&	10.3 	&	13.8 	&	16.8 	&	5.0 	&	6.6 	&	0.87 	\\
111	&	1.92 	&	23.4 	&	3.4 	&	37.7 	&	21.2 	&	63.2 	&	8.0 	&	39.8 	&	1.11 	\\
112	&	0.45 	&	20.9 	&	1.3 	&	4.0 	&	4.0 	&	5.2 	&	2.1 	&	7.8 	&	0.81 	\\
113	&	1.61 	&	30.6 	&	13.3 	&	40.6 	&	30.5 	&	113.6 	&	5.1 	&	70.1 	&	0.85 	\\
114	&	1.93 	&	29.4 	&	19.3 	&	59.1 	&	33.8 	&	152.6 	&	6.4 	&	75.4 	&	0.83 	\\
115	&	0.58 	&	20.1 	&	1.4 	&	4.4 	&	4.7 	&	5.3 	&	2.8 	&	6.0 	&	1.00 	\\
116	&	1.21 	&	21.9 	&	4.0 	&	12.3 	&	11.8 	&	17.7 	&	5.4 	&	10.4 	&	0.83 	\\
117	&	1.21 	&	41.4 	&	6.6 	&	20.3 	&	41.9 	&	103.5 	&	2.9 	&	114.6 	&	0.96 	\\
118	&	1.32 	&	20.9 	&	5.6 	&	17.1 	&	11.6 	&	22.2 	&	6.2 	&	11.3 	&	0.83 	\\
119	&	0.75 	&	22.6 	&	5.0 	&	15.2 	&	7.7 	&	23.1 	&	3.2 	&	22.6 	&	0.86 	\\
120	&	0.57 	&	29.7 	&	3.6 	&	11.0 	&	10.2 	&	29.0 	&	1.9 	&	49.1 	&	0.88 	\\
121	&	1.36 	&	27.9 	&	8.8 	&	27.0 	&	21.4 	&	63.0 	&	4.7 	&	42.1 	&	0.85 	\\
122	&	0.66 	&	30.1 	&	2.6 	&	8.1 	&	12.0 	&	21.8 	&	2.1 	&	32.5 	&	0.86 	\\
123	&	0.91 	&	20.3 	&	5.2 	&	16.0 	&	7.6 	&	19.8 	&	4.4 	&	14.4 	&	0.74 	\\
124	&	1.43 	&	24.6 	&	8.9 	&	27.1 	&	17.4 	&	48.8 	&	5.7 	&	27.2 	&	0.84 	\\
125	&	0.60 	&	27.0 	&	2.7 	&	8.1 	&	8.8 	&	17.6 	&	2.2 	&	25.7 	&	0.89 	\\
126	&	1.04 	&	34.6 	&	3.9 	&	11.9 	&	25.1 	&	42.4 	&	2.9 	&	46.0 	&	0.85 	\\
127	&	1.60 	&	22.5 	&	8.3 	&	25.4 	&	16.5 	&	38.5 	&	6.9 	&	17.5 	&	0.95 	\\
128	&	0.88 	&	24.5 	&	4.0 	&	12.1 	&	10.6 	&	21.7 	&	3.5 	&	19.7 	&	0.93 	\\
129	&	0.52 	&	22.0 	&	2.4 	&	7.5 	&	5.1 	&	10.8 	&	2.3 	&	14.9 	&	0.89 	\\
130	&	1.34 	&	23.2 	&	6.5 	&	19.9 	&	14.5 	&	31.9 	&	5.6 	&	17.9 	&	0.90 	\\
131	&	2.08 	&	30.4 	&	14.4 	&	44.1 	&	38.8 	&	121.5 	&	6.7 	&	57.7 	&	0.97 	\\
132	&	0.53 	&	22.3 	&	2.1 	&	6.4 	&	5.3 	&	9.5 	&	2.3 	&	13.0 	&	0.77 	\\
133	&	1.22 	&	24.2 	&	8.5 	&	26.1 	&	14.5 	&	45.7 	&	4.9 	&	29.3 	&	0.93 	\\
134	&	0.90 	&	26.7 	&	6.0 	&	18.3 	&	13.0 	&	39.0 	&	3.3 	&	37.6 	&	0.73 	\\
135	&	1.29 	&	25.1 	&	6.9 	&	21.2 	&	16.3 	&	39.8 	&	5.0 	&	25.1 	&	1.02 	\\
136	&	1.18 	&	26.2 	&	4.1 	&	12.7 	&	16.3 	&	25.9 	&	4.4 	&	18.7 	&	0.85 	\\
137	&	1.39 	&	24.0 	&	9.0 	&	27.6 	&	16.2 	&	47.6 	&	5.6 	&	26.8 	&	0.80 	\\
138	&	0.77 	&	37.1 	&	7.4 	&	22.7 	&	21.4 	&	93.3 	&	2.0 	&	146.0 	&	0.87 	\\
139	&	2.82 	&	43.3 	&	26.0 	&	79.5 	&	106.7 	&	443.8 	&	6.4 	&	220.9 	&	0.72 	\\
140	&	1.11 	&	22.1 	&	7.9 	&	24.2 	&	10.9 	&	35.1 	&	4.9 	&	22.7 	&	0.88 	\\
141	&	1.79 	&	39.7 	&	32.1 	&	98.1 	&	56.8 	&	460.0 	&	4.4 	&	330.8 	&	0.78 	\\
142	&	1.66 	&	21.1 	&	7.4 	&	22.7 	&	14.9 	&	30.1 	&	7.7 	&	12.4 	&	0.74 	\\
143	&	0.70 	&	21.5 	&	3.8 	&	11.6 	&	6.5 	&	15.9 	&	3.2 	&	15.9 	&	1.21 	\\
144	&	0.81 	&	27.1 	&	4.6 	&	14.2 	&	12.1 	&	31.0 	&	2.9 	&	33.5 	&	0.82 	\\
145	&	1.73 	&	39.9 	&	24.5 	&	74.9 	&	55.6 	&	355.8 	&	4.2 	&	266.6 	&	0.89 	\\
146	&	2.57 	&	47.0 	&	36.5 	&	111.6 	&	114.7 	&	734.7 	&	5.3 	&	435.4 	&	0.91 	\\
147	&	3.72 	&	26.0 	&	30.0 	&	91.7 	&	50.8 	&	185.4 	&	13.9 	&	42.2 	&	0.80 	\\
148	&	1.40 	&	26.8 	&	14.1 	&	43.1 	&	20.3 	&	92.3 	&	5.1 	&	57.3 	&	0.97 	\\
149	&	1.18 	&	23.8 	&	8.8 	&	26.9 	&	13.4 	&	45.3 	&	4.8 	&	29.8 	&	0.85 	\\
150	&	2.64 	&	26.9 	&	11.8 	&	36.1 	&	38.4 	&	77.7 	&	9.6 	&	25.7 	&	0.84 	\\
151	&	0.73 	&	28.8 	&	2.0 	&	6.2 	&	12.3 	&	15.3 	&	2.5 	&	19.6 	&	1.30 	\\
152	&	0.65 	&	20.3 	&	2.2 	&	6.8 	&	5.4 	&	8.4 	&	3.1 	&	8.5 	&	0.95 	\\
153	&	3.16 	&	21.0 	&	6.5 	&	19.8 	&	28.2 	&	26.1 	&	14.6 	&	5.6 	&	0.80 	\\
154	&	3.17 	&	39.3 	&	50.3 	&	153.8 	&	99.0 	&	708.6 	&	7.9 	&	284.9 	&	1.21 	\\
155	&	0.99 	&	33.1 	&	8.7 	&	26.5 	&	22.0 	&	86.5 	&	2.9 	&	93.5 	&	0.75 	\\
156	&	3.12 	&	31.1 	&	29.3 	&	89.7 	&	60.8 	&	258.8 	&	9.8 	&	83.9 	&	0.87 	\\
157	&	1.66 	&	22.6 	&	10.2 	&	31.2 	&	17.2 	&	47.7 	&	7.2 	&	21.1 	&	0.66 	\\
158	&	3.48 	&	33.2 	&	23.5 	&	72.0 	&	77.3 	&	236.0 	&	10.3 	&	72.9 	&	0.87 	\\
159	&	0.87 	&	20.4 	&	4.9 	&	15.1 	&	7.3 	&	18.7 	&	4.2 	&	14.1 	&	0.78 	\\
160	&	1.67 	&	30.8 	&	11.6 	&	35.4 	&	31.9 	&	99.8 	&	5.3 	&	59.7 	&	1.24 	\\
161	&	0.89 	&	30.8 	&	1.5 	&	4.5 	&	17.1 	&	12.8 	&	2.8 	&	14.3 	&	1.00 	\\
162	&	0.79 	&	21.3 	&	3.6 	&	10.9 	&	7.2 	&	14.8 	&	3.6 	&	12.9 	&	0.78 	\\
163	&	0.75 	&	28.3 	&	2.0 	&	6.2 	&	12.1 	&	14.8 	&	2.6 	&	18.2 	&	0.86 	\\
164	&	0.54 	&	20.9 	&	1.9 	&	5.8 	&	4.7 	&	7.6 	&	2.5 	&	9.6 	&	0.87 	\\
165	&	0.89 	&	25.8 	&	3.3 	&	10.2 	&	12.0 	&	20.2 	&	3.4 	&	19.0 	&	0.87 	\\
166	&	3.90 	&	44.7 	&	58.3 	&	178.3 	&	157.8 	&	1064.9 	&	8.5 	&	396.3 	&	0.81 	\\
167	&	1.99 	&	32.7 	&	19.9 	&	60.8 	&	42.8 	&	193.4 	&	5.9 	&	103.2 	&	0.77 	\\
168	&	2.58 	&	38.3 	&	15.7 	&	48.1 	&	76.5 	&	210.8 	&	6.6 	&	101.8 	&	0.68 	\\
169	&	1.19 	&	20.2 	&	9.4 	&	28.8 	&	9.9 	&	35.2 	&	5.8 	&	19.3 	&	0.83 	\\
170	&	1.60 	&	36.9 	&	9.5 	&	29.2 	&	44.0 	&	118.5 	&	4.2 	&	88.9 	&	0.71 	\\
171	&	1.73 	&	43.7 	&	12.2 	&	37.2 	&	66.6 	&	212.1 	&	3.9 	&	174.1 	&	0.81 	\\
172	&	1.21 	&	31.9 	&	12.4 	&	37.8 	&	24.9 	&	114.9 	&	3.7 	&	98.5 	&	1.07 	\\
173	&	1.73 	&	30.5 	&	10.8 	&	33.0 	&	32.4 	&	91.4 	&	5.5 	&	52.2 	&	0.80 	\\
174	&	1.25 	&	57.1 	&	6.1 	&	18.6 	&	82.4 	&	180.3 	&	2.1 	&	266.7 	&		\\
175	&	1.15 	&	56.7 	&	4.8 	&	14.7 	&	74.5 	&	140.6 	&	2.0 	&	225.3 	&		\\
176	&	1.01 	&	26.2 	&	7.5 	&	22.8 	&	14.0 	&	46.7 	&	3.8 	&	39.3 	&		\\
177	&	0.97 	&	29.2 	&	3.3 	&	10.0 	&	16.8 	&	25.4 	&	3.2 	&	24.8 	&		\\
178	&	0.49 	&	21.0 	&	1.7 	&	5.3 	&	4.4 	&	7.0 	&	2.3 	&	9.7 	&		\\
179	&	0.59 	&	23.4 	&	2.6 	&	8.0 	&	6.5 	&	13.0 	&	2.5 	&	16.7 	&		\\
180	&	1.14 	&	25.2 	&	4.1 	&	12.6 	&	14.7 	&	23.9 	&	4.4 	&	17.1 	&		\\
181	&	2.83 	&	27.4 	&	40.2 	&	123.1 	&	42.9 	&	275.4 	&	10.1 	&	86.4 	&		\\
182	&	2.51 	&	30.3 	&	21.7 	&	66.3 	&	46.5 	&	181.2 	&	8.1 	&	70.8 	&		\\
183	&	0.65 	&	25.2 	&	3.1 	&	9.5 	&	8.2 	&	18.0 	&	2.5 	&	22.8 	&		\\
184	&	1.33 	&	25.2 	&	6.0 	&	18.3 	&	17.1 	&	34.7 	&	5.2 	&	21.3 	&		\\
\hline
\end{longtable}
\end{ThreePartTable}

\clearpage

\section{Morphology of HVCCs}
\label{morphplogy}
Zoomed maps of CO {\it J}=3--2 emission for all identified HVCCs are presented here.  
The velocity-integrated {\it l-b} maps and {\it l-V} maps in Figure \ref{fig: HVCC} were extracted from the original data cube ($F_{0}$; see \S\ref{sec: id}).

\begin{figure*}[tbh]
\centering
\includegraphics[]{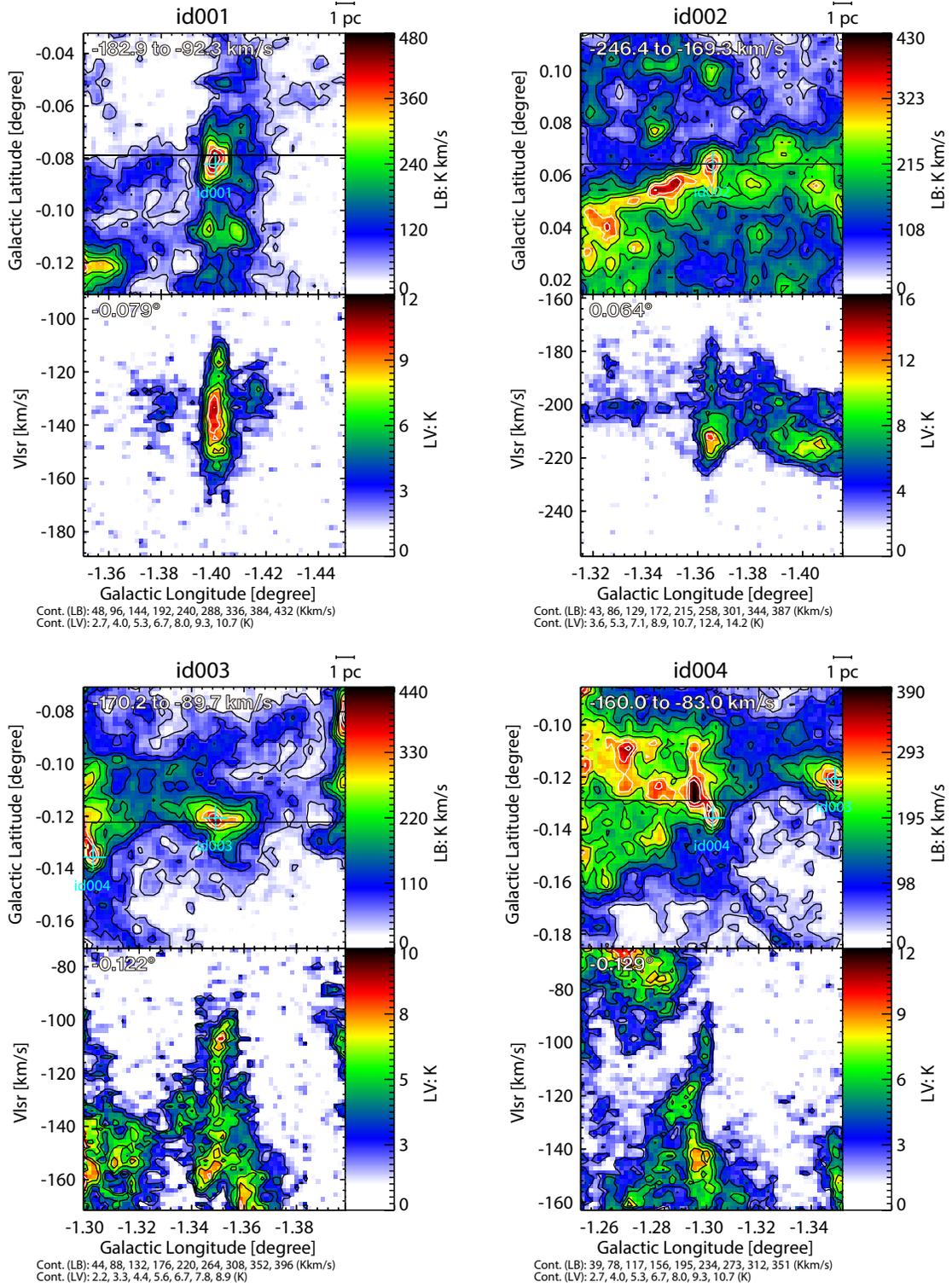}
\caption{Velocity-integrated {\it l-b} maps (top) and {\it l-V} maps of HVCCs.  The numbers in the upper panel indicate the integrated velocity range, while numbers in the bottom panel represent the lattitude slice. The numbers above each panel are the id numbers each HVCC. The blue crosses and rectangles represent the location and velocity range of the HVCC, respectively.} 
\label{fig: HVCC}
\end{figure*}

\addtocounter{figure}{-1}
\begin{figure*}[tbh]
\centering
\includegraphics[]{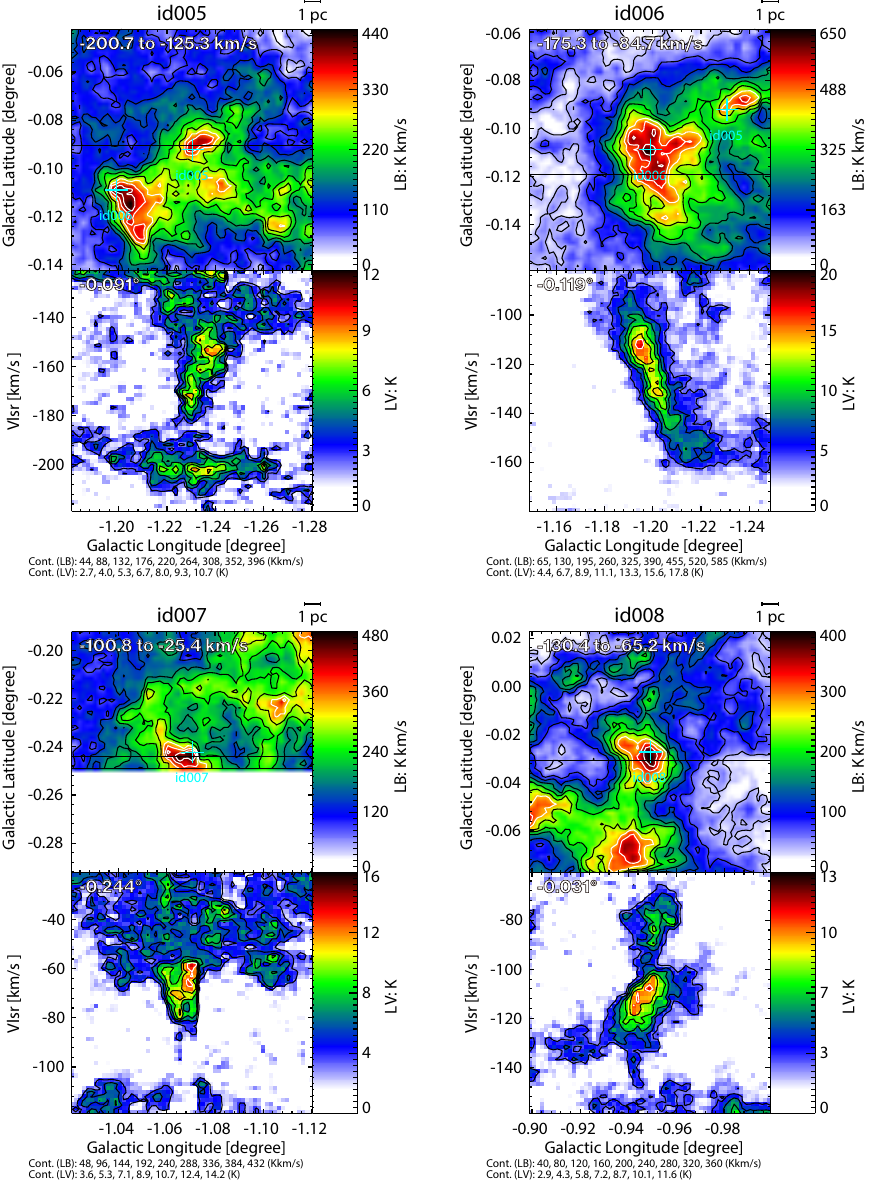}
\caption[]{Cont.}
\label{fig: HVCC}
\end{figure*}

\addtocounter{figure}{-1}
\begin{figure*}[tbh]
\centering
\includegraphics[]{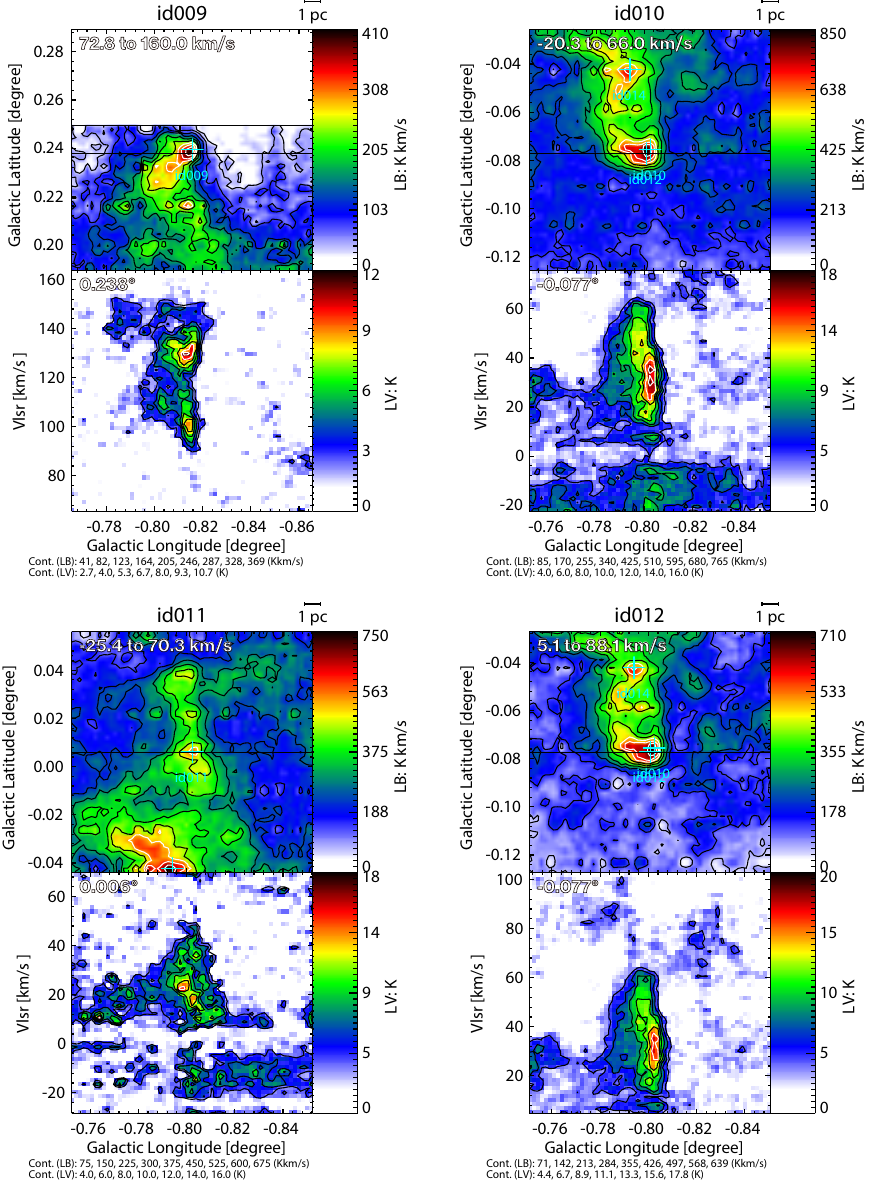}
\caption[]{Cont.}
\label{fig: HVCC}
\end{figure*}

\addtocounter{figure}{-1}
\begin{figure*}[tbh]
\centering
\includegraphics[]{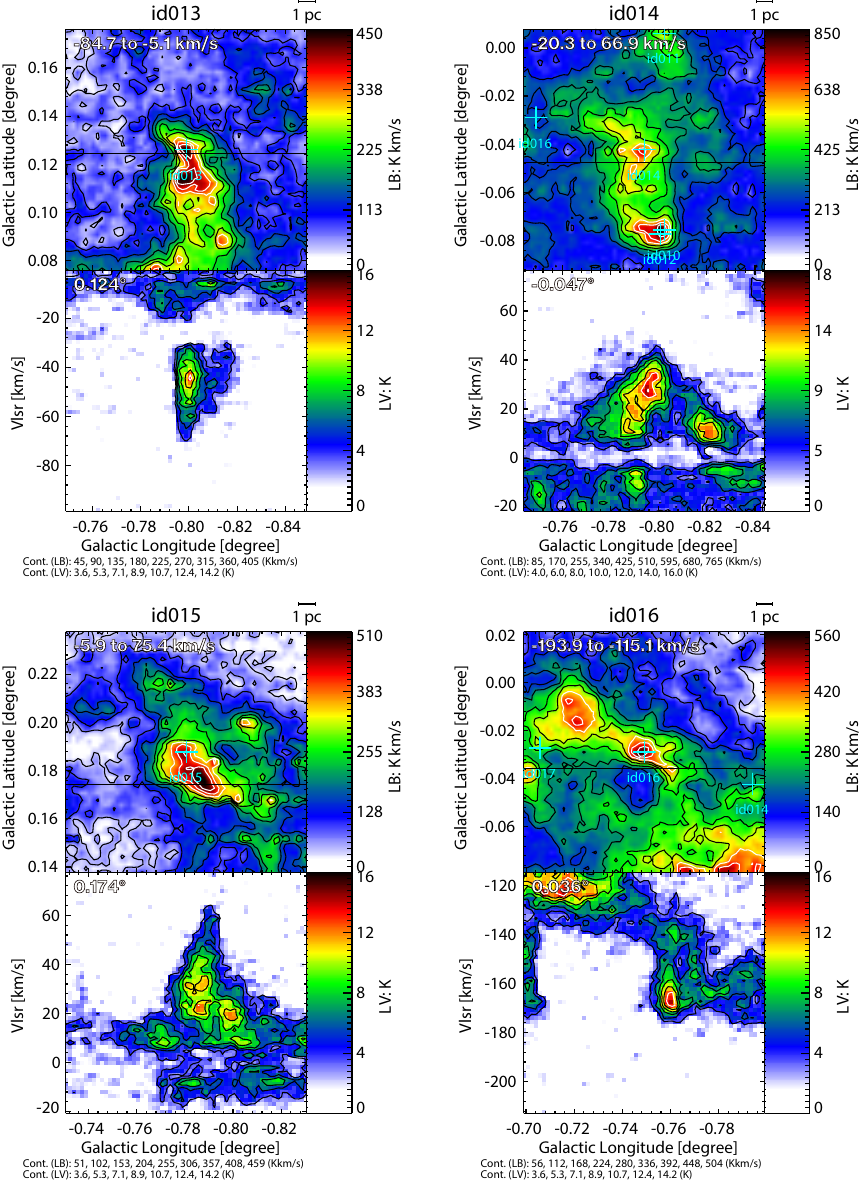}
\caption[]{Cont.}
\label{fig: HVCC}
\end{figure*}

\addtocounter{figure}{-1}
\begin{figure*}[tbh]
\centering
\includegraphics[]{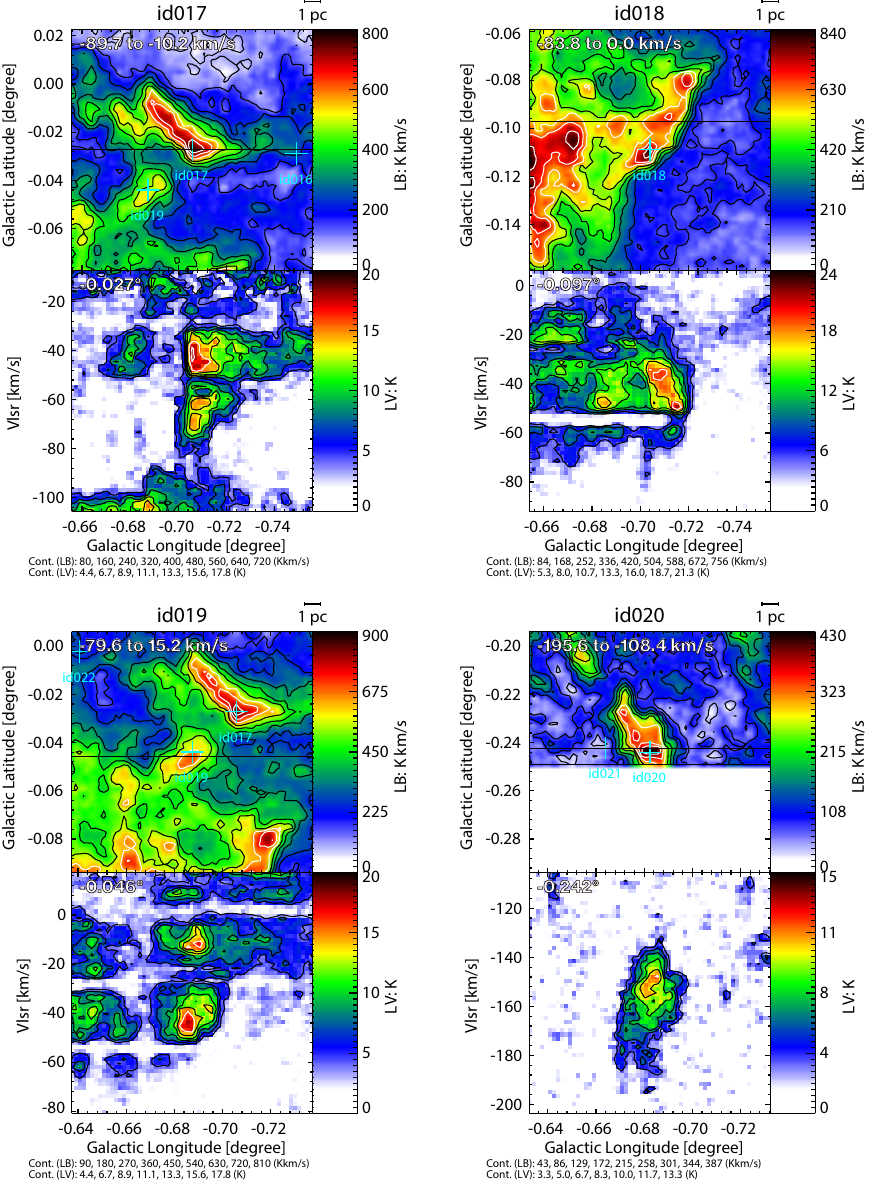}
\caption[]{Cont.}
\label{fig: HVCC}
\end{figure*}

\addtocounter{figure}{-1}
\begin{figure*}[tbh]
\centering
\includegraphics[]{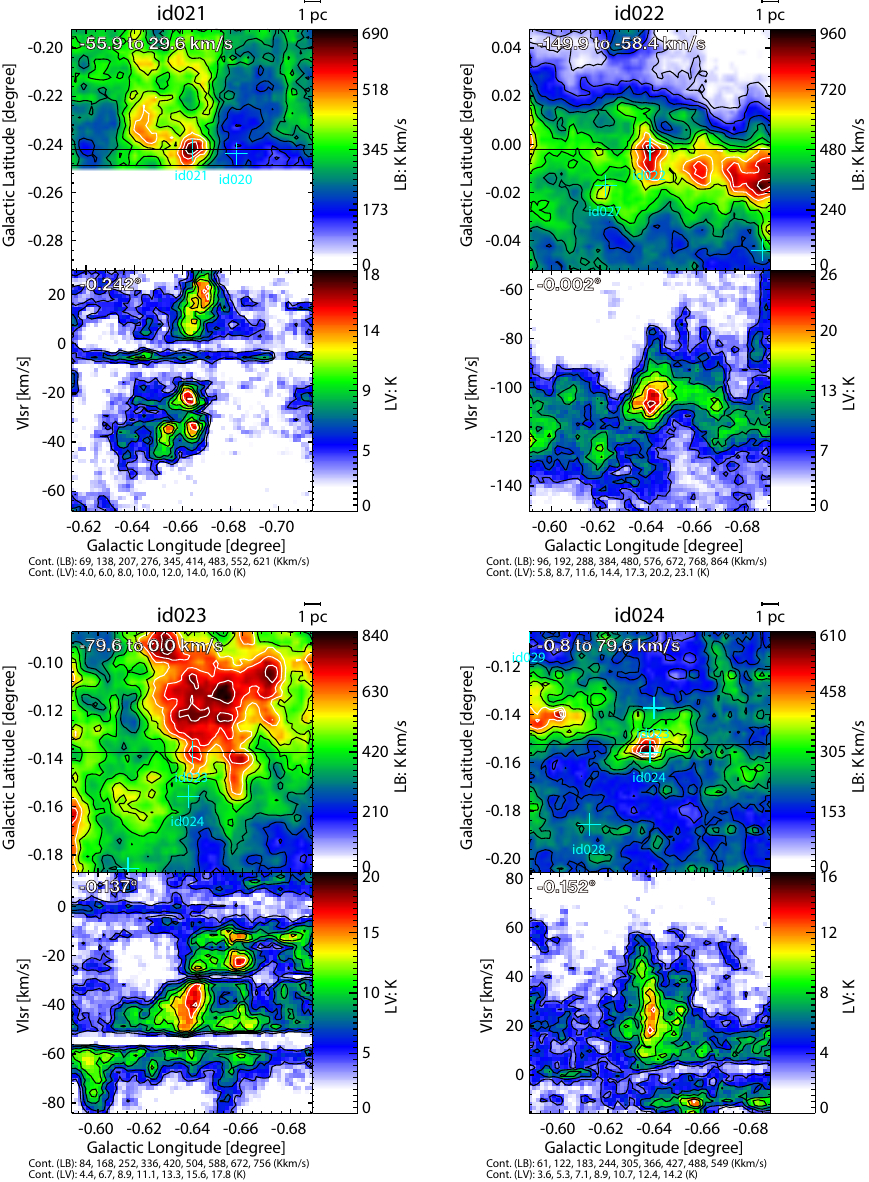}
\caption[]{Cont.}
\label{fig: HVCC}
\end{figure*}

\addtocounter{figure}{-1}
\begin{figure*}[tbh]
\centering
\includegraphics[]{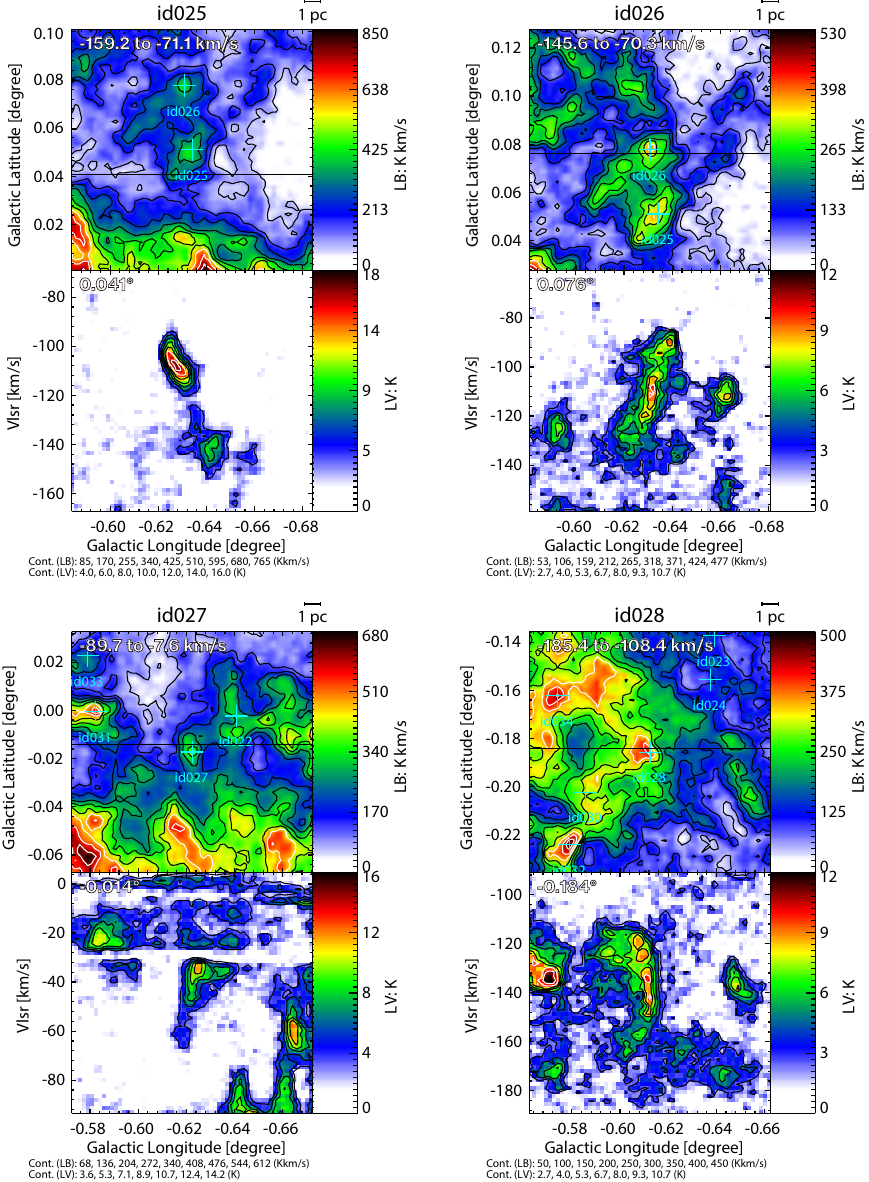}
\caption[]{Cont.}
\label{fig: HVCC}
\end{figure*}

\addtocounter{figure}{-1}
\begin{figure*}[tbh]
\centering
\includegraphics[]{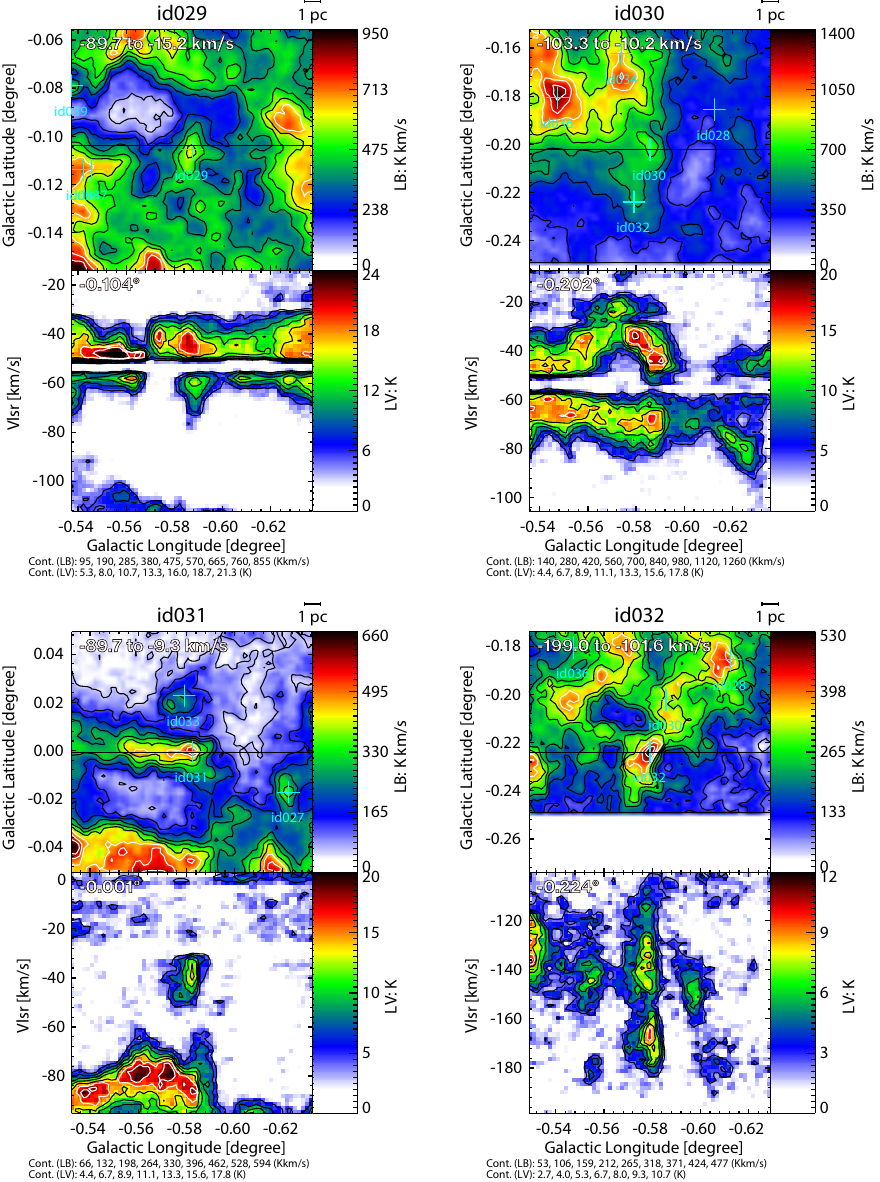}
\caption[]{Cont.}
\label{fig: HVCC}
\end{figure*}

\addtocounter{figure}{-1}
\begin{figure*}[tbh]
\centering
\includegraphics[]{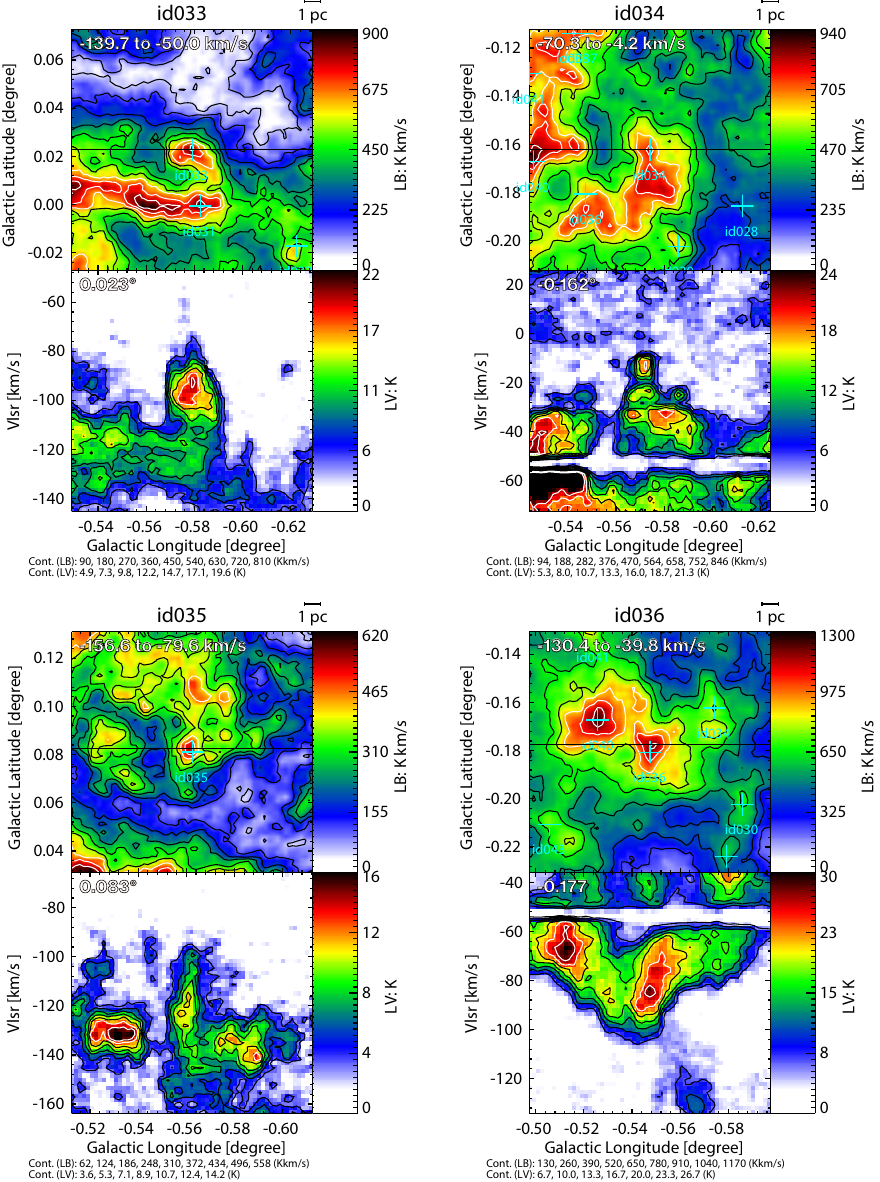}
\caption[]{Cont.}
\label{fig: HVCC}
\end{figure*}

\addtocounter{figure}{-1}
\begin{figure*}[tbh]
\centering
\includegraphics[]{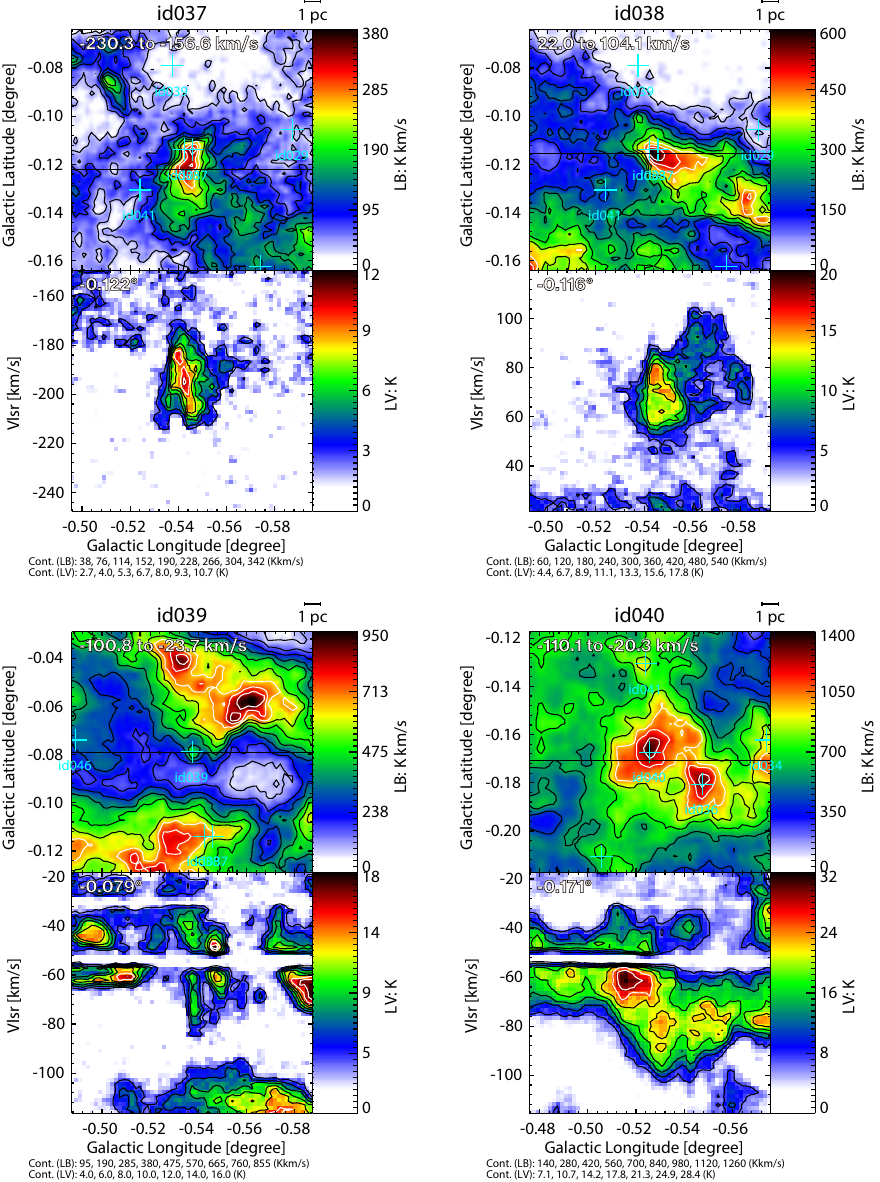}
\caption[]{Cont.}
\label{fig: HVCC}
\end{figure*}

\addtocounter{figure}{-1}
\begin{figure*}[tbh]
\centering
\includegraphics[]{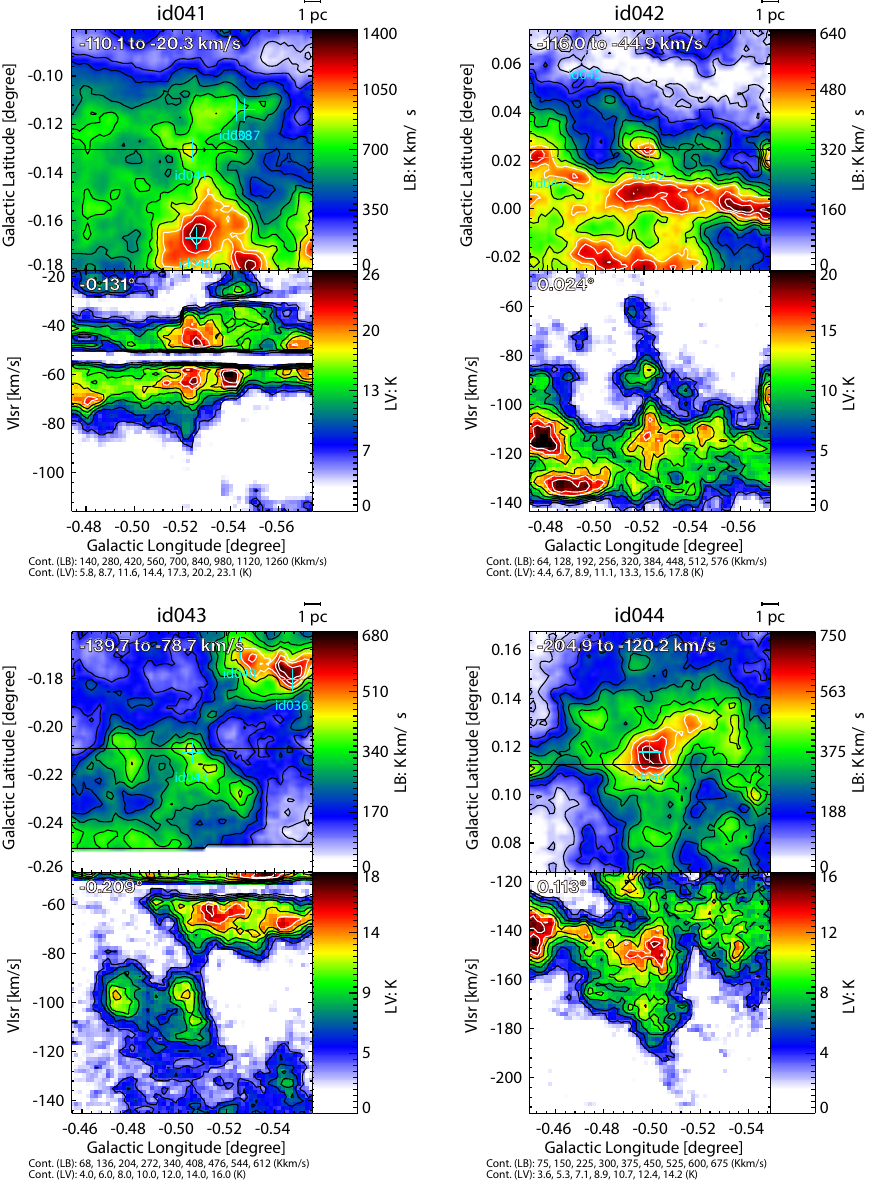}
\caption[]{Cont.}
\label{fig: HVCC}
\end{figure*}

\addtocounter{figure}{-1}
\begin{figure*}[tbh]
\centering
\includegraphics[]{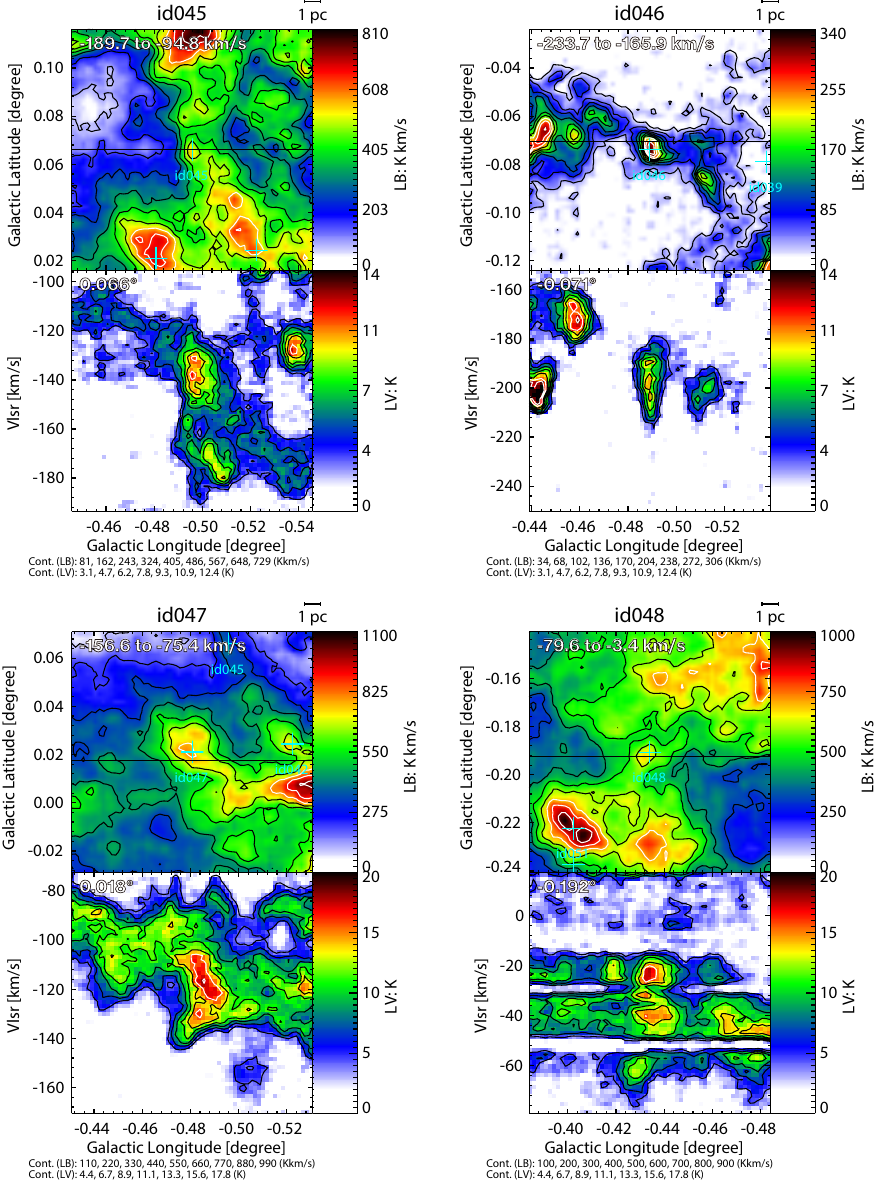}
\caption[]{Cont.}
\label{fig: HVCC}
\end{figure*}

\addtocounter{figure}{-1}
\begin{figure*}[tbh]
\centering
\includegraphics[]{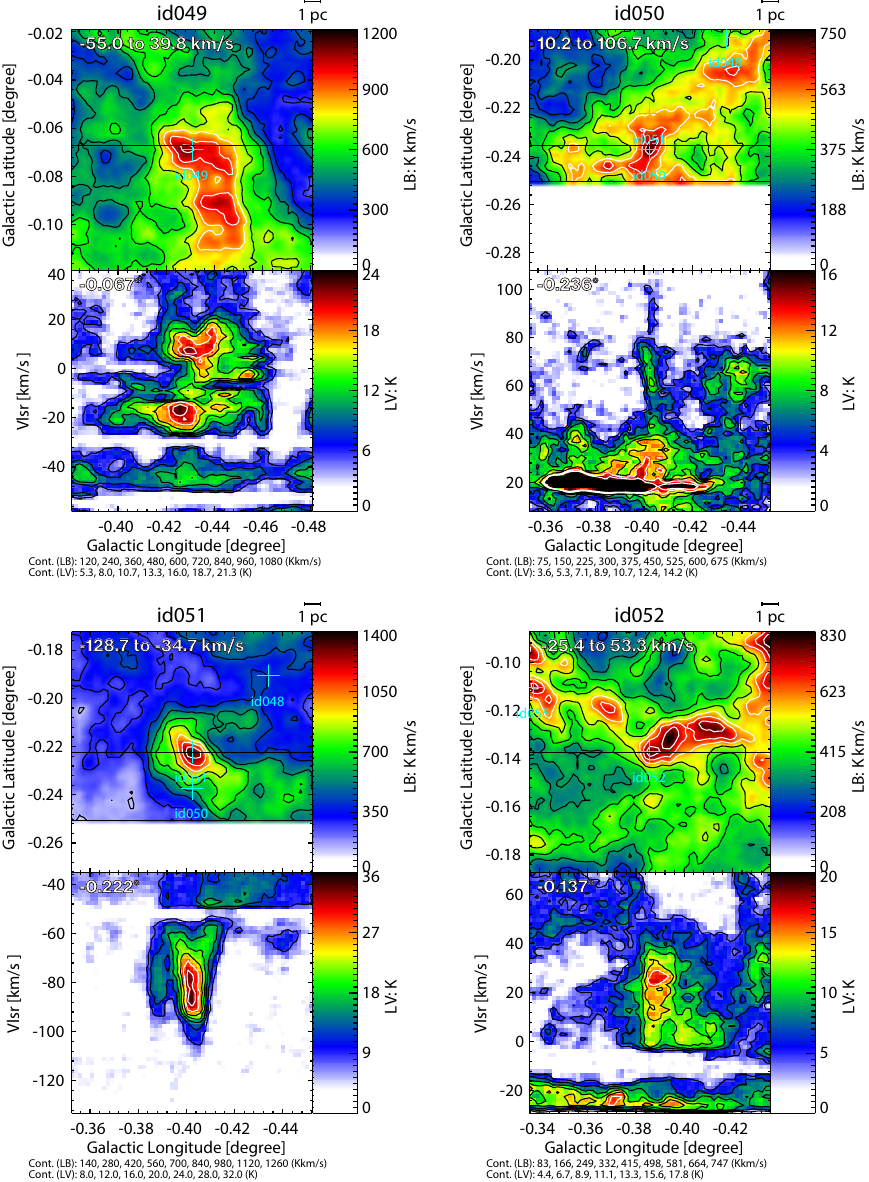}
\caption[]{Cont.}
\label{fig: HVCC}
\end{figure*}

\addtocounter{figure}{-1}
\begin{figure*}[tbh]
\centering
\includegraphics[]{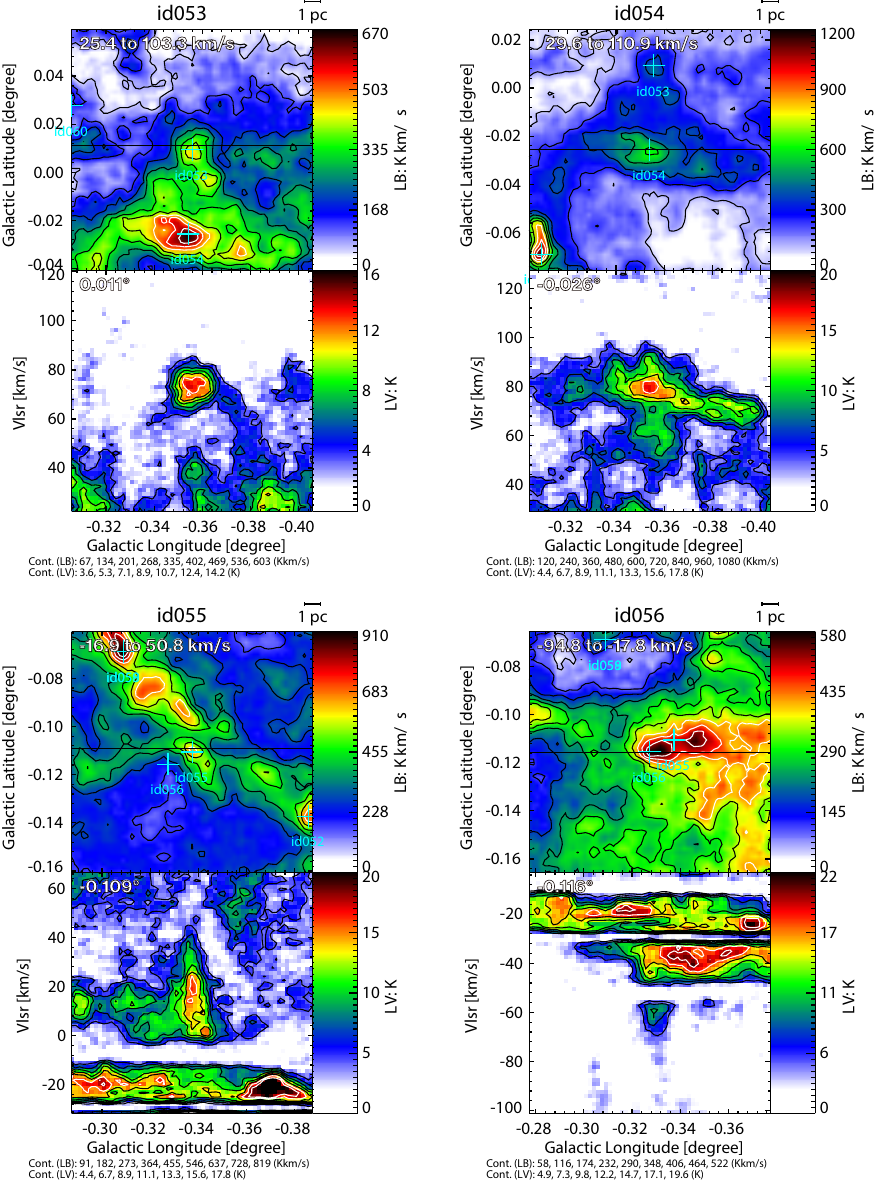}
\caption[]{Cont.}
\label{fig: HVCC}
\end{figure*}

\addtocounter{figure}{-1}
\begin{figure*}[tbh]
\centering
\includegraphics[]{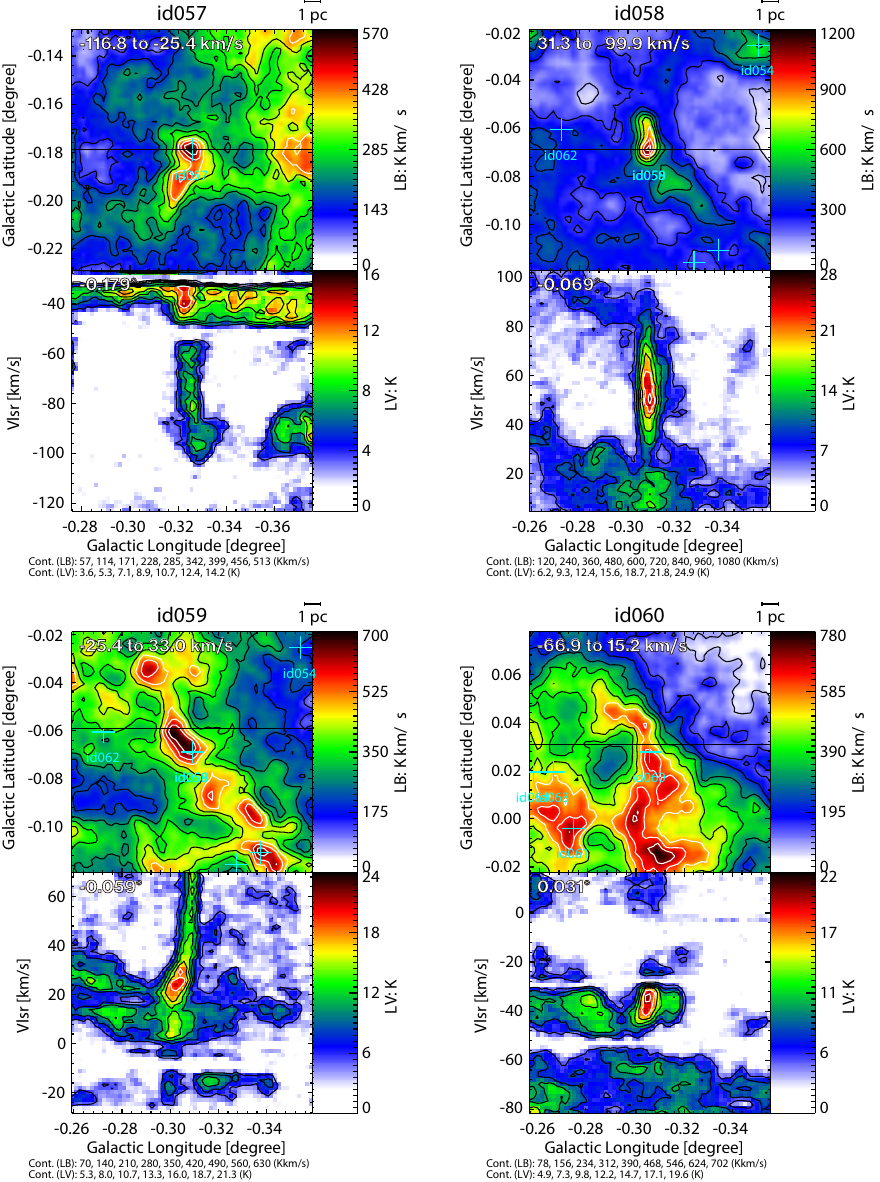}
\caption[]{Cont.}
\label{fig: HVCC}
\end{figure*}

\addtocounter{figure}{-1}
\begin{figure*}[tbh]
\centering
\includegraphics[]{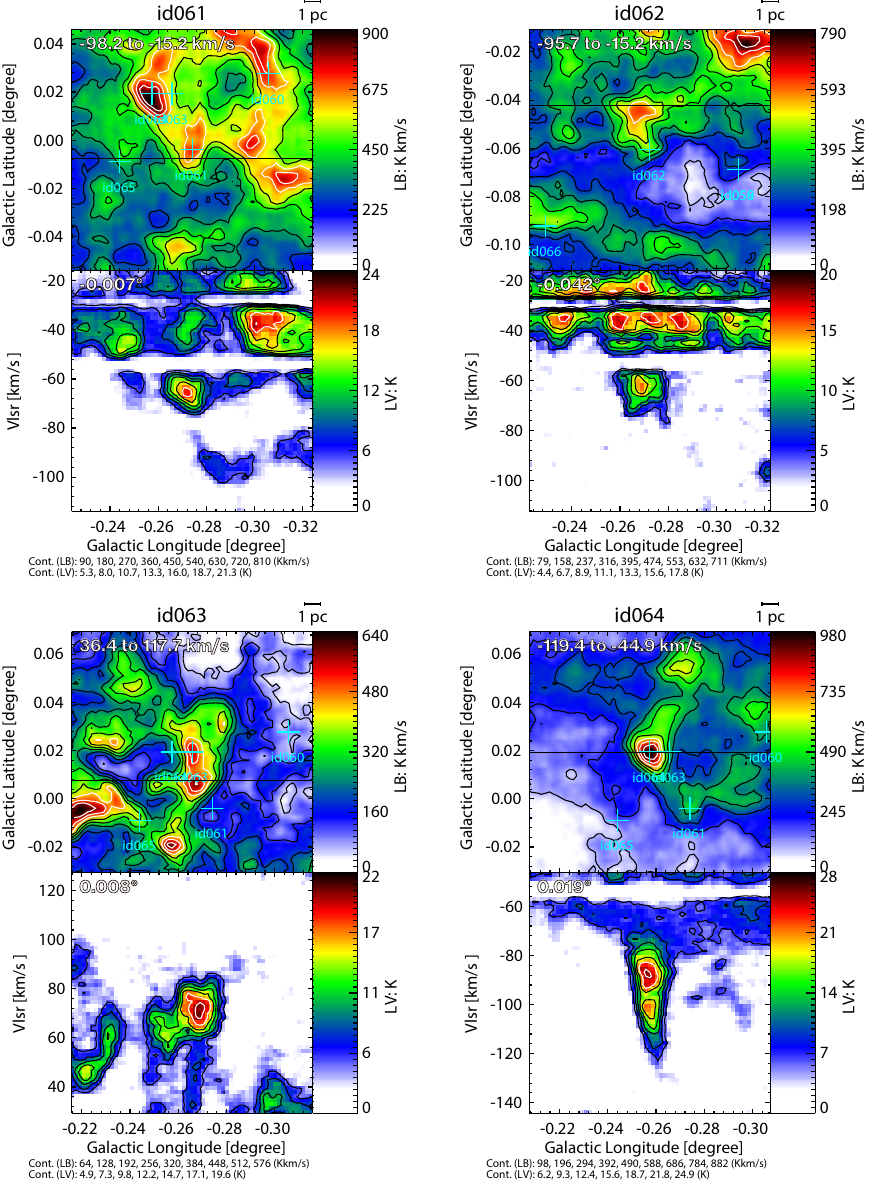}
\caption[]{Cont.}
\label{fig: HVCC}
\end{figure*}

\addtocounter{figure}{-1}
\begin{figure*}[tbh]
\centering
\includegraphics[]{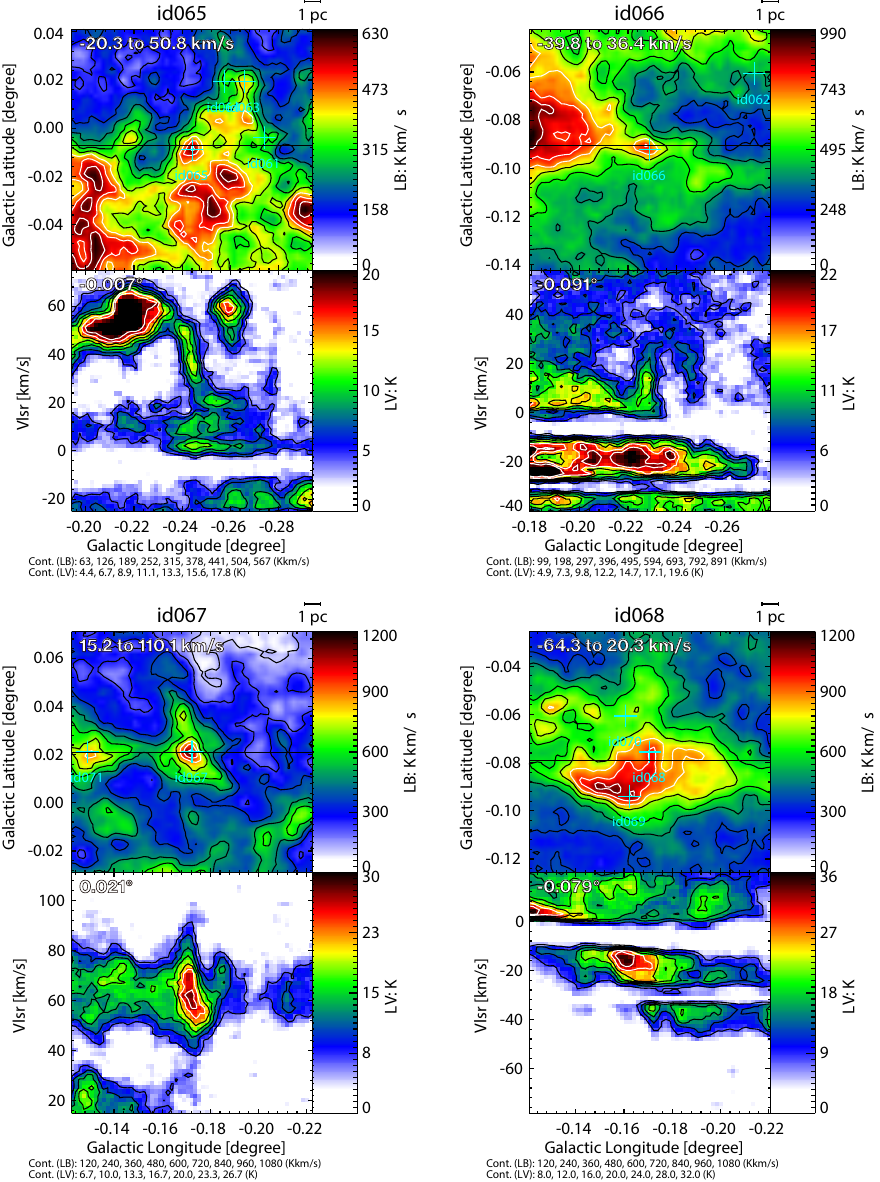}
\caption[]{Cont.}
\label{fig: HVCC}
\end{figure*}

\addtocounter{figure}{-1}
\begin{figure*}[tbh]
\centering
\includegraphics[]{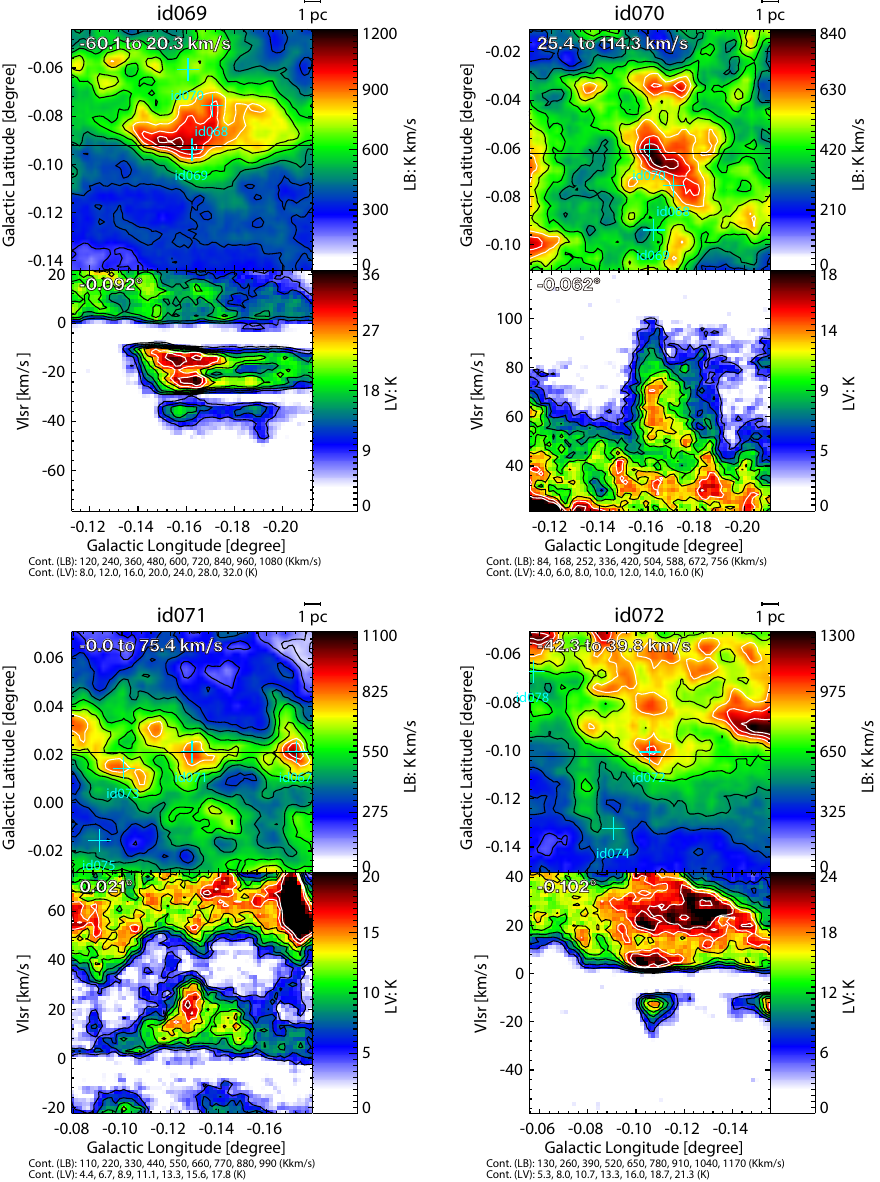}
\caption[]{Cont.}
\label{fig: HVCC}
\end{figure*}

\addtocounter{figure}{-1}
\begin{figure*}[tbh]
\centering
\includegraphics[]{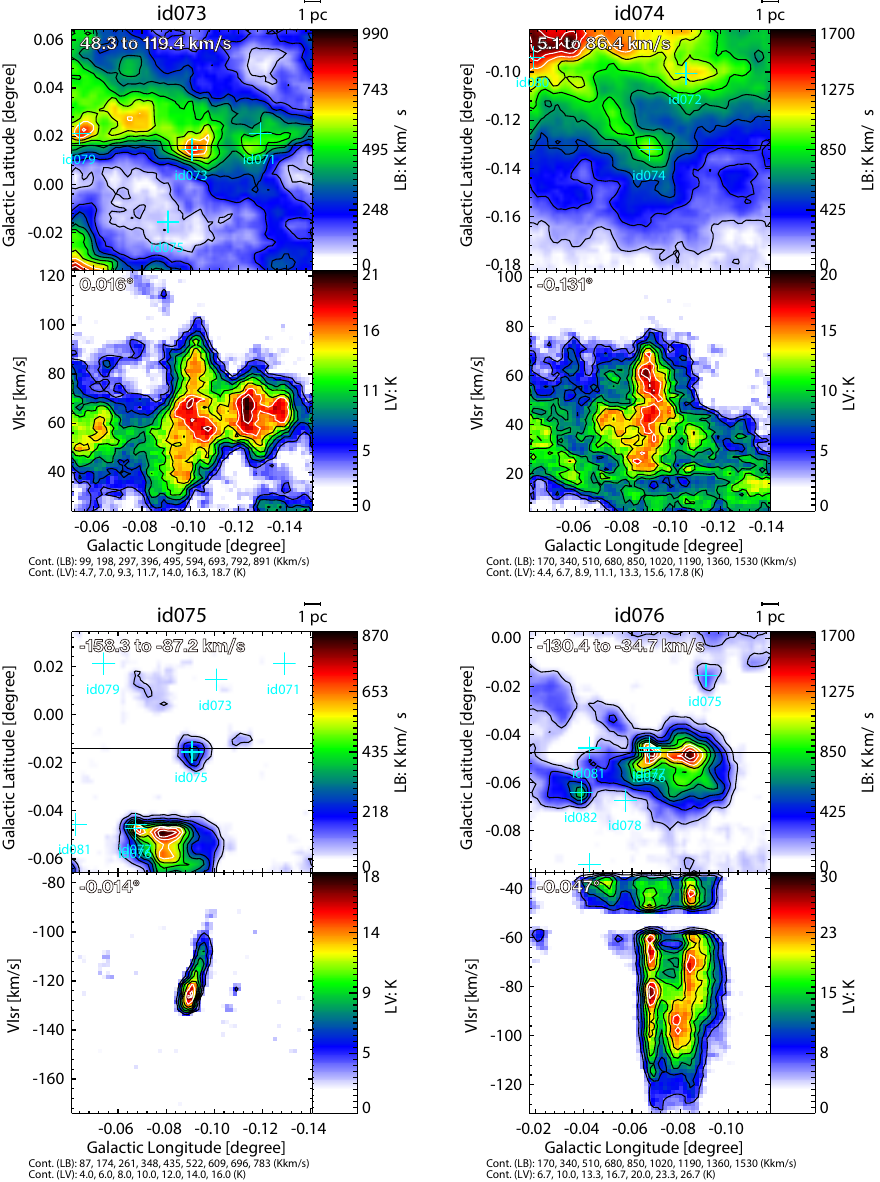}
\caption[]{Cont.}
\label{fig: HVCC}
\end{figure*}

\addtocounter{figure}{-1}
\begin{figure*}[tbh]
\centering
\includegraphics[]{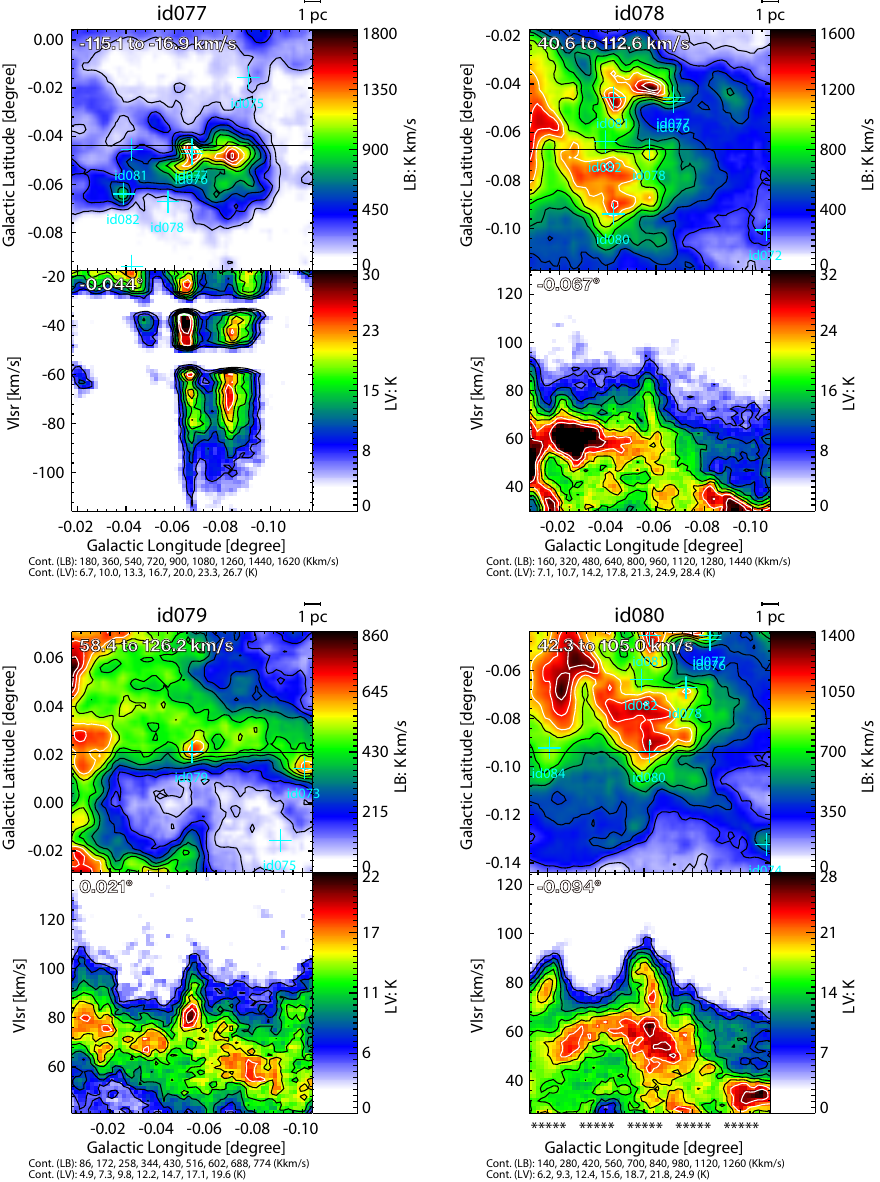}
\caption[]{Cont.}
\label{fig: HVCC}
\end{figure*}

\addtocounter{figure}{-1}
\begin{figure*}[tbh]
\centering
\includegraphics[]{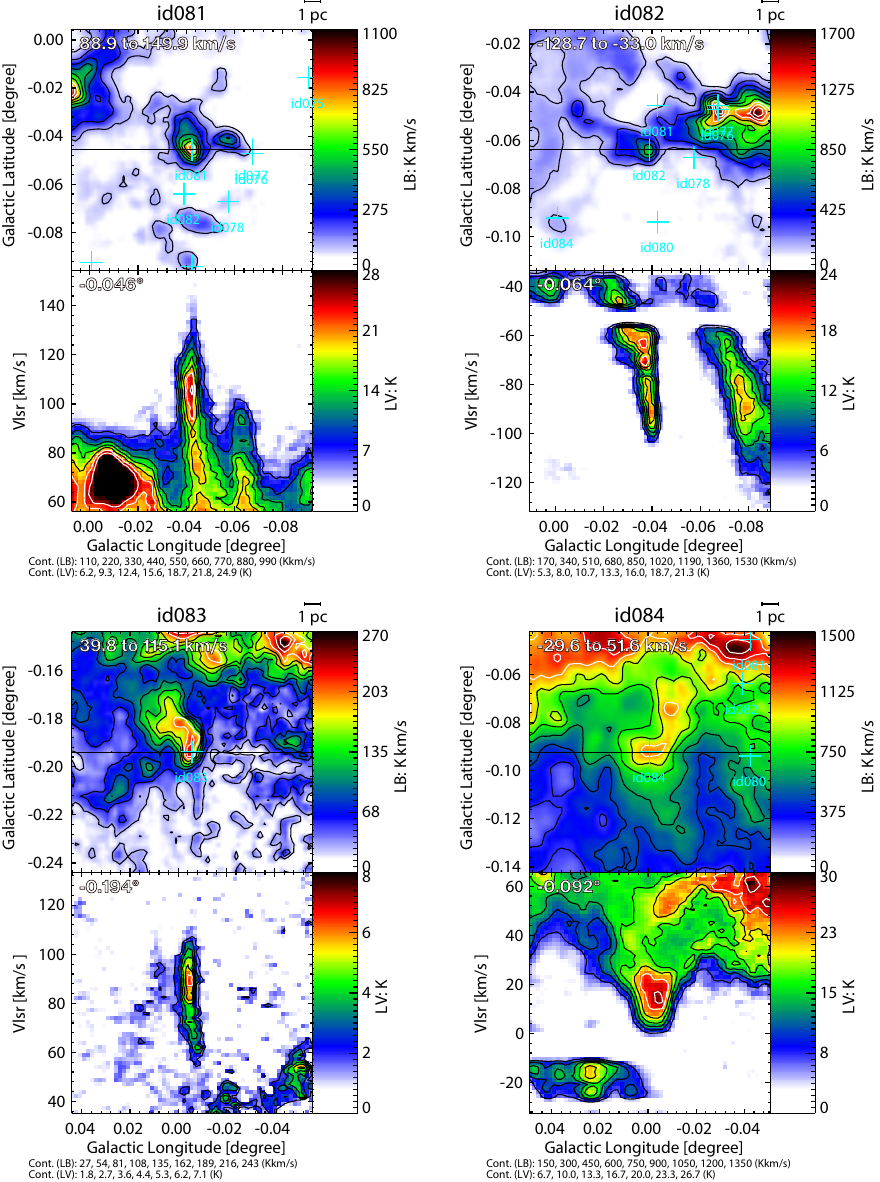}
\caption[]{Cont.}
\label{fig: HVCC}
\end{figure*}

\addtocounter{figure}{-1}
\begin{figure*}[tbh]
\centering
\includegraphics[]{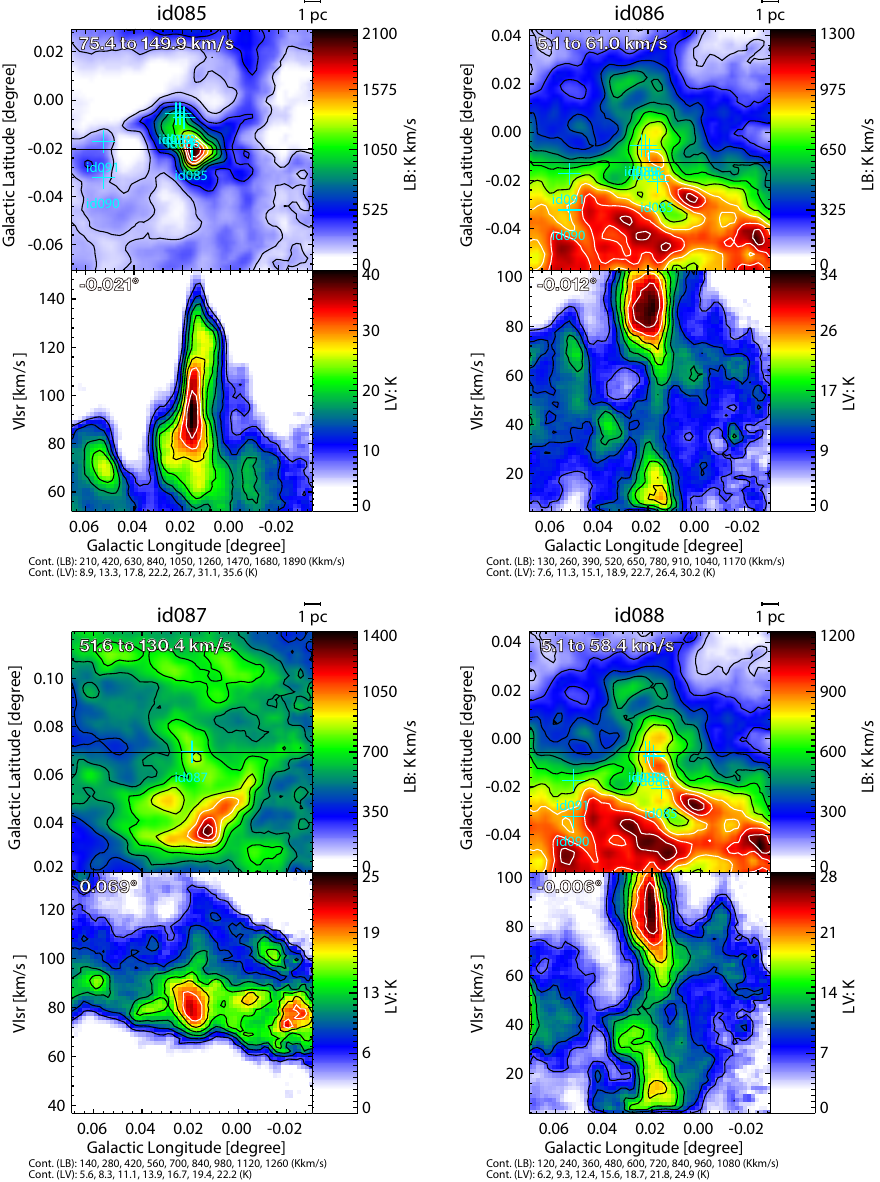}
\caption[]{Cont.}
\label{fig: HVCC}
\end{figure*}

\addtocounter{figure}{-1}
\begin{figure*}[tbh]
\centering
\includegraphics[]{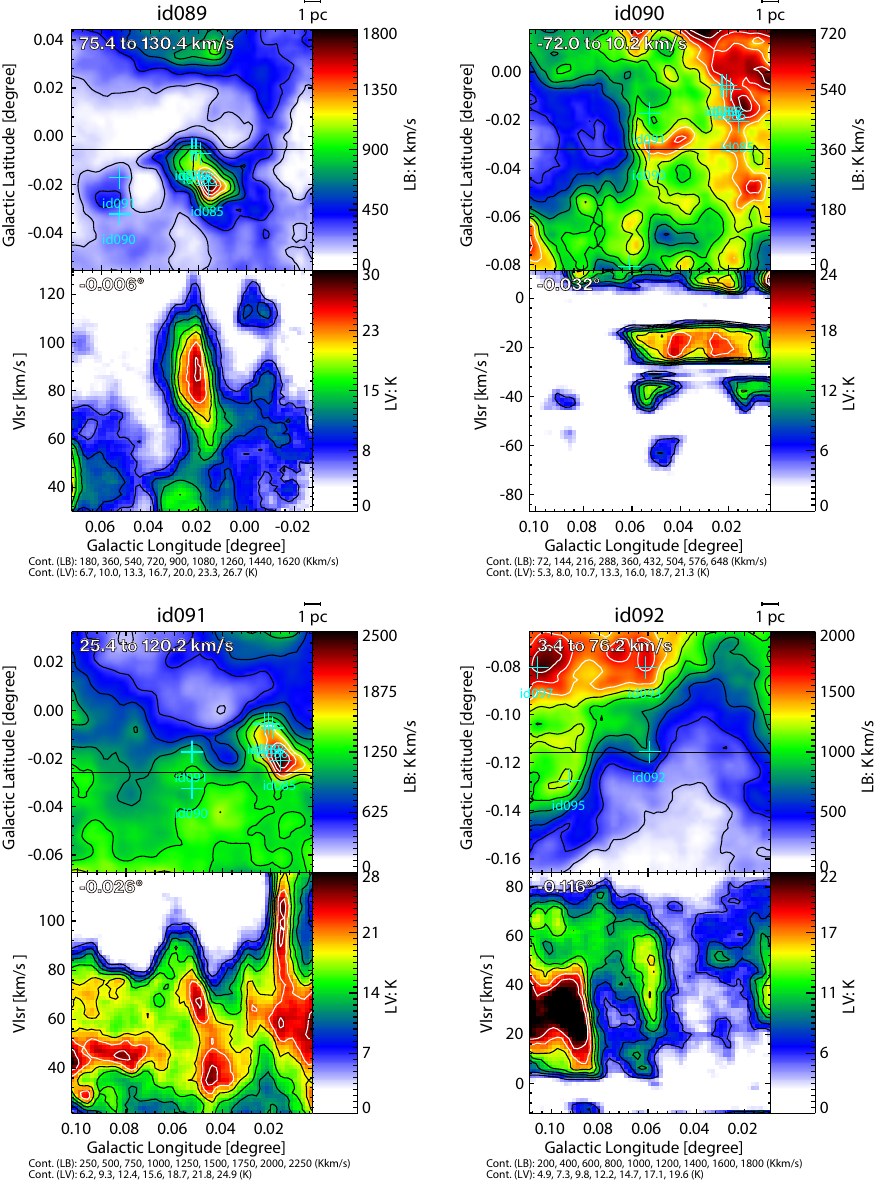}
\caption[]{Cont.}
\label{fig: HVCC}
\end{figure*}

\addtocounter{figure}{-1}
\begin{figure*}[tbh]
\centering
\includegraphics[]{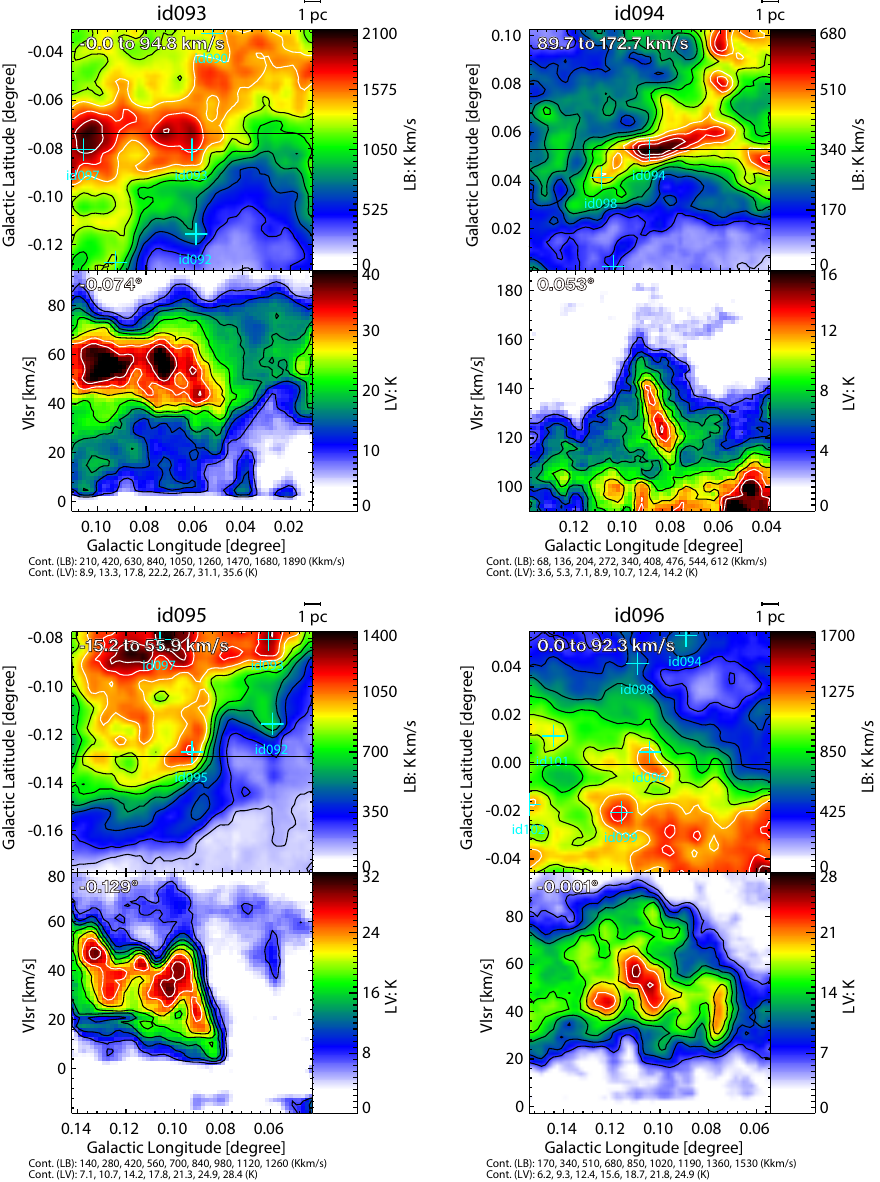}
\caption[]{Cont.}
\label{fig: HVCC}
\end{figure*}

\addtocounter{figure}{-1}
\begin{figure*}[tbh]
\centering
\includegraphics[]{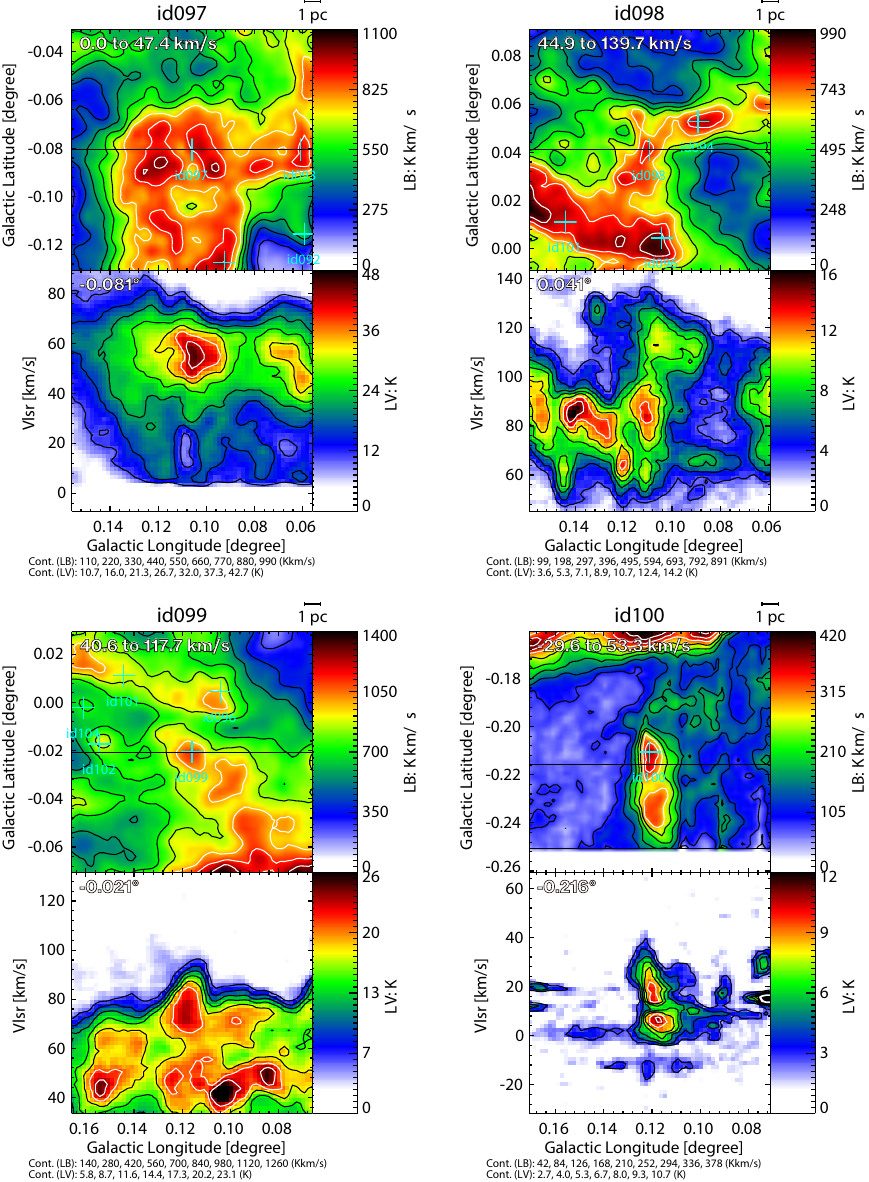}
\caption[]{Cont.}
\label{fig: HVCC}
\end{figure*}

\addtocounter{figure}{-1}
\begin{figure*}[tbh]
\centering
\includegraphics[]{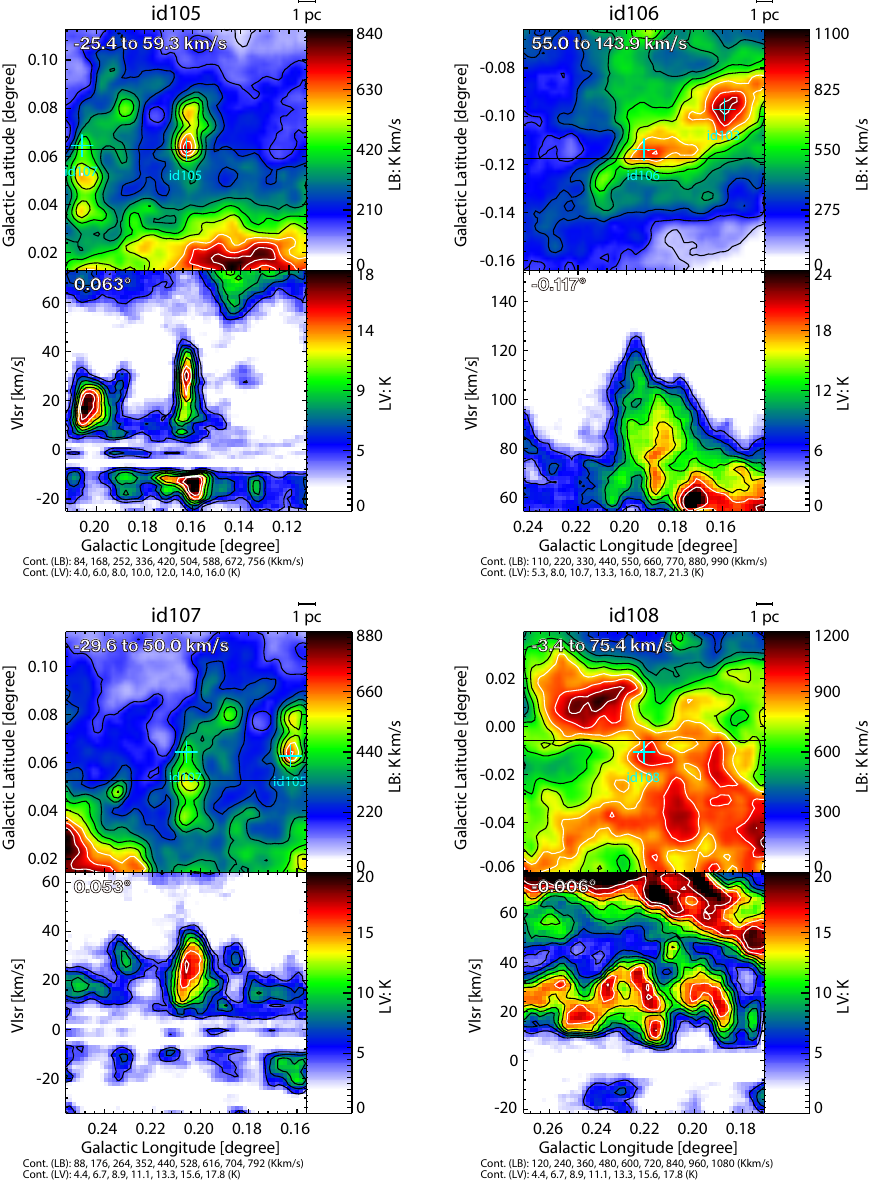}
\caption[]{Cont.}
\label{fig: HVCC}
\end{figure*}

\addtocounter{figure}{-1}
\begin{figure*}[tbh]
\centering
\includegraphics[]{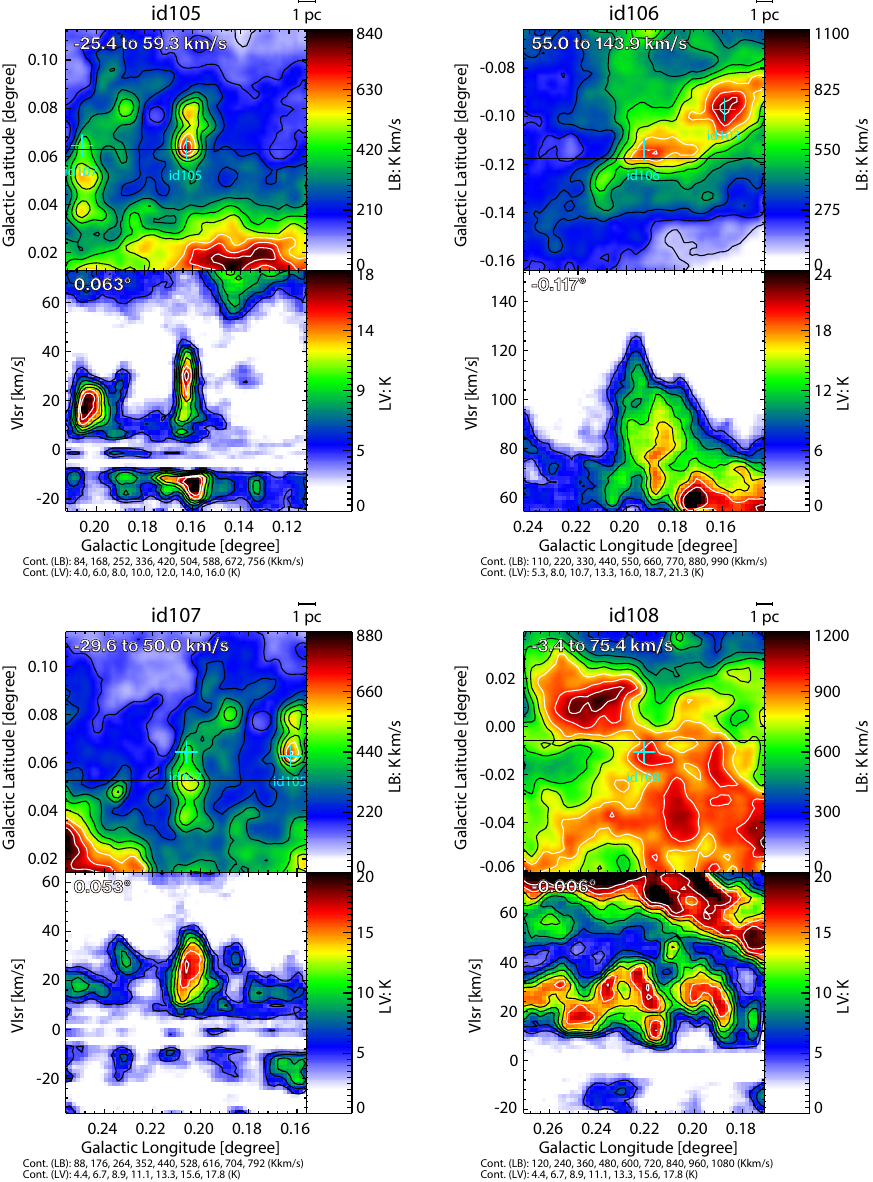}
\caption[]{Cont.}
\label{fig: HVCC}
\end{figure*}

\addtocounter{figure}{-1}
\begin{figure*}[tbh]
\centering
\includegraphics[]{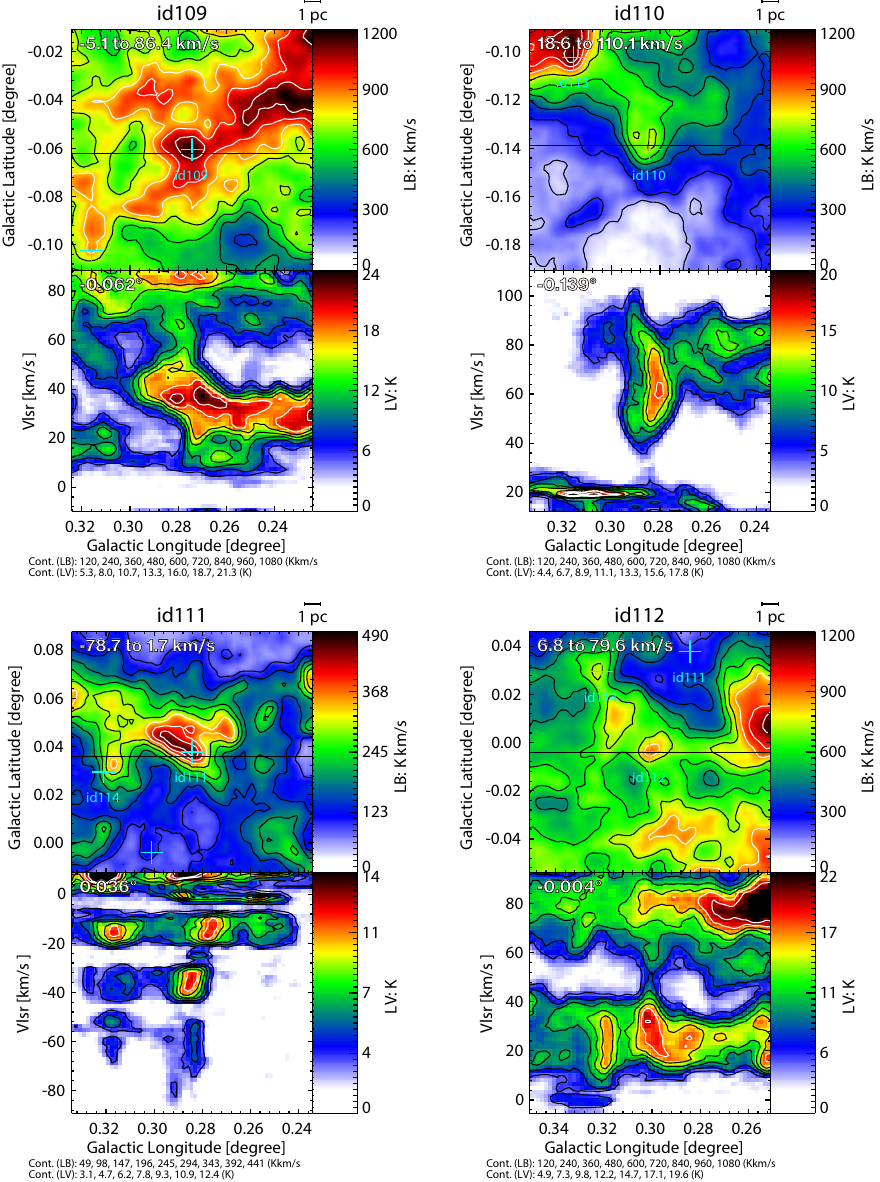}
\caption[]{Cont.}
\label{fig: HVCC}
\end{figure*}

\addtocounter{figure}{-1}
\begin{figure*}[tbh]
\centering
\includegraphics[]{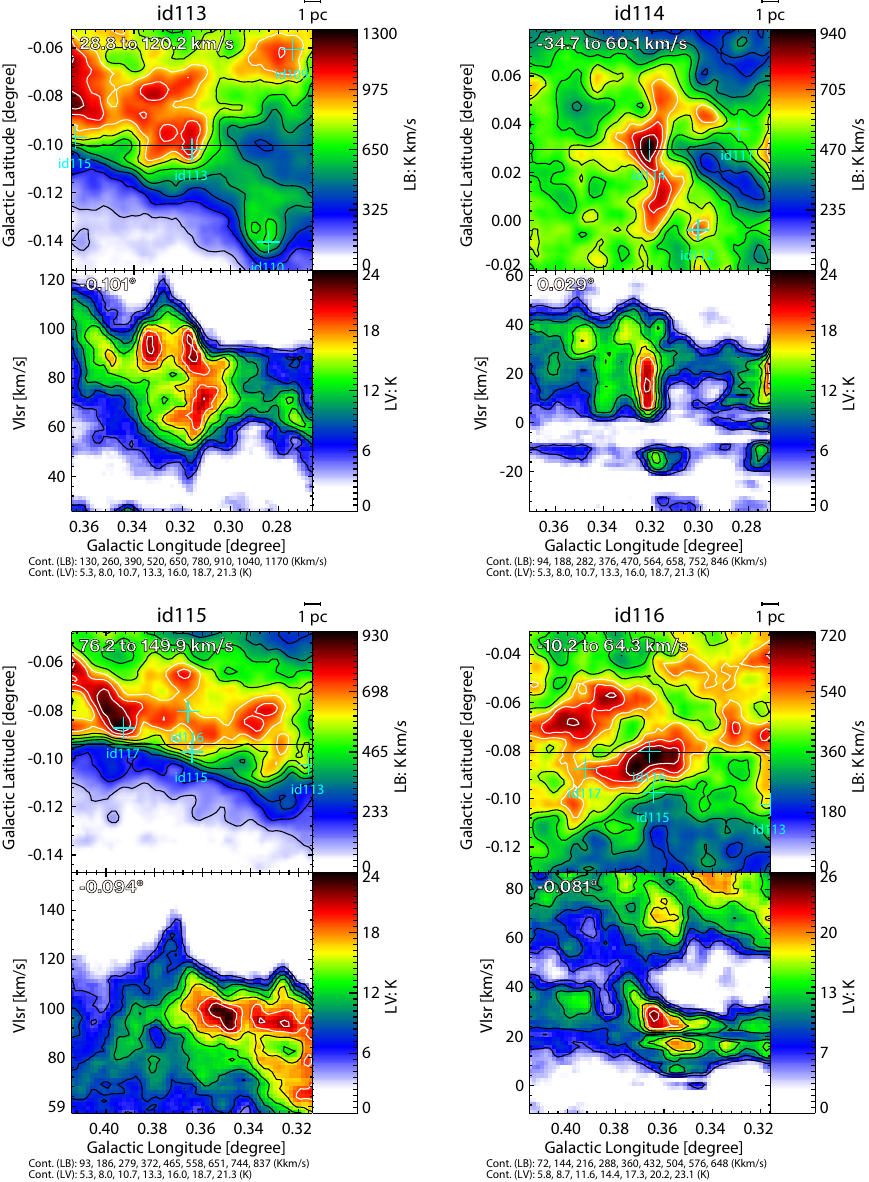}
\caption[]{Cont.}
\label{fig: HVCC}
\end{figure*}

\addtocounter{figure}{-1}
\begin{figure}[p]
\centering
\includegraphics[]{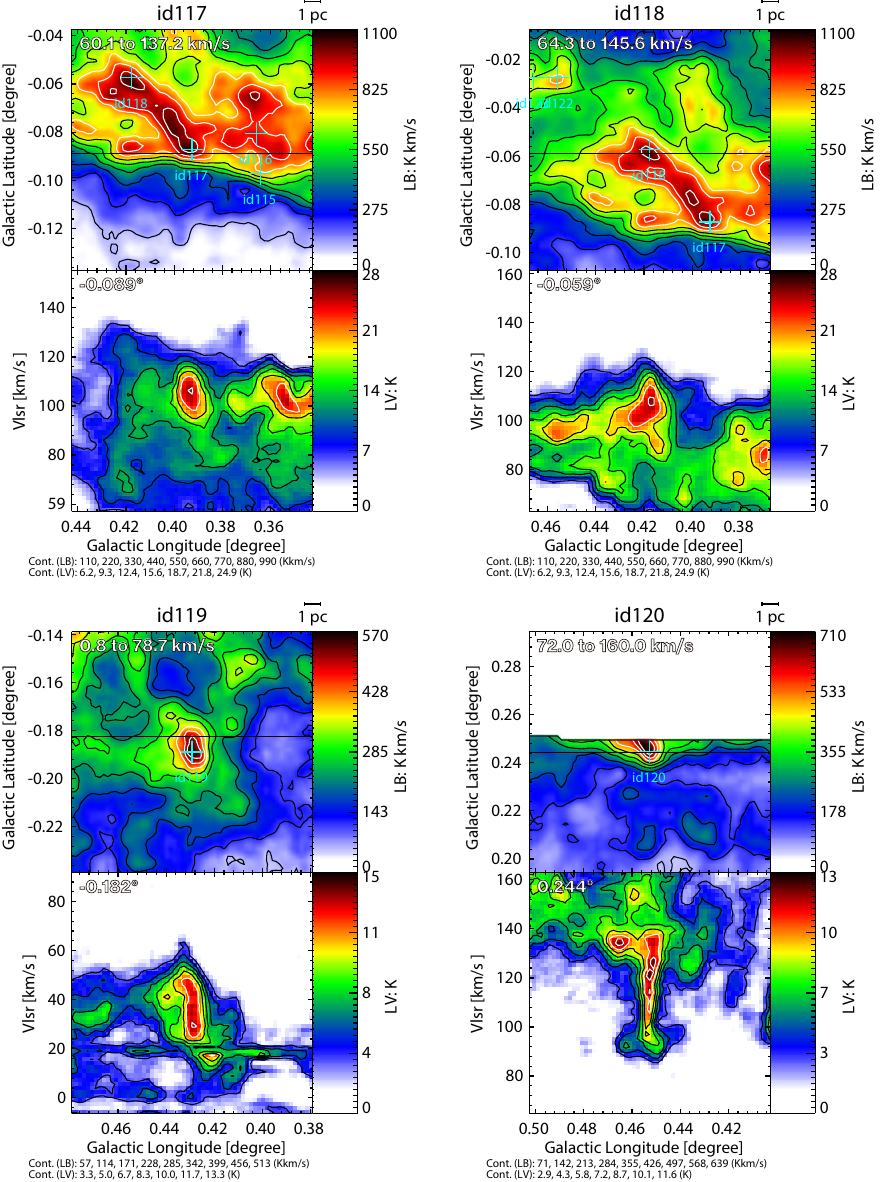}
\caption[]{Cont.}
\label{fig: HVCC}
\end{figure}

\addtocounter{figure}{-1}
\begin{figure}[p]
\centering
\includegraphics[]{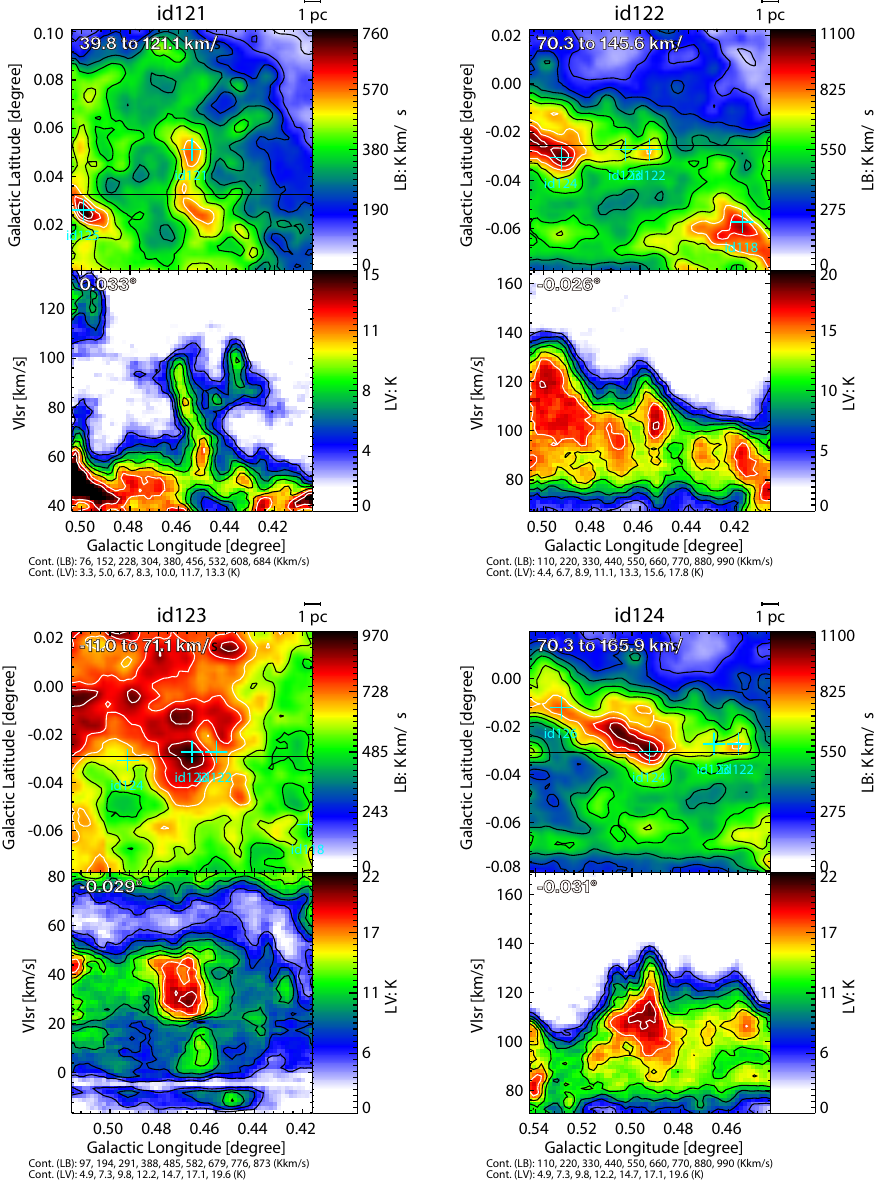}
\caption[]{Cont.}
\label{fig: HVCC}
\end{figure}

\addtocounter{figure}{-1}
\begin{figure}[p]
\centering
\includegraphics[]{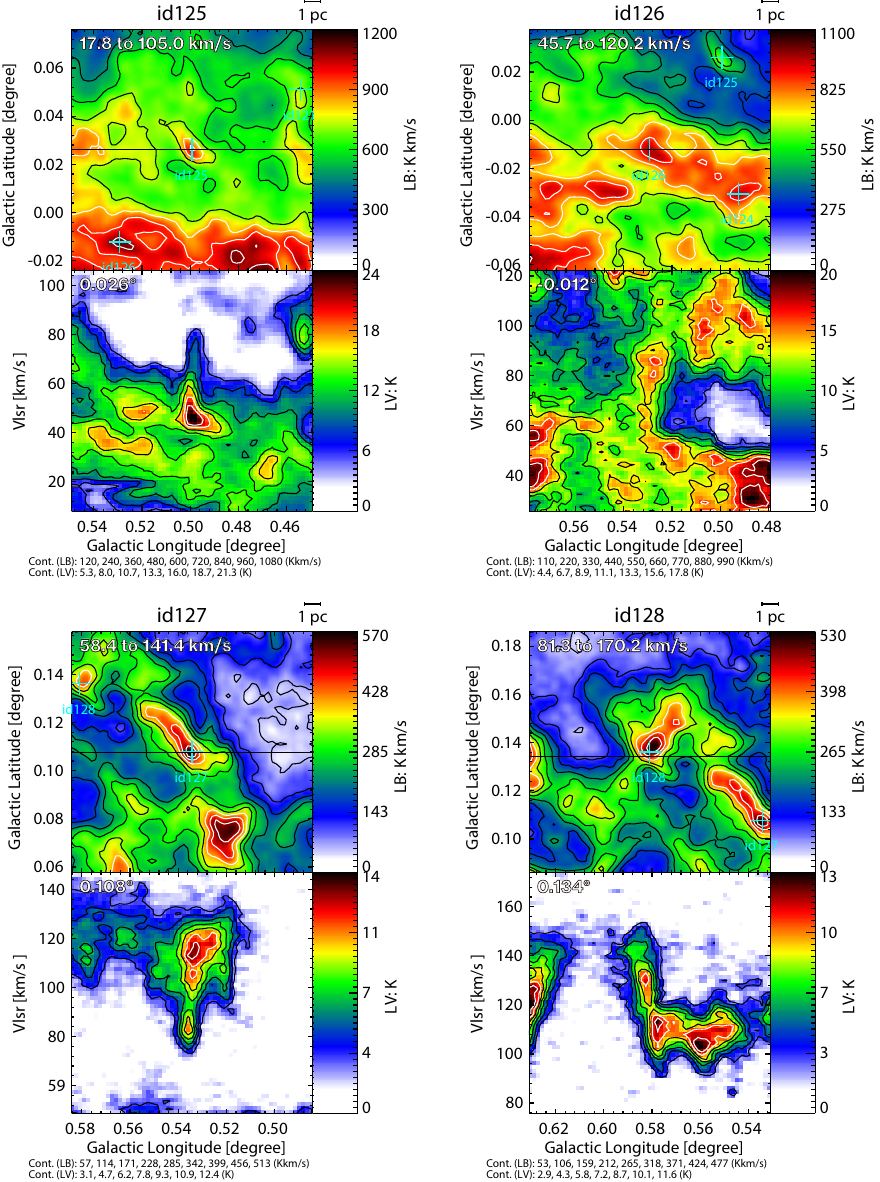}
\caption[]{Cont.}
\label{fig: HVCC}
\end{figure}

\addtocounter{figure}{-1}
\begin{figure}[p]
\centering
\includegraphics[]{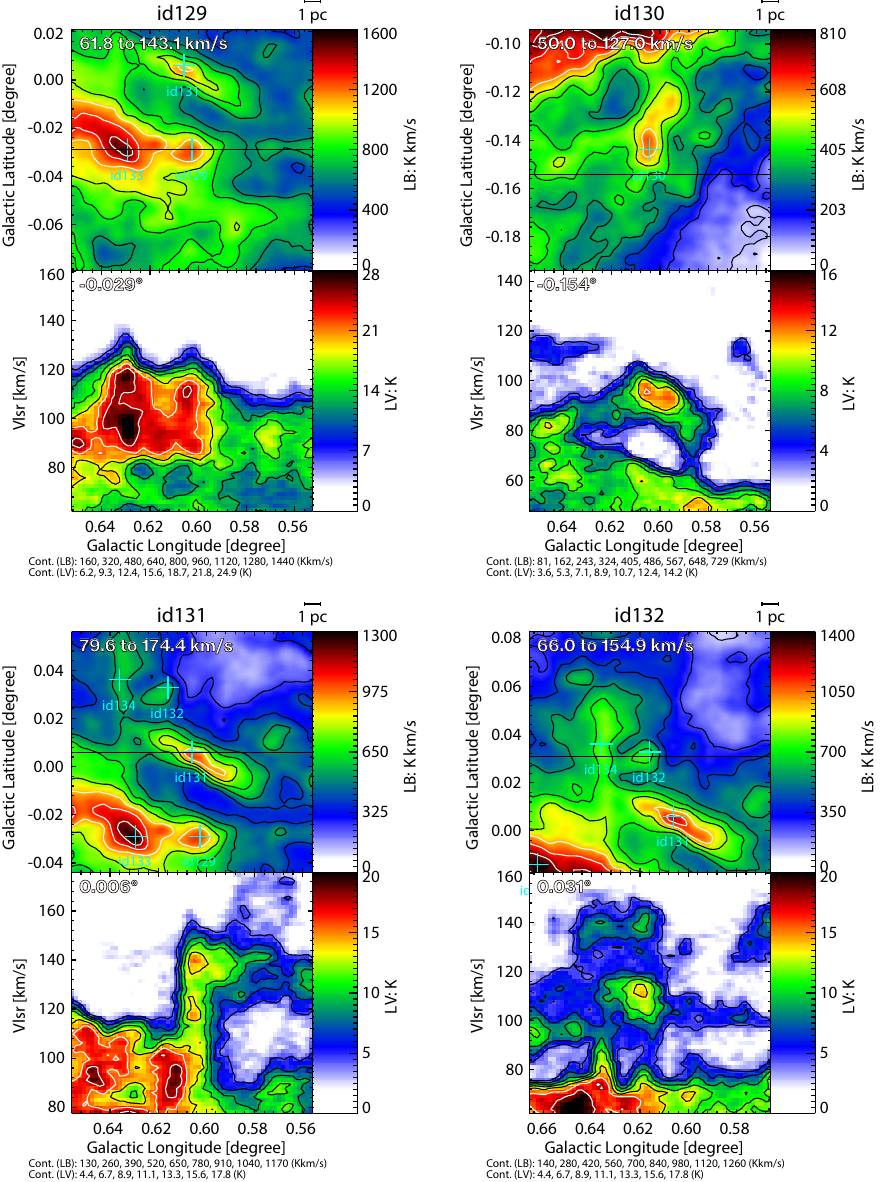}
\caption[]{Cont.}
\label{fig: HVCC}
\end{figure}

\addtocounter{figure}{-1}
\begin{figure}[p]
\centering
\includegraphics[]{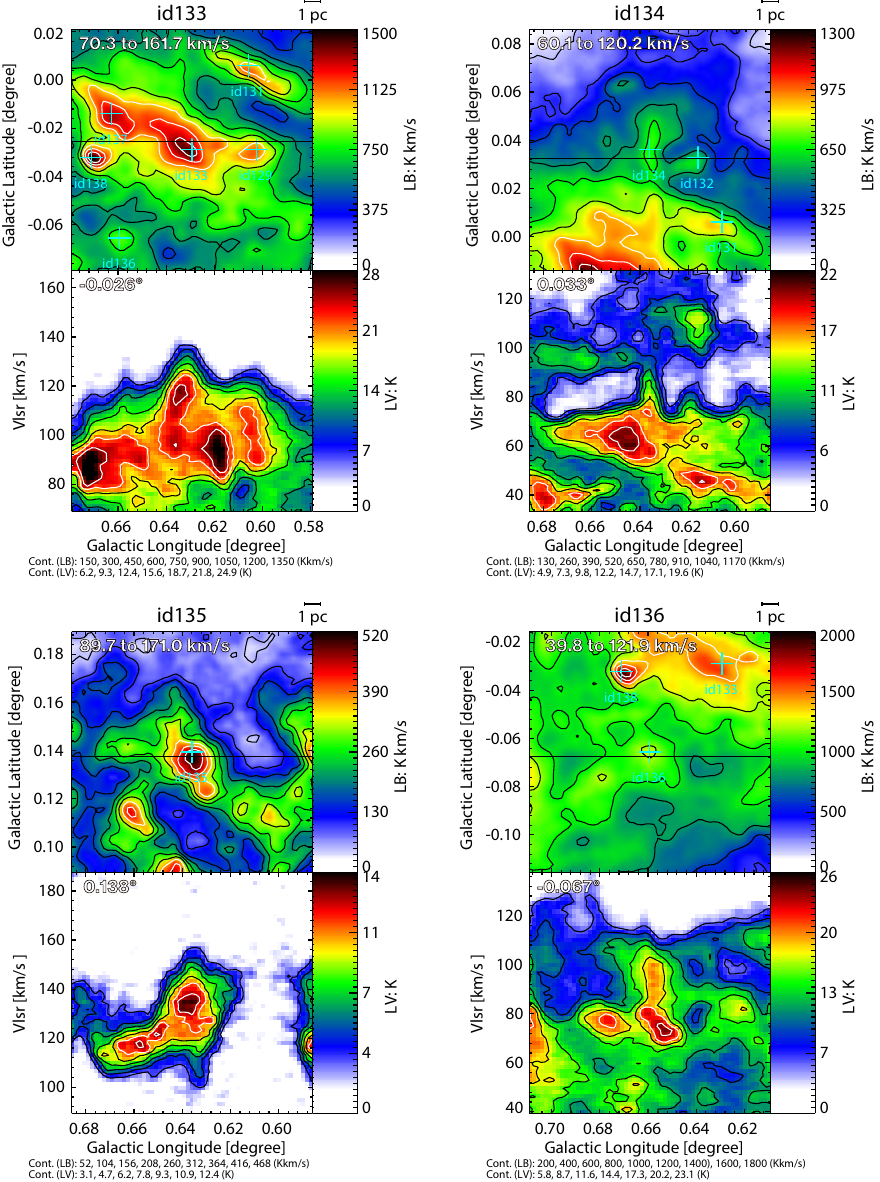}
\caption[]{Cont.}
\label{fig: HVCC}
\end{figure}

\addtocounter{figure}{-1}
\begin{figure}[p]
\centering
\includegraphics[]{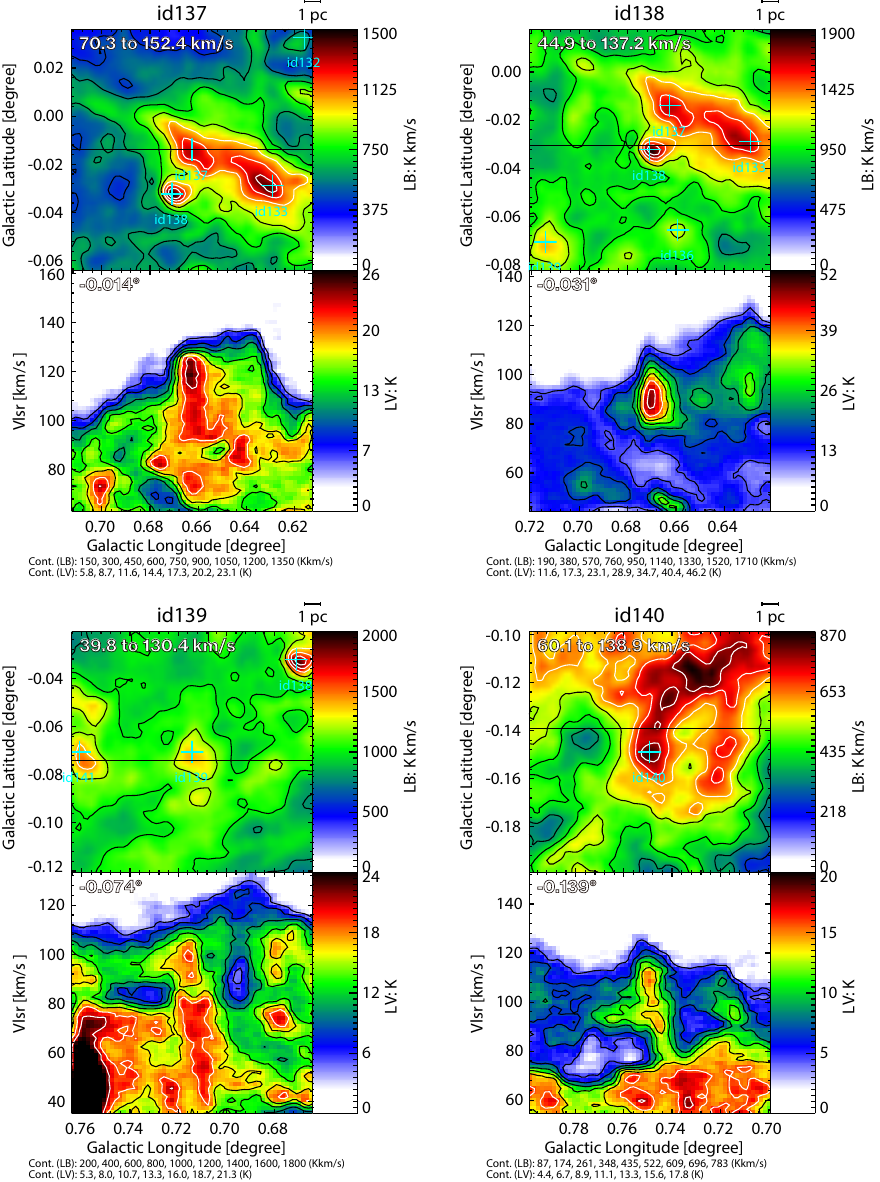}
\caption[]{Cont.}
\label{fig: HVCC}
\end{figure}

\addtocounter{figure}{-1}
\begin{figure}[p]
\centering
\includegraphics[]{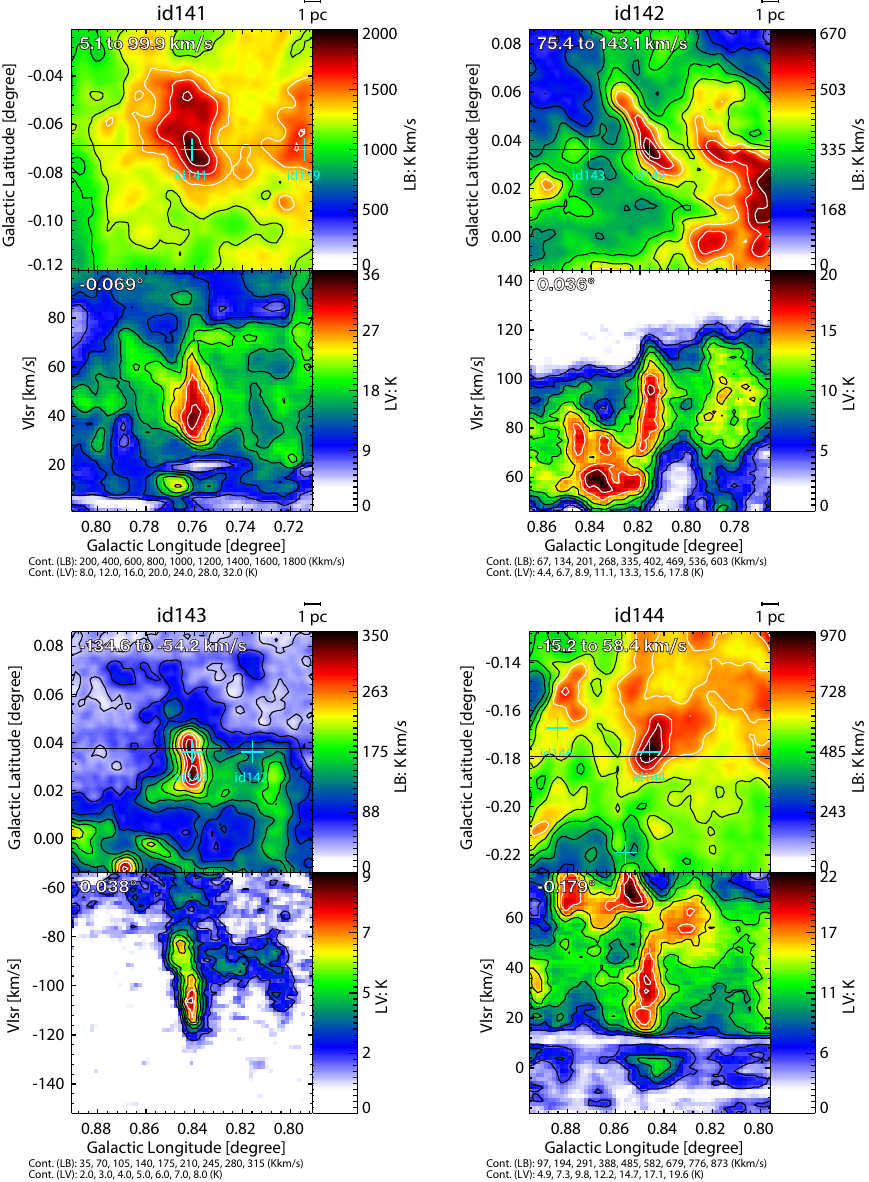}
\caption[]{Cont.}
\label{fig: HVCC}
\end{figure}

\addtocounter{figure}{-1}
\begin{figure}[p]
\centering
\includegraphics[]{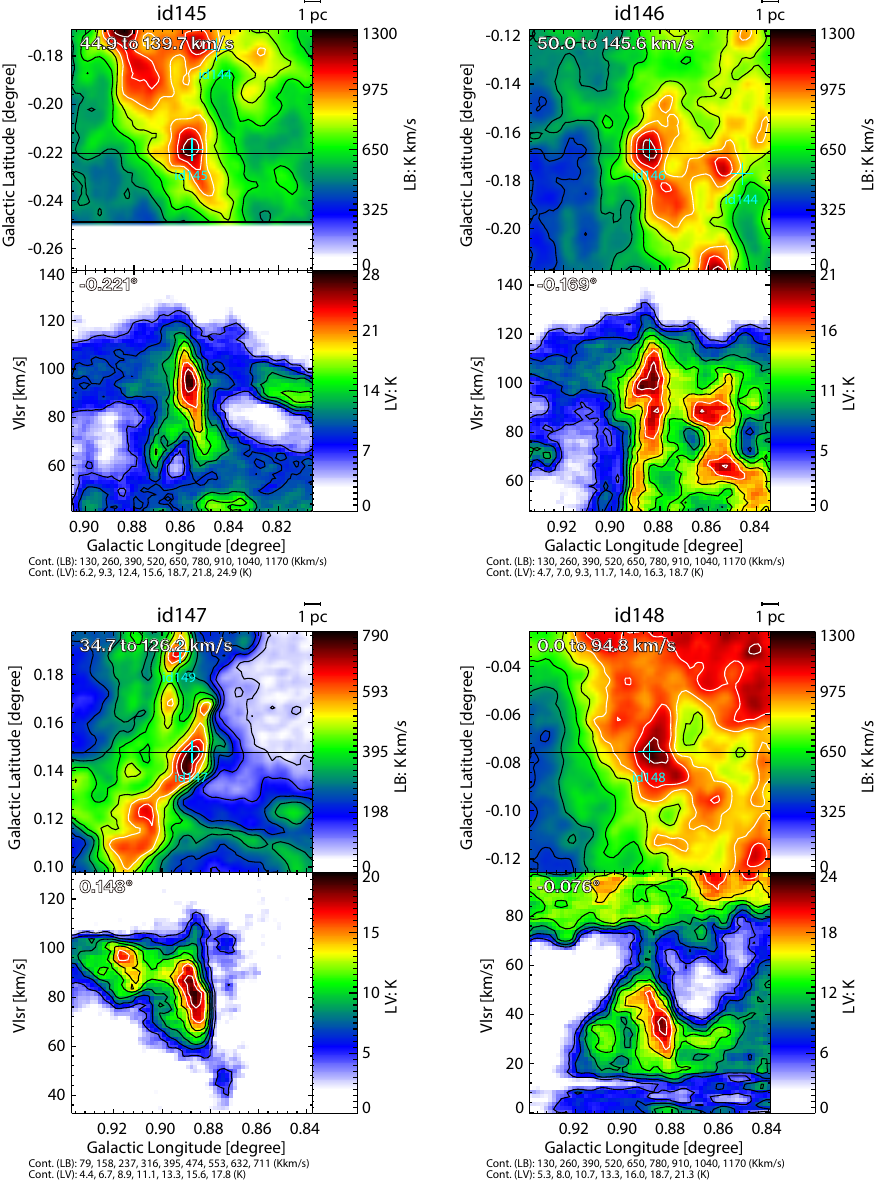}
\caption[]{Cont.}
\label{fig: HVCC}
\end{figure}

\addtocounter{figure}{-1}
\begin{figure}[p]
\centering
\includegraphics[]{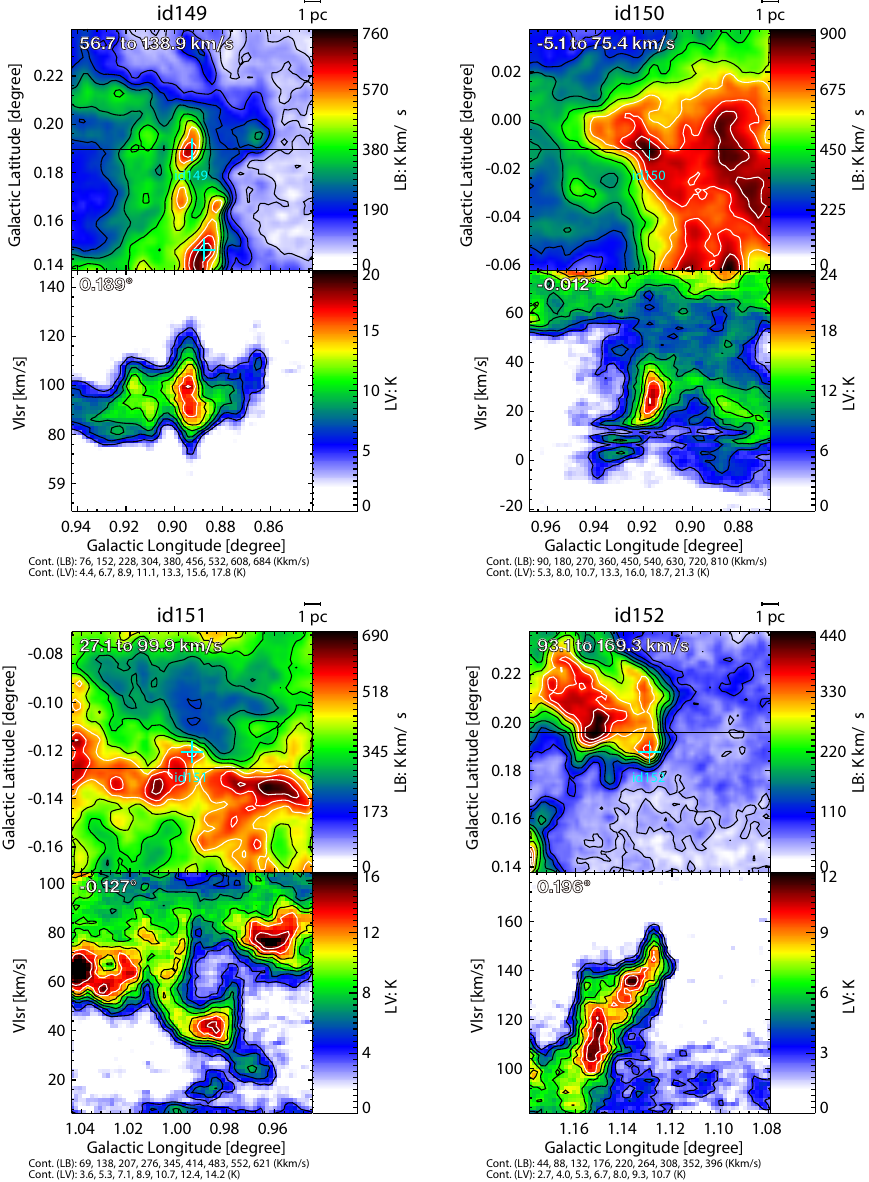}
\caption[]{Cont.}
\label{fig: HVCC}
\end{figure}

\addtocounter{figure}{-1}
\begin{figure}[p]
\centering
\includegraphics[]{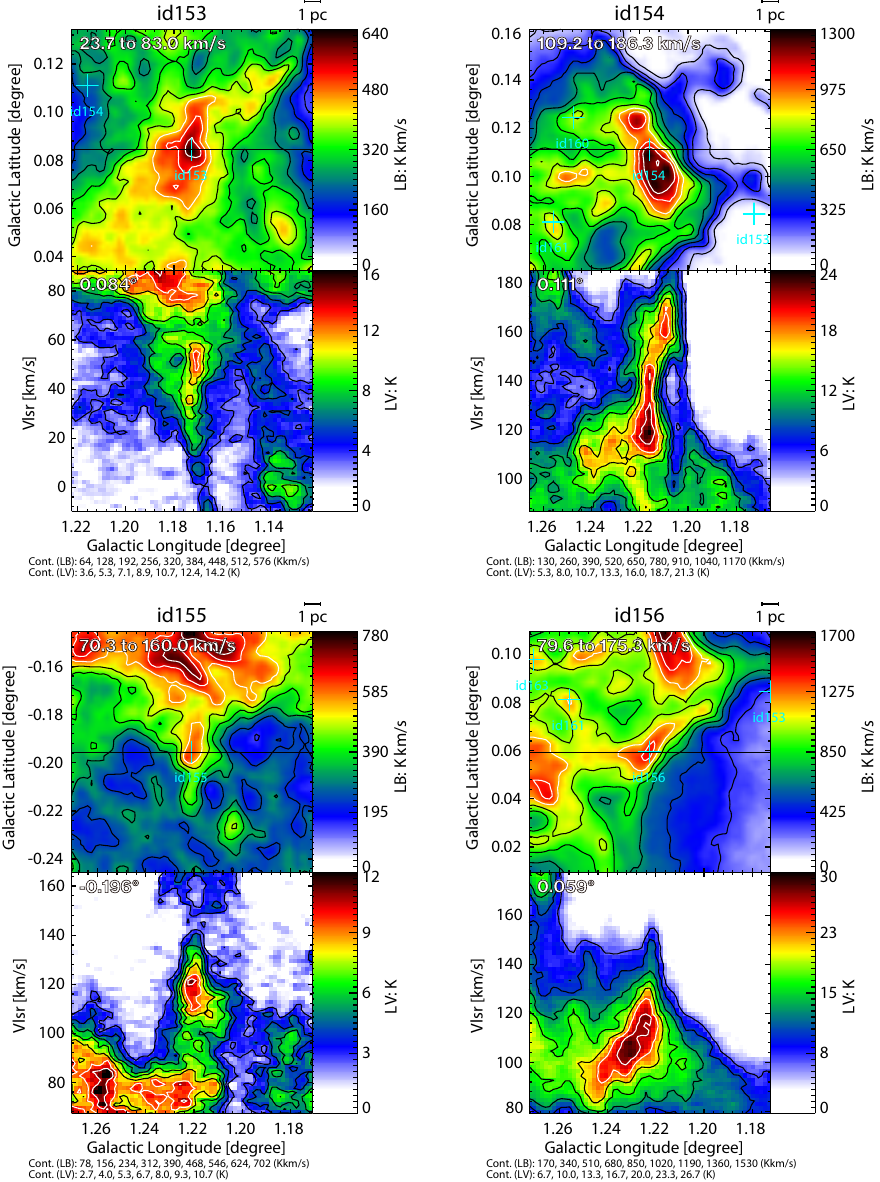}
\caption[]{Cont.}
\label{fig: HVCC}
\end{figure}

\addtocounter{figure}{-1}
\begin{figure}[p]
\centering
\includegraphics[]{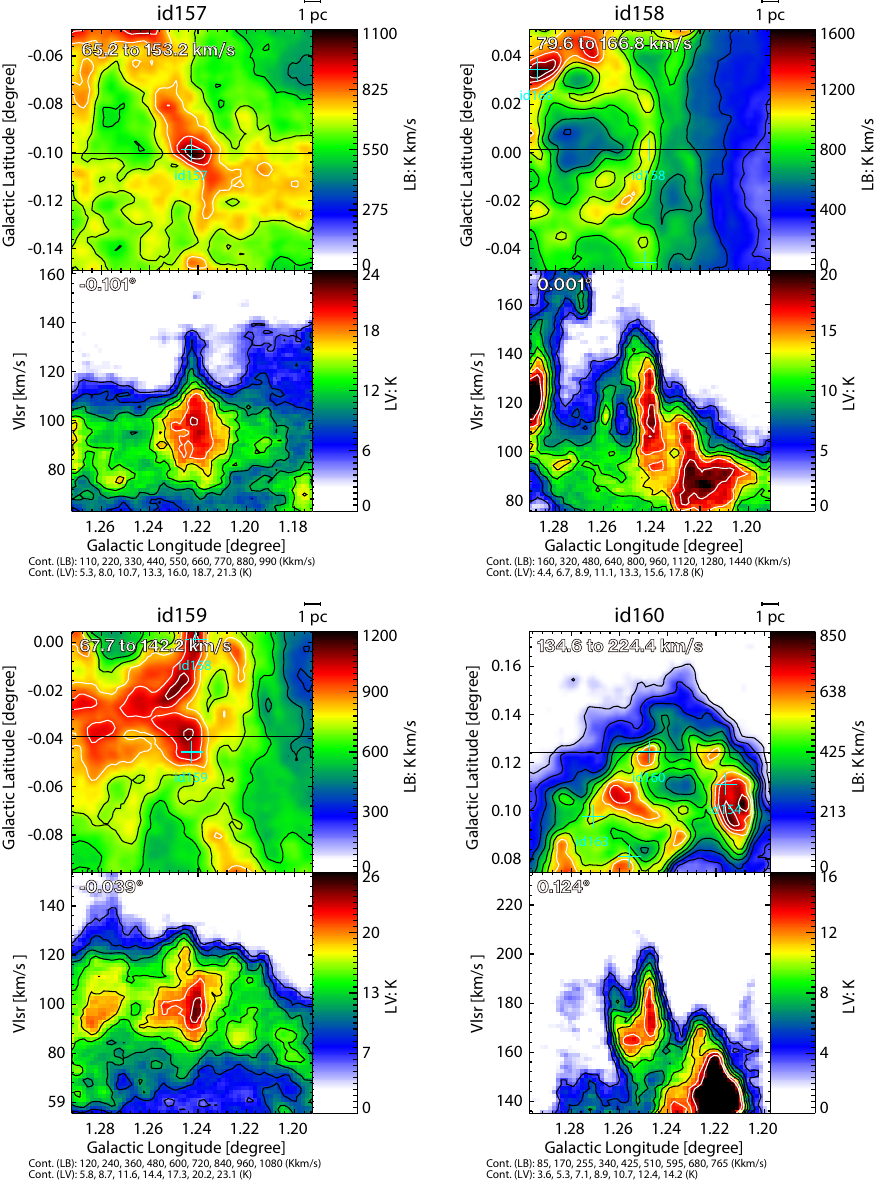}
\caption[]{Cont.}
\label{fig: HVCC}
\end{figure}

\addtocounter{figure}{-1}
\begin{figure}[p]
\centering
\includegraphics[]{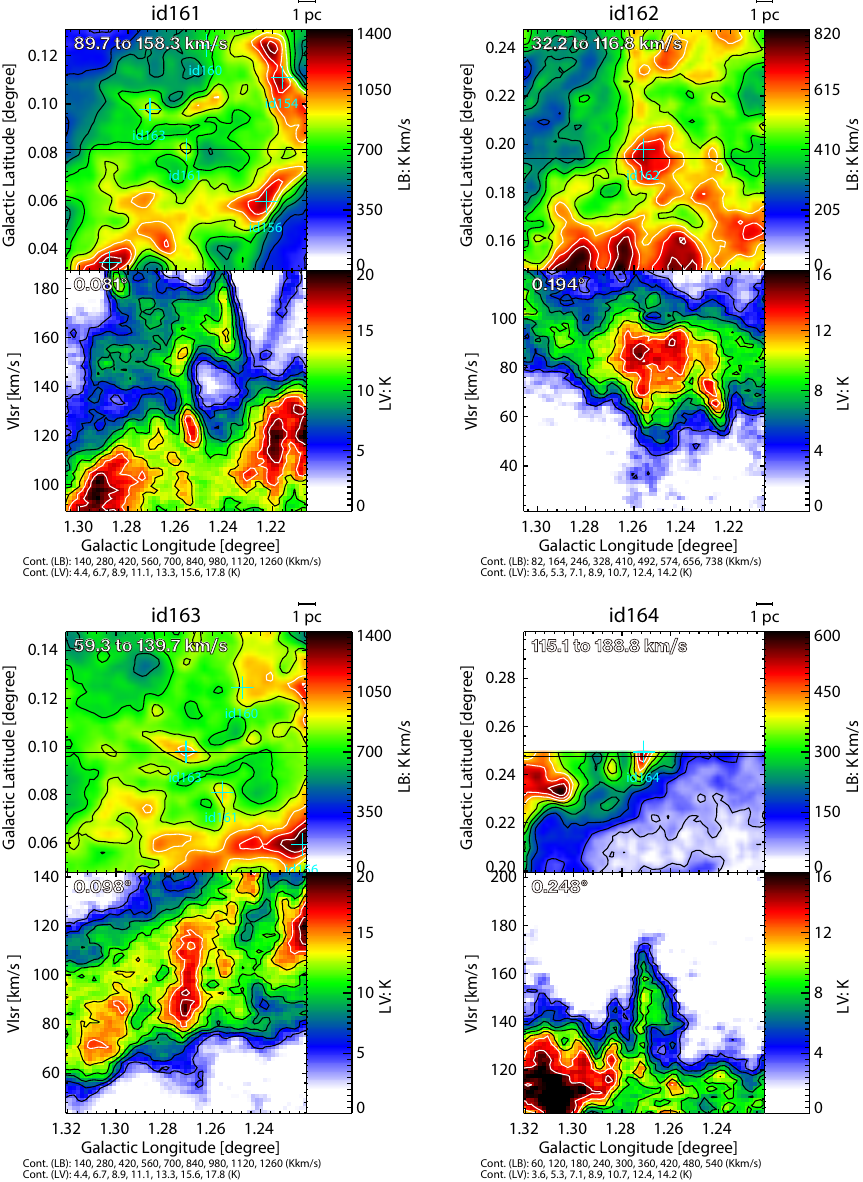}
\caption[]{Cont.}
\label{fig: HVCC}
\end{figure}

\addtocounter{figure}{-1}
\begin{figure}[p]
\centering
\includegraphics[]{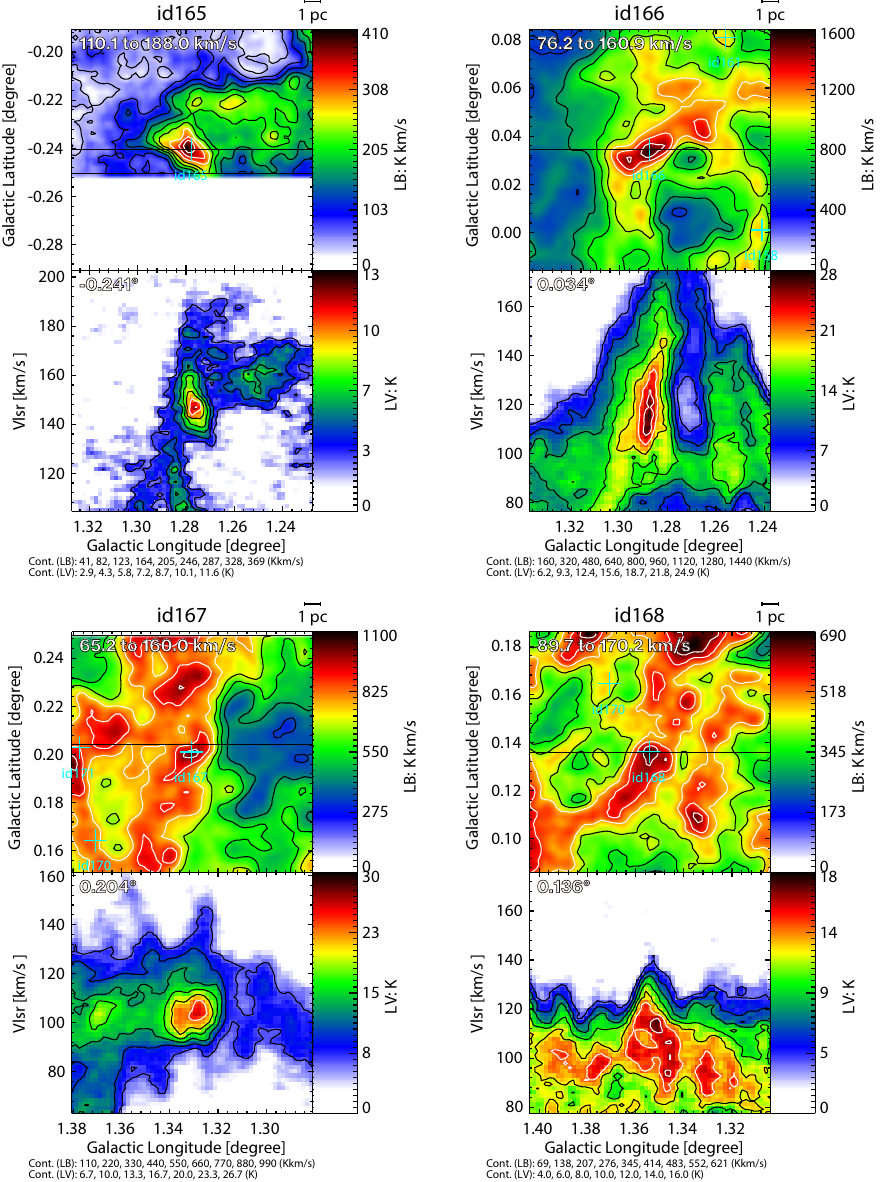}
\caption[]{Cont.}
\label{fig: HVCC}
\end{figure}

\addtocounter{figure}{-1}
\begin{figure}[p]
\centering
\includegraphics[]{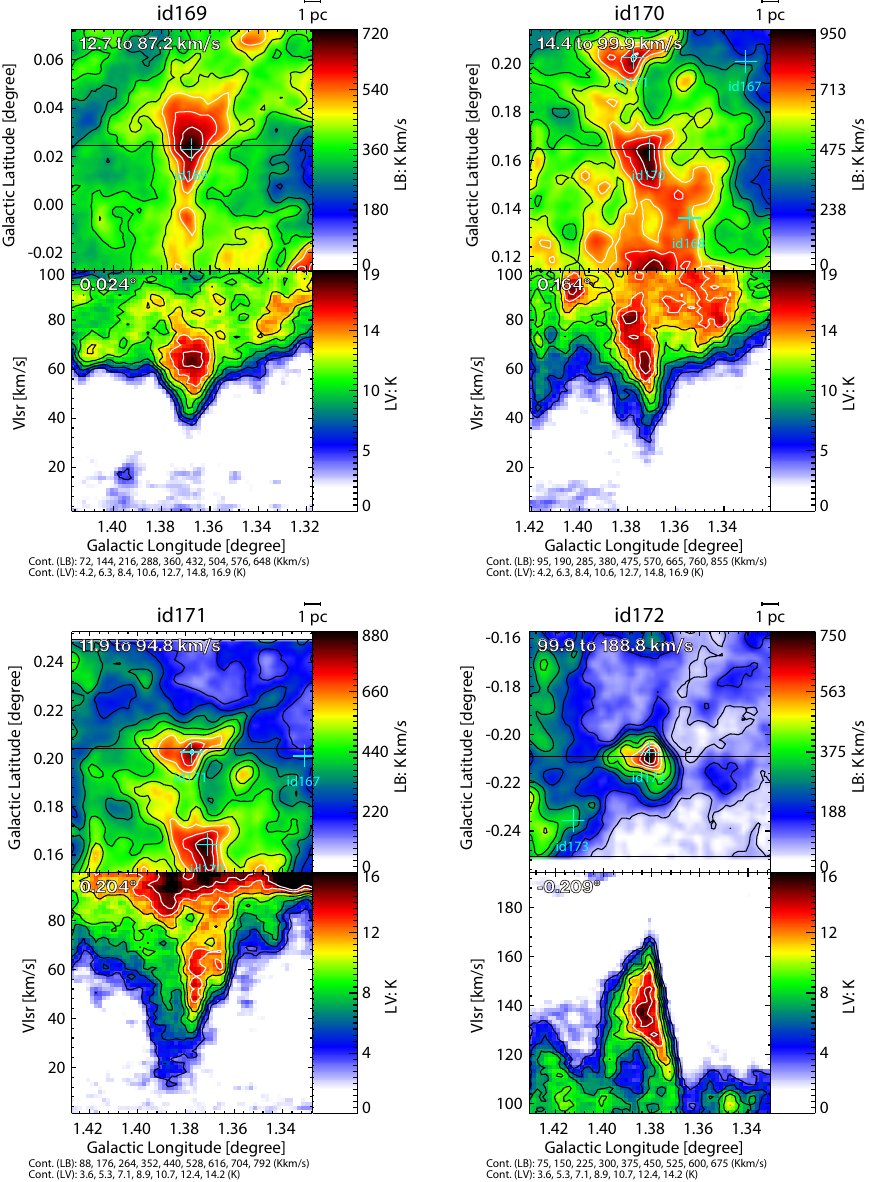}
\caption[]{Cont.}
\label{fig: HVCC}
\end{figure}

\addtocounter{figure}{-1}
\begin{figure}[p]
\centering
\includegraphics[]{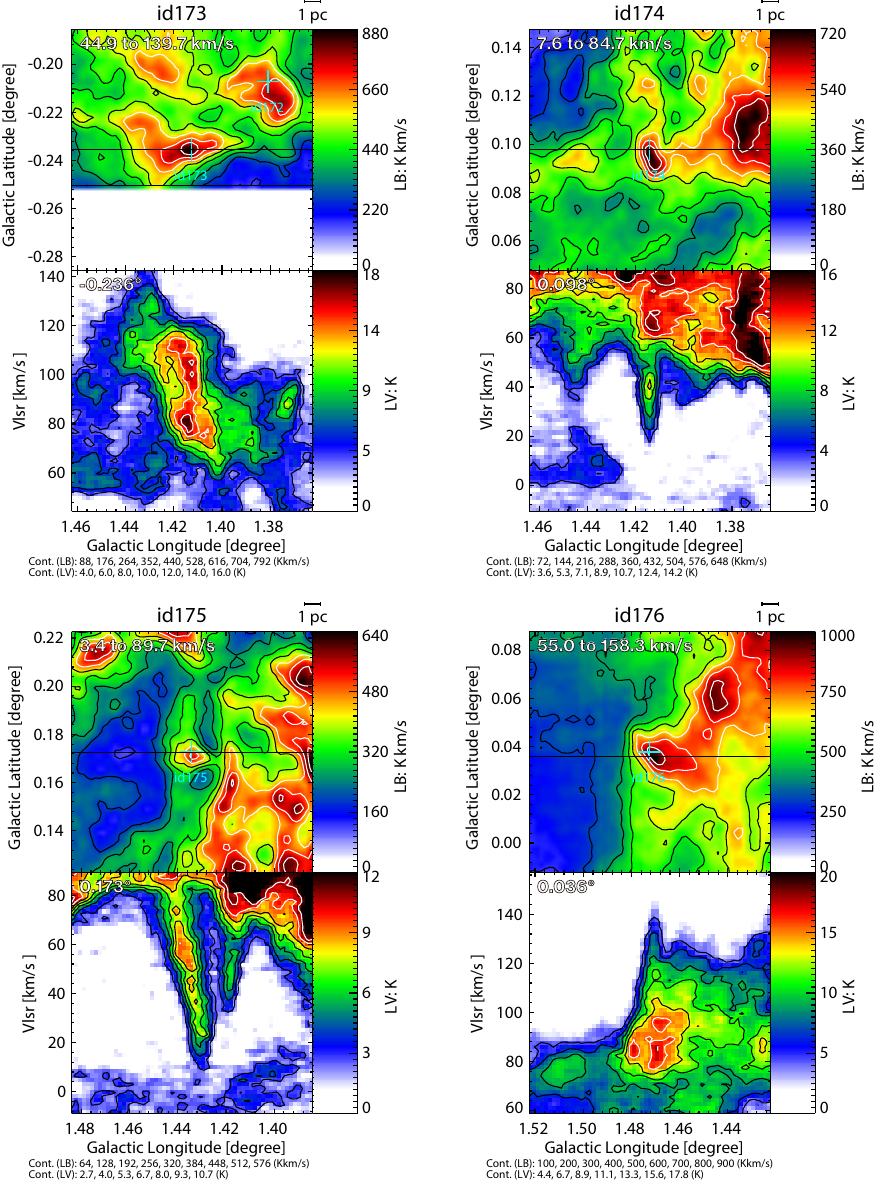}
\caption[]{Cont.}
\label{fig: HVCC}
\end{figure}

\addtocounter{figure}{-1}
\begin{figure}[p]
\centering
\includegraphics[]{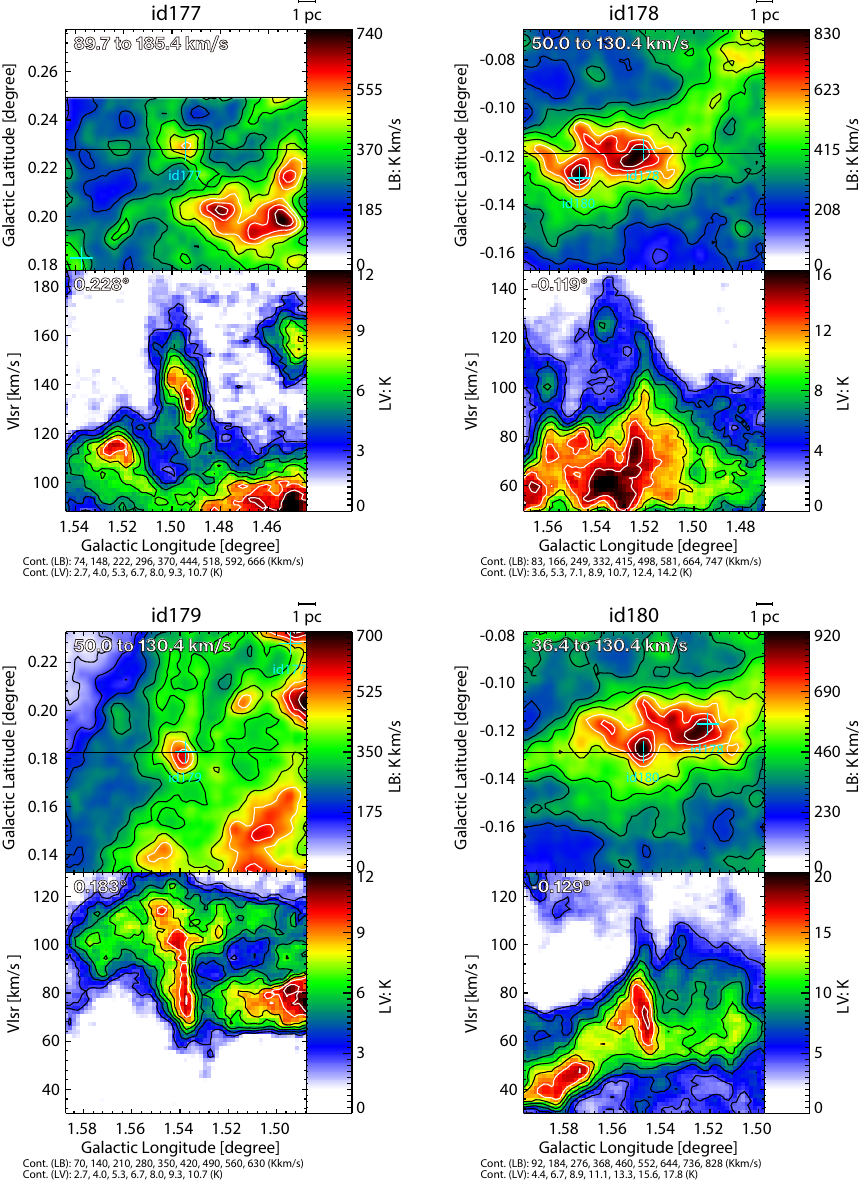}
\caption[]{Cont.}
\label{fig: HVCC}
\end{figure}

\addtocounter{figure}{-1}
\begin{figure}[p]
\centering
\includegraphics[]{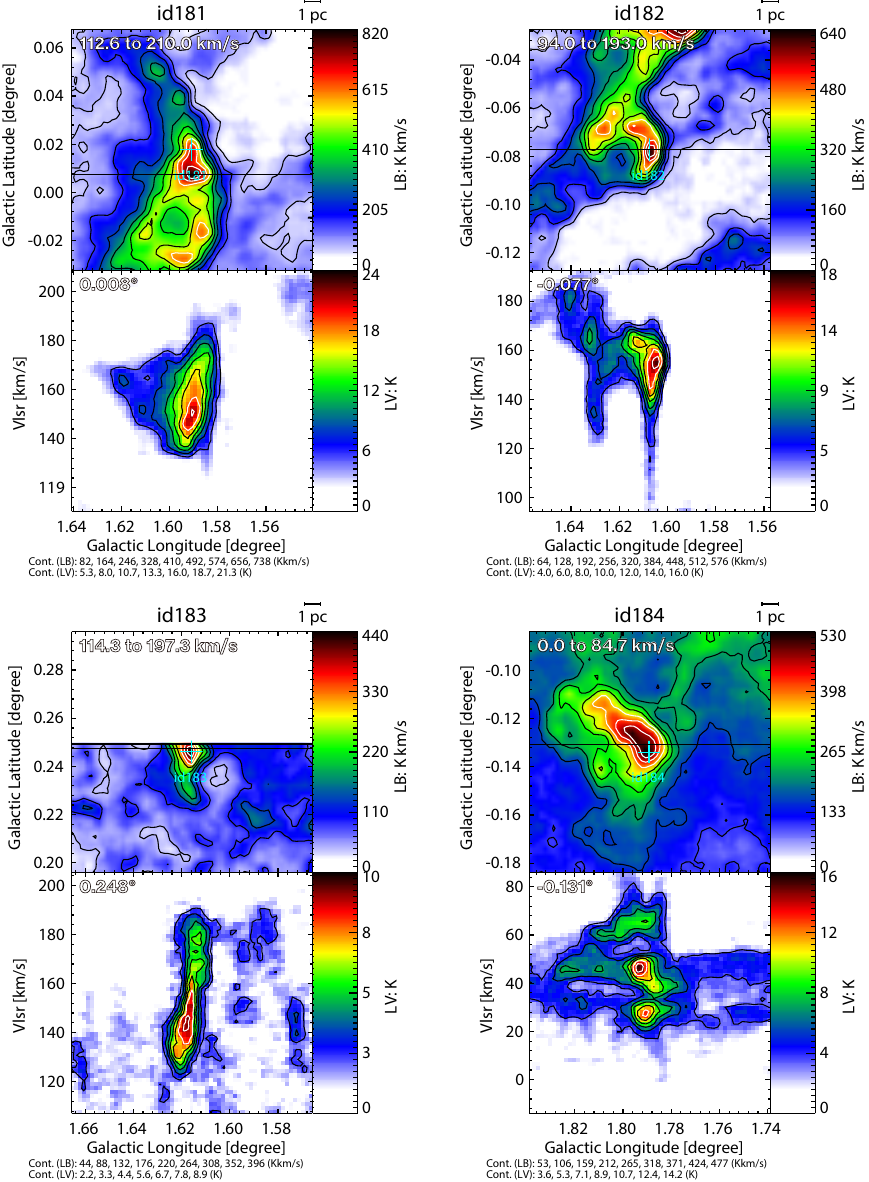}
\caption[]{Cont.}
\label{fig: HVCC}
\end{figure}


\clearpage


\begin{thebibliography}{99}
\bibitem[Arimoto et al.(1996)]{Arimoto96} Arimoto, N., Sofue, Y., \&\ Tsujimoto, T.\ 1996, \pasj, 48, 275
\bibitem[Buckle et al.(2009)]{Buckle09} Buckle, J., V., Hills, R., E., Smith, H., et al., 2009, \mnras, 399, 1026
\bibitem[Gillessen et al.(2009)]{Gillessen09} Gillessen, S., Eisenhauer, F., Trippe, S., et al.\ 2009, \apj, 692, 1075 
\bibitem[Dempsey, Thomas, \&\ Currie (2013)]{Dempsey13} Dempsey, J. T., Thomas, H. S, \&\ Currie, M. J.  2013, \apjs, 209, 8
\bibitem[Gravity Coll.(2022)]{Gravity22} Gravity Coll. 2022, A\&A, in press
\bibitem[Kaifu et al.(1972)]{Kaifu72} Kaifu, N., Kato, T., \&\ Iguchi, T., 1972, NPhS, 238, 105
\bibitem[Kumar \&\ Riffert (1997)]{Kumar97} Kumar, P. \&\ Riffert, H.  1997, \mnras, 292, 871
\bibitem[Liszt (2006)]{Liszt06} Liszt, H. S.  2006, A\&A, 447, 533 
\bibitem[Matsumura et al.(2012)]{Matsumura12} Matsumura, S., Oka, T., Tanaka, K., Nagai, M., Kamegai, K., \& Hasegawa, T. 2012, \apj, 756, 87
\bibitem[Nagai(2008)]{Nagai08} Nagai, M. 2008, PhD thesis, the University of Tokyo
\bibitem[Nomura et al.(2018)]{Nomura18} Nomura, M., Oka, T., Yamada, M., et al. 2018, \apj, 859, 29
\bibitem[Oka et al.(1998a)]{Oka98a} Oka, T., Hasegawa, T., Hayashi, M., Handa, T., \&\ Sakamoto, S. 1998, \apj, 493, 730
\bibitem[Oka et al.(1998b)]{Oka98b} Oka, T., Hasegawa, T., Sato, F., Tsuboi, M., \&\ Miyazaki, A., 1998, \apjs, 118, 455
\bibitem[Oka et al.(1999)]{Oka99} Oka, T., White, G., J., Hasegawa, T., et al., 1999, \apj, 515, 249
\bibitem[Oka et al.(2001a)]{Oka01a} Oka, T., Hasegawa, T., Sato, F., Tsuboi, M., and Miyazaki, A., 2001, \pasj, 53, 787
\bibitem[Oka et al.(2001b)]{Oka01b} Oka, T., Hasegawa, T., Sato, F., et al. 2001, \apj, 562, 348
\bibitem[Oka et al.(2007)]{Oka07} Oka, T., Nagai, M., Kamegai, K., Tanaka, K., and Kuboi, N., 2007 \pasj, 59, 15
\bibitem[Oka et al.(2008)]{Oka08} Oka, T., Hasegawa, T., White, G. J., et al. 2008, PASJ, 60, 429
\bibitem[Oka et al.(2011)]{Oka11} Oka, T., Nagai, M., Kamegai, K. et al., 2011. \apj, 732, 120
\bibitem[Oka et al.(2012)]{Oka12} Oka, T., Onodera, Y., Nagai, M., et al., 2012. \apjs, 201, 14
\bibitem[Oka et al.(2016))]{Oka16} Oka, T., Mizuno, R., Miura, K., \&\ Takekawa, S. 2016, \apj, 816, L7
\bibitem[Oka et al.(2017)]{Oka17} Oka, T., Tsujimoto, S., Iwata, Y., Nomura, M., \&\ Takekawa, S. 2017, NatAs, 1, 709
\bibitem[Parsons et al.(2018)]{Parsons18} Parsons, H., Dempsey, J., T., Thomas, H., S., et al. 2018, \apjs, 234, 22
\bibitem[Sashida et al.(2013)]{Sashida13} Sashida, T., Oka, T., Tanaka, K., et al., 2013, \apj, 774, 10
\bibitem[Seta et al.(1998)]{Seta98} Seta, M., Hasegawa, T., Dame, T. M., et al. 1998, ApJ, 505, 286
\bibitem[Seta et al.(2004)]{Seta04} Seta, M., Hasegawa, T., Sakamoto, S., et al. 2004, AJ, 127, 1098
\bibitem[Scoville(1972)]{Scoville72} Scoville, N., Z., 1972, \apjl, 175, L127
\bibitem[Sofue(1995)]{Sofue95} Sofue, Y., 1995, \pasj, 47, 527
\bibitem[Solarz et al.(2015)]{Solarz15} Solarz, A., Pollo, A., Takeuchi, T. T. 2015, A\&A, 582, 58
\bibitem[Solomon et al.(1987)]{Solomon87} Solomon, P. M., Rivolo, A. R., Barrett, J., \&\ Yahil, A. 1987, \apj, 319, 730
\bibitem[Shetty e al.(2012)]{Shetty12} Shetty, R., Beaumont, C. N., Burton, M. G., et al. 2012, \mnras, 425, 720
\bibitem[Sunada et al.(2000)]{Sunada00} Sunada, K., Yamaguchi, C., Nakai, N., et al.\ 2000, \procspie, 4015, 237
\bibitem[Takekawa et al.(2017)]{Takekawa17} Takekawa, S., Oka, T., Iwata, Y., Tokuyama, S., and Nomura, M., 2017, \apjl, 843, 11
\bibitem[Takekawa et al.(2019a)]{Takekawa19a} Takekawa, S., Oka, T., Iwata, Y., Tsujimoto, S., \&\ Nomura, M. 2019, \apj, 871, 1 
\bibitem[Takekawa et al.(2019b)]{Takekawa19b} Takekawa, S., Oka, T., Tokuyama, S., et al. 2019, PASJ, 71, 21
\bibitem[Takekawa et al.(2020)]{Takekawa20} Takekawa, S., Oka, T., Iwata, Y., Tsujimoto, S., \&\ Nomura, M. 2020, \apj, 890, 167
\bibitem[Tokuyama et al.(2019)]{Tokuyama19} Tokuyama, S., Oka, T., Takekawa, S., et al., 2019, PASJ, 71, S19
\bibitem[Tsujimoto et al.(2018)]{Tsujimoto18} Tsujimoto, S., Oka, T., Takekawa, S., et al., 2018, \apj, 856, 91
\bibitem[Tsujimoto et al.(2021)]{Tsujimoto21} Tsujimoto, S., Oka, T., Takekawa, S., et al., 2021, \apj, 910, 61
\bibitem[Williams \&\ Perry(1994)]{Williams94} Williams, R., J., R., Perry, J., J., 1994, \mnras, 269, 538
\bibitem[Yamada et al.(2017)]{Yamada17} Yamada, M., Oka, T., Takekawa, S., et al. 2017, \apj, 834, L3
\bibitem[Yamaguchi et al.(2000))]{Yamaguchi00} Yamaguchi, C., Sunada, K., Iizuka, Y., Iwashita, H., and Noguchi, T., 2000, \procspie, 4015, 614
\bibitem[Yokozuka et al.(2021)]{Yokozuka21} Yokozuka, H., Oka, T., Takekawa, S., Iwata, Y., \&\ Tsujimoto, S. 2021, \apj, 908, 246
\bibitem[Young \&\ Scoville(1991)]{Young91} Young, J. S., \&\ Scoville, N. Z, 1991, ARA\&A, 29, 581
\end{thebibliography}
\end{document}